\documentclass[aps,prl,preprint,longbibliography,groupedaddress]{revtex4-1}

\usepackage{bbm}
\usepackage{psfrag}
\usepackage{graphicx}
\usepackage{mathptm}
\usepackage{float}
\usepackage{amsmath}
\usepackage[verbose]{geometry}
 \usepackage{ulem}
\usepackage{relsize}
\usepackage{color}
\usepackage{hyperref}
\def\tup{\widetilde{u}^\prime}

\def\tup2{\widetilde{u}_2^\prime}

\def\up2{\widetilde{u_2^{\prime}}}

\def\tu2p{u_2^{\prime +}}

\begin{document}

\title
{
{ Transitional natural convection with conjugate heat transfer \\ over smooth and rough walls}
}

\author{Paolo Orlandi and Sergio Pirozzoli}
\affiliation{
"Sapienza" Universit\`a di Roma, Dipartimento di Ingegneria Meccanica e Aerospaziale, Dipartimento di Ingegneria Meccanica e Aerospaziale, Via Eudossiana 16, Roma (Italy)
}
\date{\today}

\begin{abstract}
We study turbulent natural convection in enclosures with conjugate heat transfer. The simplest way to increase the heat transfer in this flow is through rough surfaces. In numerical simulations often the constant temperature is assigned at the walls in contact with the fluid, which is unrealistic in laboratory experiments. The DNS (Direct Numerical Simulation), to be of help to experimentalists, should consider the heat conduction in the solid walls together with the turbulent flow between the hot and the cold walls. Here the cold wall, $0.5h$ thick (where $h$ is the channel half-height) is smooth, and the hot wall has two- and three-dimensional elements of thickness $0.2h$ above a solid layer $0.3h$ thick.  The independence of the results on the box size has been verified. A bi-periodic domain $4h$ wide allows to have a sufficient resolution with a limited number of grid points. It has been found that, among the different kind of surfaces at a Rayleigh number $Ra \approx 2 \cdot 10^6$, the one with staggered wedges has the highest heat transfer. A large number of simulations varying the $Ra$ from $10^3$ to $10^7$ were performed to find the different ranges of the Nusselt number ($Nu$) relationship  as a function of $Ra$. Flow visualizations allow to explain the differences in the $Nu(Ra)$ relationship.  Two values of the thermal conductivity were chosen, one corresponding to copper and the other ten times higher. It has been found that the Nusselt number behaves as $Nu=\alpha Ra^\gamma$, with $\alpha$ and $\gamma$ independent on the solid conductivity, and dependent on the roughness shape. 
\end{abstract}

\maketitle

\section{Introduction}

Rayleigh-Benard convection is the basic flow to understand several phenomena
occurring in nature and in technological applications. This flow
has been largely studied theoretically, experimentally and numerically.
It is impossible to reference  the large number of research papers, these can
be found in the review papers by \citet{chilla_12}
and by \citet{lohse_10}.
Before dealing with turbulent flows it is worth               
understanding the transition from a laminar conducting condition with
constant layers of temperature parallel to the cold and hot walls,
to a condition where the mutual interaction
between temperature and velocity fields generates complex flow structures.
In practical applications the solid walls are thick and made
of materials with thermal conductivity depending on the applications of interest~\citep{incropera_11}.
At low conductivity, the temperature decreases considerably in the solid,
while at high conductivity a mild decrease of the temperature
in the solid is followed by a large
reduction in the fluid. In laminar flows the linear profiles of temperature 
in the solid and in the fluid have slopes linked to the thermal diffusivity 
of the solid and of the fluid.  This condition is desired in the insulation
of buildings, in particular for those covered by glass surfaces. 
A thin gap between the two glass layers prevent
any motion of the air. 

In other circumstances 
the increase of the heat transfer is desired,
therefore, once the solid material has been chosen, the heat transfer
can be controlled by acting on the shape of the surface between solid and
fluid or by letting the fluid to move inside the gap.             
In both ways changes on the flow structures are promoted. 
The different shapes of the thermal and flow structures
reported in the review paper by \citet{bodenschatz_00}
suggest that, in the presence of smooth walls, for a box with large 
aspect ratio $\Gamma=l/H$ ($H$ is
the gap thickness and $l$ the lateral dimensions of the plates)
the first transition generates longitudinal rollers.
The transitional Rayleigh number ($Ra=g\alpha\Delta T H^3/k\nu$)
in the experiments agrees 
with the value ${ Ra_c}=1707$ found theoretically by~\citet{chandrasekhar_61}.
In the expression of $Ra$ $\nu$ is the kinematic viscosity, $k$
the thermal diffusivity of the fluid, $g$ the gravity acceleration,
$\alpha$ the thermal expansion coefficient and $\Delta T$ the
temperature differences of the two surfaces in contact with
the fluid separated by the distance $H$.
The  spanwise size of the rollers is about $2H$, suggesting that in 
direct numerical simulations (DNS) in order
to capture one  convective cell the box size should have at least an
aspect ratio $\Gamma=2$.  In these circumstances the DNS 
require limited computational resources to have
the sufficient resolution to reproduce the transition from
a laminar to an unsteady or to  a fully turbulent regime.
However it is necessary to demonstrate that the flow structures
for $Ra > Ra_c$ do not vary with $\Gamma$, and that 
in fully turbulent conditions ($Ra >> Ra_c$), the flow statistics
are not affected.

In the fully turbulent regime theoretical and experimental results
found that the Nusselt number should vary as $Nu=Q/(k\Delta T/H)\approx Ra^n$
($Q$ is the total heat flux), with the exponent $n$ varying with 
the Rayleigh number. 
The present study is focused on the transitional and on the first
part of the fully turbulent regime, that is for $Ra_c< Ra < 10^{6}$. 
For smooth walls \citet{chandrasekhar_61} reported the data of  
Silveston's experiment~\citep{silveston_58}, confirming the agreement between
the experimental transitional Rayleigh  and the theoretical $Ra_c=1707$ value.
This value was found for rigid boundaries,  corresponding to the most
unstable wave length $a=3.117H$.
By increasing the Rayleigh number, $Nu$ was found 
to grow as $1+(Ra-Ra_c)/Ra_c)^{1/2}$.
In Chandrasekhar's book pictures of the convective cells were also shown, 
these, being generated between two large circular 
plates were  affected by the side walls. 
However, it was clear that the width of the cell
was about twice the dept of the layer. The picture of the cells
reported in figure 15 of Chandrasekhar's book were not so clear as
those given by in figure 5 of \citet{bodenschatz_00},
where the straight rolls at a certain value of $\epsilon=(Ra-Ra_c)/Ra_c$ 
showed a further instability
of wave length smaller than the size of the cell.

From these papers dealing with smooth walls it was clear  the mechanism 
of multiple instabilities leading to a fully
turbulent state.  Much less is known 
in the presence of rough surfaces. 
Numerical simulations in two spatial dimensions
\citep{shishkina_11} and in a cylindrical container \citep{stringano_06}
in cylindrical container did not fully clarify the effects of the
shape of the roughness surfaces on the flow structures. In fact 
they did not consider surfaces of different shapes. 
In numerical and laboratory experiments flow visualization 
can provide insight on the dependence of the flow structures on the shape
of the surface. In real experiments the visualizations require 
particular care, especially near solid walls. In numerical
experiments instead any quantity can be easily visualized and
the relative statistics can be calculated. From previous
studies it was found a strong flow anisotropy near the surface which
reduces moving towards the  centre of the container.
Temperature and velocity spectra may provide insight on 
the flow anisotropy, in different regions of the container. 
In cylindrical containers, frequency spectra do not allow to have
a direct link with the flow structures, although these were used
by \citet{camussi_99} to compare
their results with those in the experiments. This was emphasized
in the review article of \citet{kadanoff_01} on
the structures and scaling in turbulent heat flows.
On the other hand, one-dimensional longitudinal and transverse
spectra can be calculated at different distances from the walls by using
a computational box with two periodic directions. 
The global effect
of the shape of the surface should affect the $Nu(Ra)$ relationship.
\citet{roche_01} observed the ultimate
regime ($Nu\approx Ra^{1/2}$) for $Ra>10^{12}$ by inserting a $V$-shaped 
surface in a cylindrical container. So far, this regime has never been clearly
observed in the presence of smooth walls. \citet{du_00} 
compared the $Nu(Ra)$ behavior 
in a cylindrical cell in the presence
of smooth and rough surfaces, and for $10^9 < Ra < 10^{11}$ they 
observed the same $Nu\approx Ra^{2/7}$ law,  however with an increase 
of more than $76\%$ in the presence of rough walls.
Through flow visualizations they found that the dynamics
of the thermal plumes over rough surfaces is different from
that near smooth surfaces. 
\citet{stringano_06} performed numerical simulations 
for a set-up very similar to that used by \citet{du_00}, and observed
a different $Nu=cRa^n$ relationship for smooth and rough surfaces. 
In particular, for $Ra< 10^7$ they observed 
the same behavior for smooth and grooved walls, 
whereas at higher $Ra$ a steeper growth (with $n=0.37$) was observed 
for rough walls than for smooth walls (having $n=0.31$).
Flow visualizations near the rough surfaces 
confirmed a plume dynamics similar to that of
\citet{du_00}.
\citet{ciliberto_99}  considered a rectangular contained 
with copper walls covered with glass spheres, and
obtained the same value of $n$ as for smooth copper plates, but
smaller multiplicative constant $c$.  The reason
should be related to the insulating material of the roughness elements.  
From these results and from figure 1 of \citet{stringano_06} it is 
clear that more investigations are necessary to better understand
the influence of the shape and of the material of rough surfaces.

The present study is limited to the transitional regime.
Numerical simulations have been performed
in a box with two homogeneous directions in order to
have the spectra which together with flow visualizations
allow to get a clear picture of the flow and thermal structures.
The study has been extended
to flows with the hot wall corrugated by two- and three-dimensional
elements. For this purpose
an efficient numerical method to reproduce flows past complex 
geometries has been used. This tool allows to understand the 
complex flow physics in the
presence of rough walls through the evaluation of any 
flow quantity. The numerical method 
developed by \citet{orlandi_06}
has been validated by comparing the results with those
obtained by \citet{burattini_08}
in experiments designed to reproduce the numerical simulations.
The numerical and laboratory experiments considered turbulent
flows in channels with one rough wall with 
two-dimensional transverse square bars with different
values of $w/k$ ($w$ is the separation distance between square bars
with height $k$) and different values of the Reynolds number.
The method has been extended
to simulate conjugate heat transfer including the heat diffusion in the
solid layer with smooth and rough walls Orlandi etal 2005 and \citet{orlandi_16}.
At high Reynolds numbers part of the viscous,
together with the non-linear terms, can be  treated explicitly 
to save computational time.
At low $Re$ numbers the viscous
restriction $\Delta t/(Re\Delta x^2)<1/2$ is penalising the
efficiency of the simulation, therefore the implicit
treatment of the viscous terms is mandatory. 
However, at low $Re$ coarse grids can be used.            
Dealing with complex surfaces  the resolution is dictated by
the requirement to resolve the wall corrugation with a
sufficient number of points.

\section{ Physical model and numerical method }

\subsection{ Governing equations  }

The momentum and continuity equations for  the perturbation
velocity components and temperature for
incompressible flows and within the Boussinesq approximation,
are reported by \citet{chilla_12}     

\begin{equation}
\frac{\partial U_i}{\partial t} 
+\frac{\partial U_i U_{j}}{\partial x_{j}} =
-\frac{1}{\rho_0 }\frac{\partial P}{\partial x_i}
+\nu\frac{\partial^2 U_i}{\partial x_j^2}
+g\alpha(T-T_{ref})\delta_{i2}
;\ \ \ \ \ \ \ \ \ \frac{\partial U_j}{\partial x_j}= 0 ,
\label{eq1}
\end{equation}
where \noindent $U_{i}$ are the components
of the velocity vector in the $i$ directions,
$p$ is the pressure variations about the hydrostatic
equilibrium profile, $x_1$, $x_{2}$ and $x_{3}$ are the streamwise,
wall-normal and spanwise directions respectively, $\rho_0$ is the
reference density determined by $-\partial \rho_0/\partial x_2+\rho_0 g=0$,
$g$ the gravity acceleration, $\nu$ the kinematic viscosity,
$\alpha$ the isobaric thermal expansion coefficient, and $T$
the total  temperature which is obtained by 

\begin{equation}
\frac{\partial T}{\partial t} + 
\frac{\partial T U_{j}}{\partial x_{j}} =
\frac{\partial}{\partial x_{j}} 
\kappa_M
\frac{\partial T}{\partial x_{j}} \
\label{eq2}
\end{equation}
here $\kappa_M$ is the thermal diffusivity.
In this study not only the fluid is considered
but also the solid, therefore $\kappa_M$ can be either
the fluid thermal diffusivity $\kappa_F=k/(\rho_0 C_p)$,
or the thermal diffusivity of the material
used for the solid walls, say $\kappa_S$.
Dimensionless variables are used by introducing as
reference quantities the distance between
the upper smooth wall and the center of the box $h$, 
the free-fall velocity $U_f=\sqrt{g \alpha \Delta T h}$,
and the reference pressure $\rho_0 U_f^2$.
By introducing the dimensionless temperature $\theta=(T-T_{ref})/\Delta T$
the governing equations become

\begin{equation}
\frac{\partial u_i}{\partial t} 
+\frac{\partial u_i u_j}{\partial x_{j}} =
-\frac{\partial p}{\partial x_i}
+\frac{1}{Re}\frac{\partial^2 u_i}{\partial x_j^2}
+\theta\delta_{i2}
;\ \ \ \ \ \ \ \ \ \frac{\partial u_j}{\partial x_j}= 0 ,
\label{eq3}
\end{equation}

\begin{equation}
\frac{\partial \theta}{\partial t} + 
\frac{\partial \theta u_j}{\partial x_{j}} =
\frac{\partial}{\partial x_{j}} 
\frac{1}{RePr_M}
\frac{\partial \theta}{\partial x_{j}} \
\label{eq4}
\end{equation}
\noindent where $Re=h U_f/\nu$, and $2\Delta T$ is the assigned 
temperature difference between the
outer boundary of the lower hot wall and of the upper cold wall, 
respectively at $x_2=1.5h$ and at $x_2=-1.5h$.
The Reynolds number is linked to the Rayleigh number through 
the relation $Ra=Re^2$
only for perfectly conducting walls or with zero thickness.
In the region with the fluid $Pr_M=Pr_F=1$ has been assumed. 
For the materials of the solid
two different values of $Pr_M=Pr_S=\kappa_S/\kappa_F$ have been
set, one for the copper $Pr_S=0.134$ and the other for
a material ten times more conductive, with $Pr_S=0.0134$.
In the case of smooth walls and for
the surface covered by wedges, simulations
have also been performed with low conductivity materials $Pr_S=1.34$.
The total heat flux $Q$ (which is constant along the $x_2$ direction)
is given by the sum of the turbulent
$Q_T=<u_2\theta^\prime>$ and the conductive
contribution $Q_F=\frac{1}{RePr_M}
\frac{\partial <\theta>}{\partial x_{2}}$. In the present
condition $Q$ is also equal to $Q_F=\frac{1}{RePr_S}
\frac{\partial <\theta>}{\partial x_{2}}$ evaluated in the 
thick upper solid wall.
The overall physical box  has dimensions
$L_1=4h, L_2=3h$ and $L_3=4h$.
The upper (smooth) smooth wall is at $x_2=h$, overlaid by a 
layer of material with thickness $0.5h$.
The lower wall is rough, with the plane of crests at $x_2=-h$,
the roughness thickness is $0.2h$, with a further layer of material  
underneath with thickness $0.3h$.

At $t=0$, the non-dimensional temperature $\theta$ has been set equal
to $\pm 1$ in the two solid layers from $x_2=\pm 1.5$ to
$x_2=\pm 1.$. In the fluid layer the
profile of $\theta$ is linear. Random disturbances have
been assigned to the velocity field with amplitude
equal to $0.02 U_f$. After a certain time, 
depending on $Re$, the disturbances 
may grow or decrease.
For laminar flows a large-scale cell 
forms, and the amplitude of the velocity fluctuations tends to zero.
In the transitional regime, after a transient when
the initial random fluctuations as well as  the $rms$ values
decrease, the fluctuations 
grow exponentially for a short time lapse.
The transitional Reynolds number is evaluated by extrapolating to
zero the values of the growth rate at different $Re$.

The aspect ratio in the homogeneous directions can affect
the results. \citet{kerr_96}, in a range of $Ra$
number similar to that here considered used a
domain $L=L_1=L_3=12h$, wider than used here. This aspect
ratio in the presence  of roughness elements should require too many
grid points, if at least $10$ points are
used to represent the surface of each element.
The present simulations are devoted to investigating
how the shape of the heated rough surface changes the heat transfer, 
the velocity and temperature structures  with respect to
those generated by smooth walls. Therefore, it is mandatory
to prove that for smooth walls the  results of
simulations in a wider domain do not change with respect to those
performed in the square domain of size $L=4h$.

The system of equations \eqref{eq3}, \eqref{eq4}
with periodic conditions in the direction $x_1$ and $x_3$
have been discretized by  a finite difference scheme 
combined with the immersed boundary technique
described by \citet{orlandi_16}.
The equations were integrated in time 
with a third-order Runge Kutta low-storage
scheme for the explicit nonlinear terms.
In the previous simulations
an implicit Crank Nicholson procedure based
on a factorization of the wide banded matrices
was used. For parallel implementation the computational box 
was subdivided in layers parallel 
to the walls by limiting
the use of a large number of processors. On the
other hand, by subdividing the domain in
rectangular pencils a greater number of processors can be
used. When the factorization is applied the CPU time necessary to
transfer data among the processors increases.
The implicit treatment of the viscous terms is required 
at low $Re$ to avoid the viscous stability condition
$\Delta t/(Re \Delta x^2)<2$, which is more restrictive than the 
Courant-Friedrichs-Lewy (CFL) restriction $|U_i\Delta t/\Delta x_i|<1$.
However, at low $Re$, a coarse grid reduces the amount
of data transfer among the processors.
At high $Re$, a fine grid is necessary to resolve all
the scales and  the computational time is kept reasonable by using 
an explicit scheme for the viscous terms in the momentum equations.
For the $\theta$ equation the viscous terms in the homogeneous
directions are handled explicitly. Instead the term in the $x_2$ direction,
due to the large variations of the temperature gradients at the
solid/fluid interface, is handled implicitly.

The assumption of a cold wall at the top 
generates a heat flux from the rough
to the smooth wall.
Equation~\eqref{eq4} holds in the region with
the fluid and does not require any
condition at the interface between the fluid
and the solid because it is solved
together with  the transport equation for $\theta$ in the solid layers.
In these layers $u_j=0$, and only the temperature equation is solved,
with $Pr_M=Pr_{S}$.
At the interface the heat flux in the fluid and in the solid sides
must be equal; to reach this goal
it is convenient to define the $\theta$ at the same
location of $u_2$ and the diffusivity at the
centre of the cell. To account for the difference between
fluid and solid a function $f(x_2)$ is used equal
to $1$ in the solid and to $0$ in the fluid.
The convective term and $Pr_F$ are then multiplied by $(1-f(x_2))$
and $Pr_{S}$ by $f(x_2)$, and the transport equation for $\theta$ becomes

\begin{equation}
\frac
{\partial \theta}{\partial t} + 
(1-f)\frac{\partial \theta U_{j}}{\partial x_{j}} =
\frac{\partial}{\partial x_{j}} \frac{1}{Re \; (Pr_F(1-f)+Pr_Sf)}
\frac{\partial \theta}{\partial x_{j}} 
\end{equation}

\begin{figure}
\centering
(a)
\includegraphics[width=4.0cm,angle=270,clip]{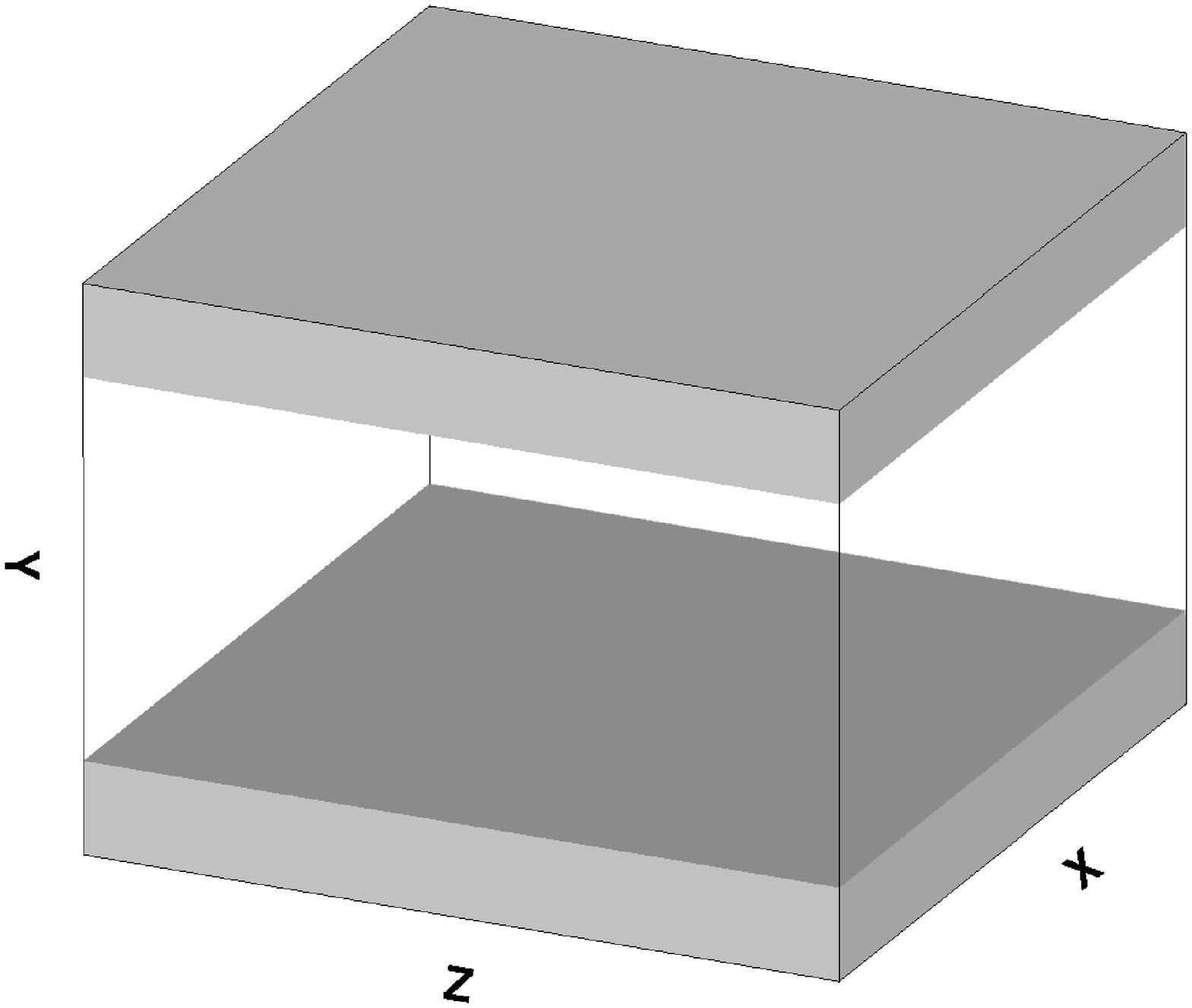}
(b)
\includegraphics[width=4.0cm,angle=270,clip]{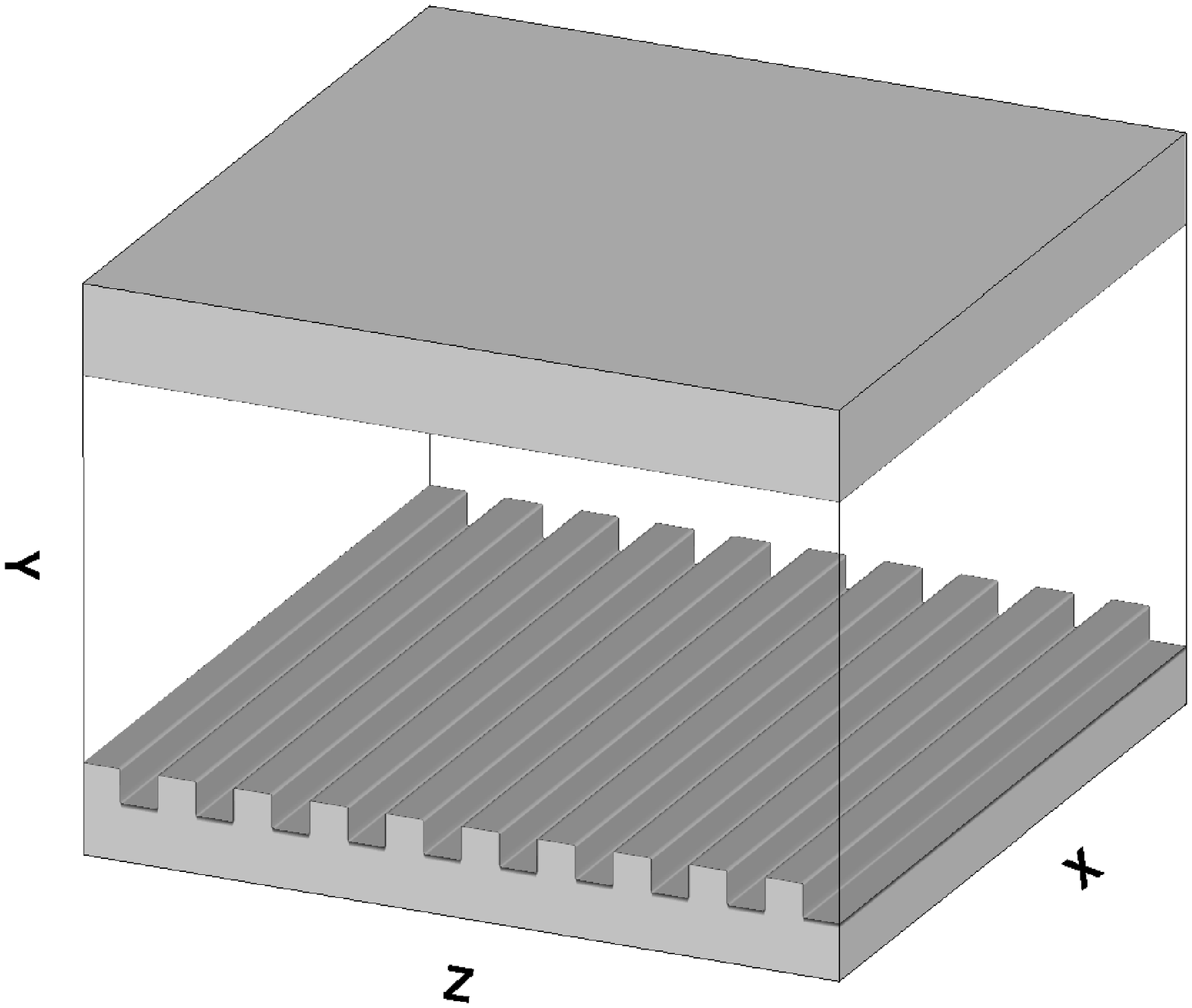} 
(c)
\includegraphics[width=4.0cm,angle=270,clip]{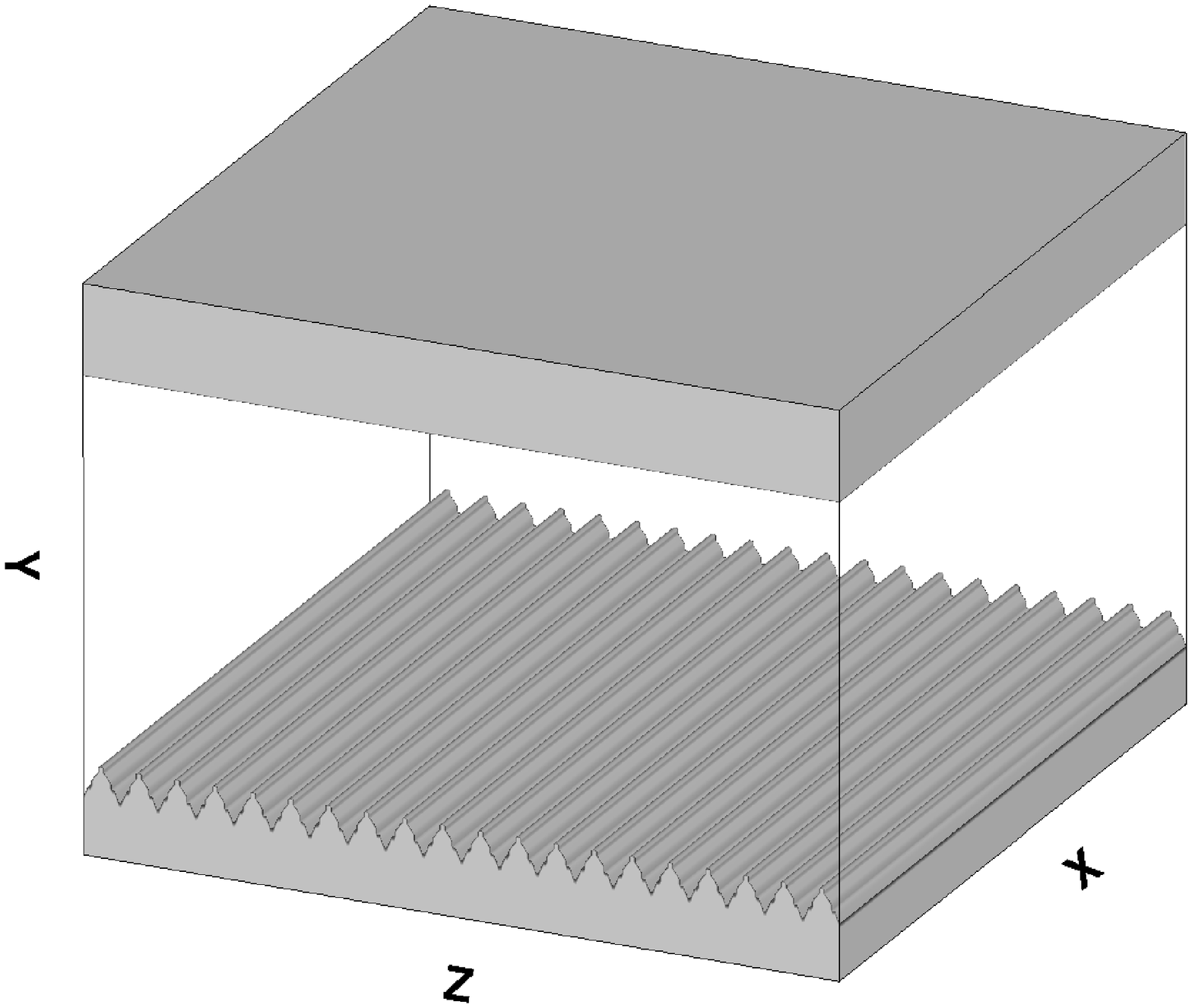} \\ \vskip1.em
(d)
\includegraphics[width=4.0cm,angle=270,clip]{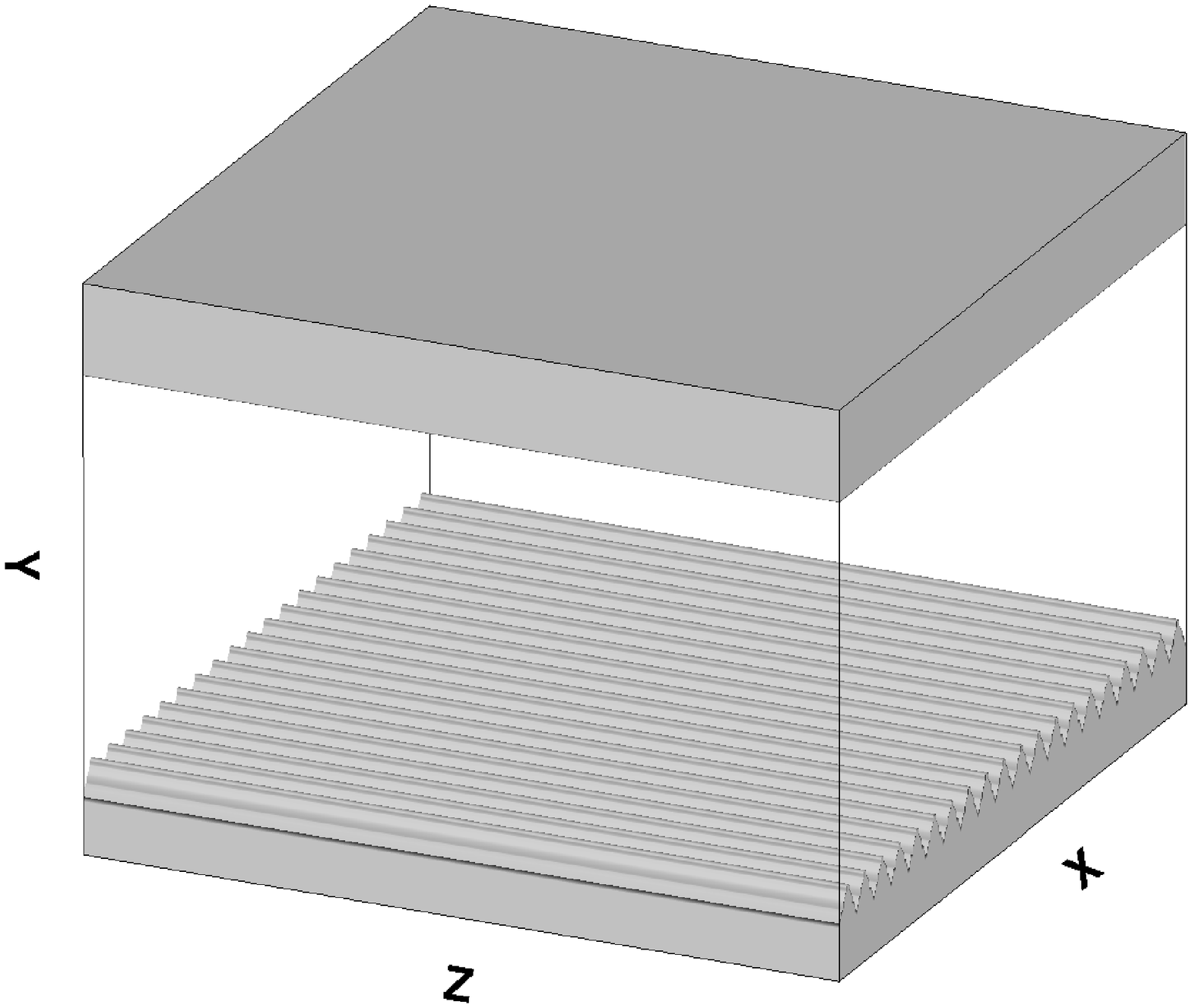}
(e)
\includegraphics[width=4.0cm,angle=270,clip]{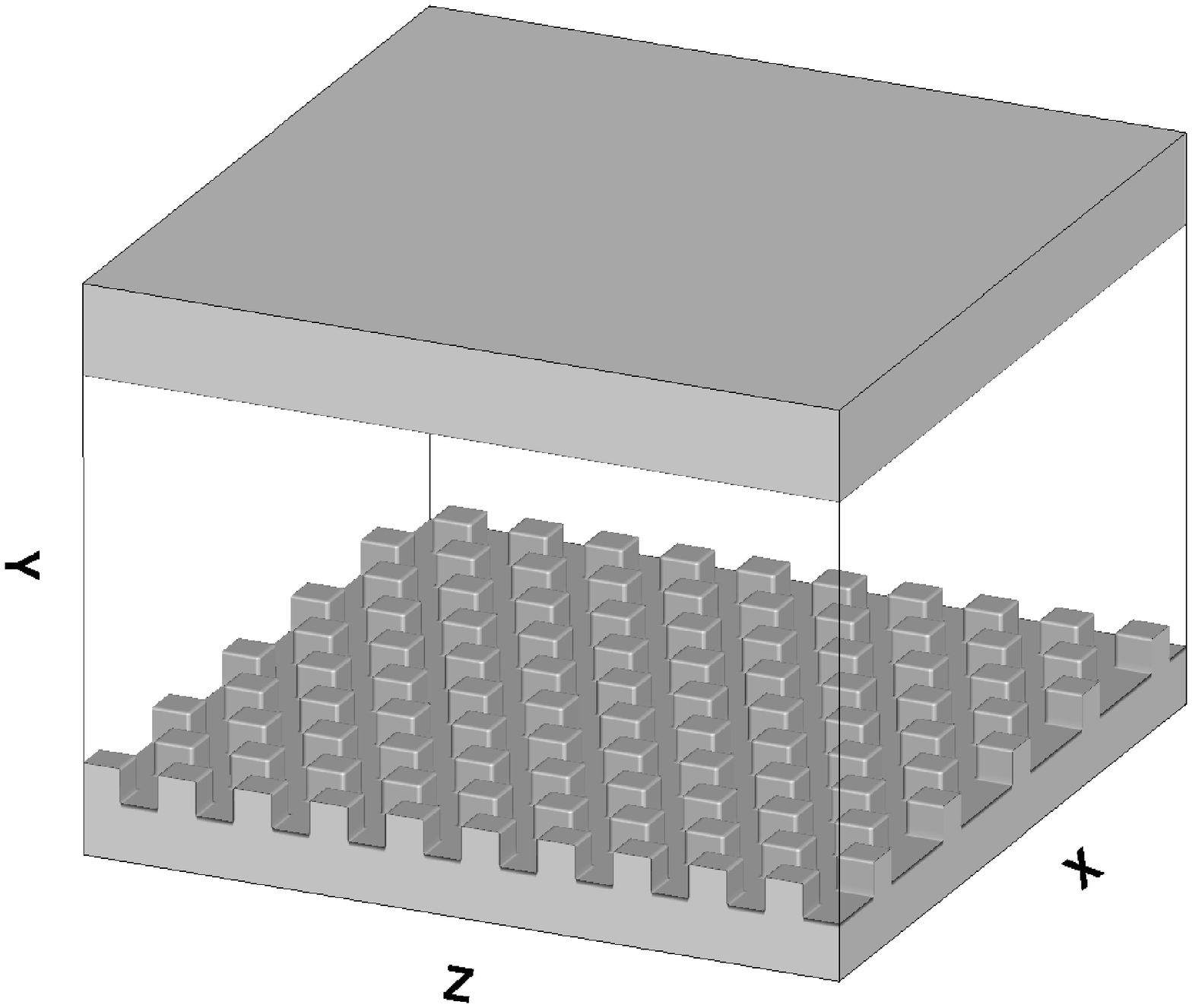} 
(f)
\includegraphics[width=4.0cm,angle=270,clip]{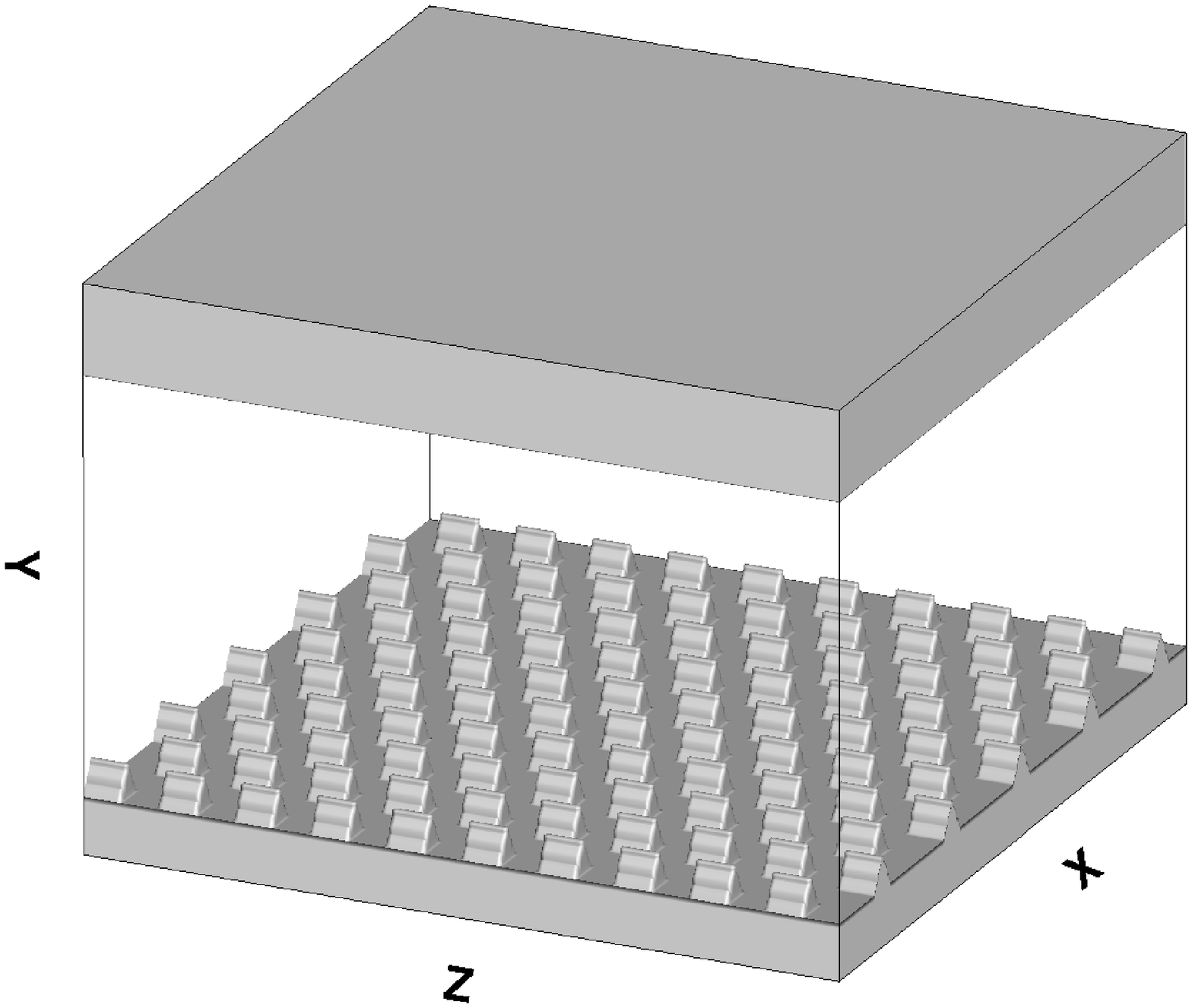} \\ \vskip1.em
\caption{Computational domain for flow over smooth walls SM 
(a), longitudinal squares bars LSB (b),
longitudinal triangular bars LTB (c),
transversal triangular bars TTB (d), 
staggered cubes bars CS (e), 
and staggered wedges WS (f). 
The grey shades denote the solid material.
}
\label{fig0}
\end{figure}

The resolution depends on the Reynolds number. 
At intermediate values of $Re$ ($Re>75$)
the computational grid is $320 \times 256 \times 320$ points 
in $x_1$, $x_2$, $x_3$ respectively. The mesh is uniform
in the horizontal directions, and non-uniform in the $x_2$ direction
to increase  the resolution
at the interface between solid and fluid. For each of the solid layers
$32$ grid points were used, and among them         
$20$ were allocated for the roughness layer.
The shape of the roughness
elements in the $x_1$ and $x_3$ directions is reproduced by $16$ grid points.
Two-dimensional roughness elements with different
orientation as well as three-dimensional elements have been 
considered to investigate the
influence of the shape of the roughness on the heat transfer. 
The different geometries are indicated as follows (also see figure~\ref{fig0}):
$LSB$ for square bars with $w/b=1$ oriented along the $x_1$ direction, 
$LTB$ for triangular bars with $w/b=0$ parallel to $x_1$, 
$TTB$ for bars of  same shape  oriented along the $x_3$,
$CS$ for staggered cubes with ratio $R_A=A_S/A_F=2$, where $A_S=b^2$
is the solid area and $A_F$ is the fluid area, and  $WS$ for a geometry with
the same layout, with cubes replaced by wedges. 
The results are
compared with those obtained with two smooth walls, and
indicated with $SM$.
For the $SM$ and the $WS$ geometry the simulations were performed
in a large range of Reynolds number to investigate the $Nu(Ra)$
dependence in the laminar, transitional and fully turbulent regime.

In the present simulations with a strong coupling of the thermal
field in the solid and in the fluid layers, the relationship
between $Re$ and $Ra$ is not unique. Hence, to compare the results
with those in literature it is necessary to relate the
Rayleigh number to the Reynolds number 
in the momentum transport equation. 
In the present work we select as the relevant temperature 
difference $\Delta \theta_F=<\theta>_R-<\theta>_S$, 
with $<\theta>_R$ and $<\theta>_S$ respectively the 
mean temperature at the plane of the crest of the roughness elements, 
and at the top wall. 
Defining the relevant Rayleigh number as 
$Ra=g\alpha\Delta T_F (2 h)^3/k\nu$, we get
$Ra=4Re^2/\Delta \theta_F$. The Nusselt 
number is also defined as $Nu=Q/(\Delta \theta_F/(Re Pr_F))$. These expressions
point out that  in the present simulations $Ra$ can not be fixed a priori,
but rather is an output of the simulations, also depending on
$Re$ and $Pr_S$.

\section{ Results }

Large effort was directed 
numerically, theoretically and experimentally to determine the
$Nu(Ra)$ relationship for smooth surfaces. However, much less effort has been 
put into looking
at their dependence in the presence of rough surfaces.
Before investigating the effects of the roughness on
the $Nu(Ra)$ relationship in the transitional regime, 
is worth demonstrating the independence of the statistics on the size
of the computational box. As previously mentioned, if an influence
occurs, it should be more appreciable for smooth walls.

\subsection{ Domain dependence }

\begin{table}[htb]
\footnotesize
 \begin{center}
 \begin{tabular*}{1.00\textwidth}
{@{\extracolsep{\fill}}cccccccc}
  \hline
$L/h$ &$ N_1\times N_2 \times N_3 $&   $Ra$ & $Nu$  & $q_1$ & $q_2$ & $q_3$& $T_\theta$ \\
\hline
4 & $ 200 \times 128 \times 200$ &   0.51588E+08 & 0.26712E+02 & 0.24338E-01 & 0.50116E-01&  0.40605E-01&  0.37564E-02 \\
4 & $ 320 \times 256 \times 320 $ &   0.51672E+08 & 0.26610E+02 & 0.24415E-01 & 0.54737E-01 & 0.46763E-01 & 0.39403E-02\\
8 & $640\times 256 \times 640$&   0.52150E+08 & 0.25694E+02 & 0.24080E-01 &  0.57262E-01& 0.55241E-01 &  0.39125E-02  \\
  \hline
 \end{tabular*}
 \end{center}
\caption{Grid and box sensitivity study: computational parameters and main results. All simulation are carried out for 
smooth walls, with $Re=3000$, $Pr_S=0.134$.
}
\label{table1}
\end{table}

To investigate whether the assumption of a box size with
lateral dimensions $L=L_1=L_3=4h$ is satisfactory to get 
good results the check has been performed at a value of $Re=3000$,
with solid layers made of copper ($Pr_S=0.134$).
At this Reynolds number the effects of the thermal plumes coexist
with those of the large-scale rollers, 
the former being linked to the resolution, and the
latter to the size of the domain.
The check consists in the comparison 
among the results of a simulation with 
$320\times 256 \times 320$ grid points and $L=4h$
and those obtained with the same resolution in a domain with
$L=8h$. To check the adequacy of the resolution a 
further simulation  with $L=4h$ and a $200\times 128 \times 200$
was performed. The results of the latter have
been used to compare the spectra and the profiles of
the $u_2$ and $\theta$ variances. Flow visualizations of the fluctuations
of the same variables are shown at the same 
distances from the wall where the spectra are calculated. 
Near the wall the thermal plumes should affect the spectra 
and the visualizations. The rollers should be  better visualised 
at the center. The simulations were initiated with 
random disturbances in the velocity components, 
and it was observed that the
rollers may change their orientation during the time evolution.

\begin{figure}
\centering
\hskip -1.5cm
\includegraphics[width=4.95cm]{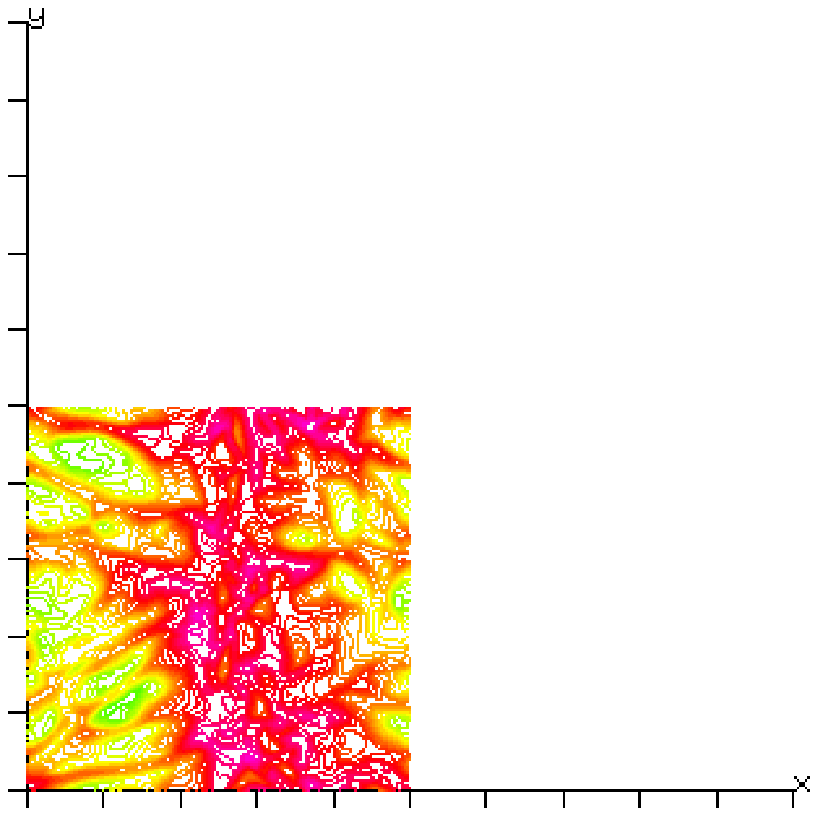}
\hskip -0.9cm
\includegraphics[width=4.95cm]{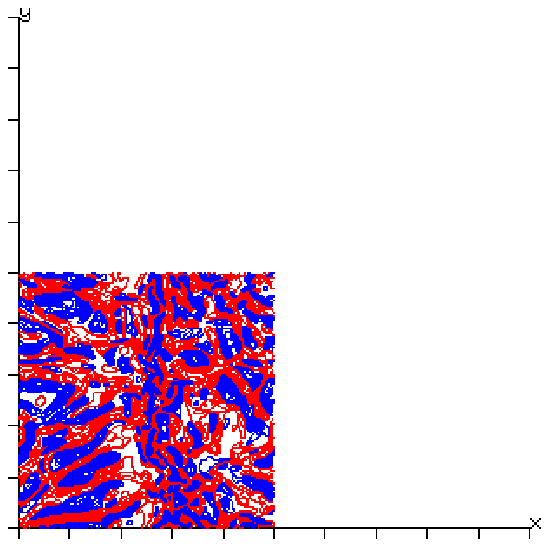}
\hskip -0.9cm
\includegraphics[width=4.95cm]{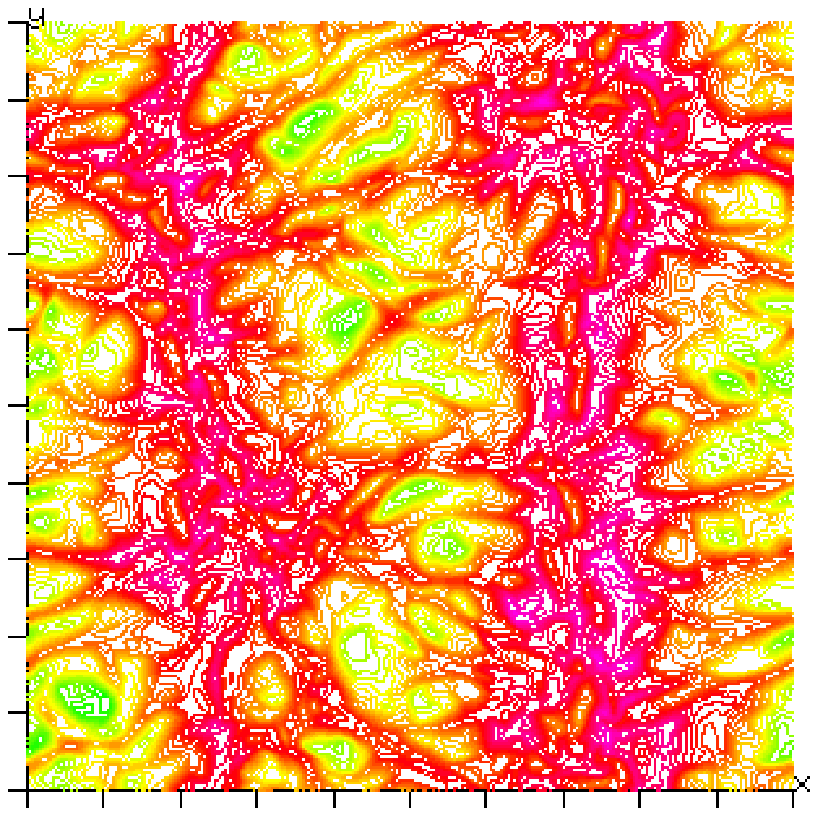}
\hskip -0.9cm
\includegraphics[width=4.95cm]{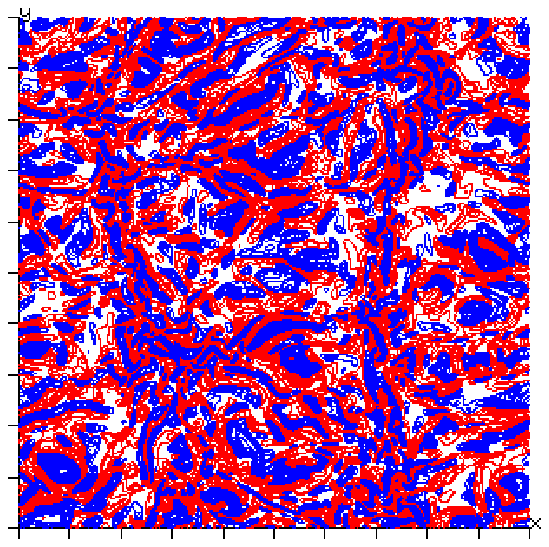}
\vskip -0.1cm
\hskip 1.25cm a) \hskip 4.25cm b) \hskip 4.25cm c) \hskip 4.25cm d)
\vskip -0.05cm
\caption{Visualizations of $\theta$ (a,c) and $u_2$ (b,d),
for $Re=3000$ and $Pr_S=0.134$ at $x_2=-0.9992$,
for box size $L=4h$ (a,b), and $L=8h$ (c,d).
Contour levels are given in intervals $\Delta u_2=0.001$ (blue negative,
red positive), $\Delta \theta =0.01$ from $\theta=0$ and colours 
from blue to red. 
}
\label{fig1}
\end{figure}

\begin{figure}
\centering
\hskip -1.5cm
\includegraphics[width=4.95cm]{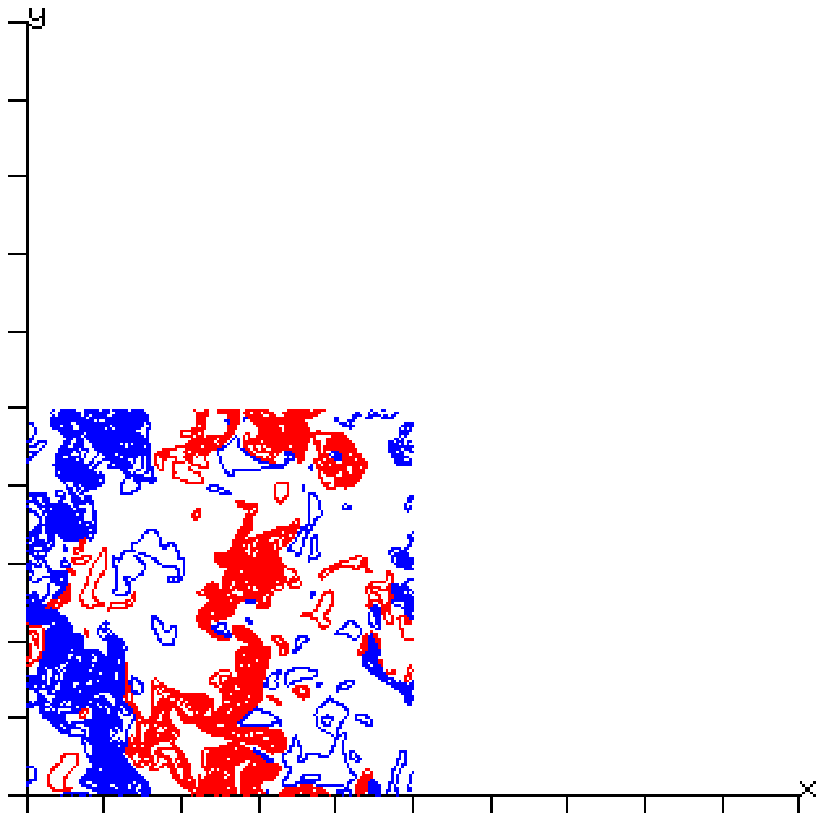}
\hskip -0.9cm
\includegraphics[width=4.95cm]{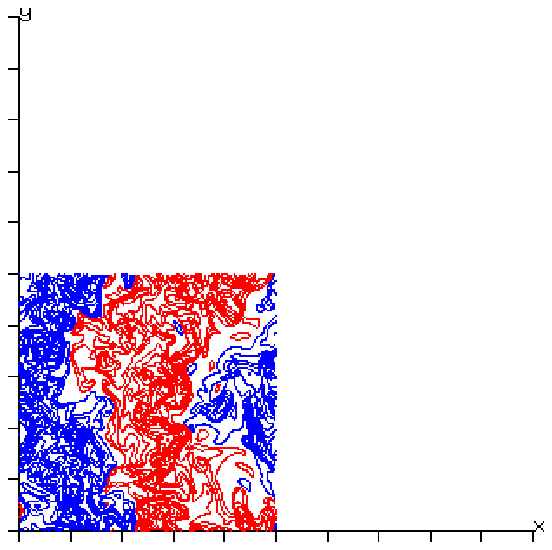}
\hskip -0.9cm
\includegraphics[width=4.95cm]{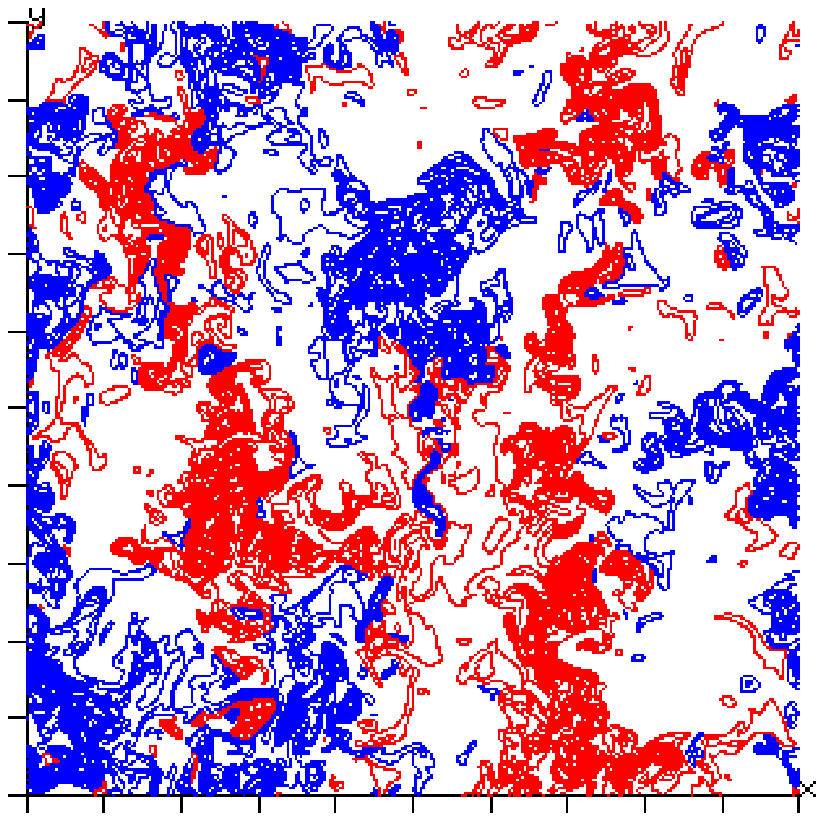}
\hskip -0.9cm
\includegraphics[width=4.95cm]{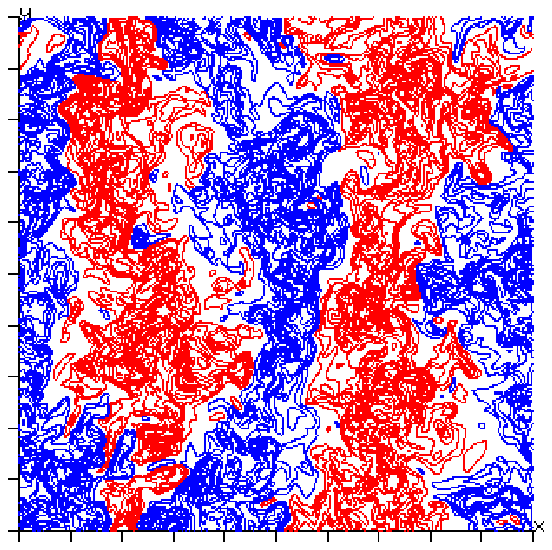}
\vskip -0.1cm
\hskip 1.25cm a) \hskip 4.25cm b) \hskip 4.25cm c) \hskip 4.25cm d)
\vskip -0.05cm
\caption{Visualizations of $\theta$ (a,c) and $u_2$ (b,d),
for $Re=3000$ and $Pr_S=0.134$ at $x_2=0$,
for box size $L=4h$ (a,b), and $L=8h$ (c,d).
Contour levels are given in intervals $\Delta u_2=0.05$ (blue negative,
red positive), $\Delta \theta =0.01$ from $\theta=-.15$ with blue for negative
and red for positive.
}
\label{fig2}
\end{figure}

The global results given in table~\ref{table1}
show a difference of $4\%$ between $Nu$ and $Ra$
in large and small boxes. In addition,
the results do not differ by changing the resolution,
with the meaning that all the flow scales are well
resolved also in the coarse simulation.
This is better proved by looking
at the spectra in Kolmogorov units. Greater but still  marginal
differences are observed in the values of the  integrated variances,
$q_i=\int <u_i^2> dx_2$, $T_\theta = \int <\theta^{\prime 2}> dx_2$. 
The $<\ >$ indicates averages in  time and in the two homogeneous
directions. 
The greatest difference in the values of $q_3$ are due to the fact
that the simulation with $L=8$ hadn't reached yet a
statistically steady state. This comparison was aimed
at looking at the instantaneous structures, as
shown in figure~\ref{fig1} (near the rough wall) and 
in figure~\ref{fig2} (at the channel centerline).
Figure \ref{fig1}a, indeed shows that one of the two rollers
well depicted by $\theta$ contours in figure \ref{fig1}c for
the large box, is also captured in the small box. The plumes with
red contours of $\theta$ are elongated and correspond to
regions of positive $u_2$. 
The elongated plumes which emanate
from the interface between the rollers are more clearly 
visualized by contours of $\theta^\prime$ demonstrating
that near the wall the  contribution 
$\theta^\prime u_2$ to the turbulent heat transfer component
starts to grow. On the side of these positive
thermal plumes there are thermal structures of circular
shape due to the descending fluid. Both structures contribute
to the  component of the turbulent heat flux $Q_T=<u_2\theta^\prime>$.
Due to the different evolution the fields at a given
time in the small domain cannot exactly reproduce one
fourth of the field in the large domain. However, the qualitative
comparison between figure \ref{fig1}a and figure \ref{fig1}c, and
between figure \ref{fig1}b and figure \ref{fig1}d shows that
the flow characteristics near the wall are captured by the simulations
in the box of size $4h$.
The rollers are well depicted in  $x_1-x_3$ planes at the
center of the box, as it is shown in figure \ref{fig2}.

\begin{figure}
\centering
\hskip -1.5cm
\psfrag{ylab}[][]{$E^*_{{\theta}l},\ \ E^*_{{u_2}l} $}
\psfrag{xlab}[][]{$k^* $}
\includegraphics[width=8.55cm]{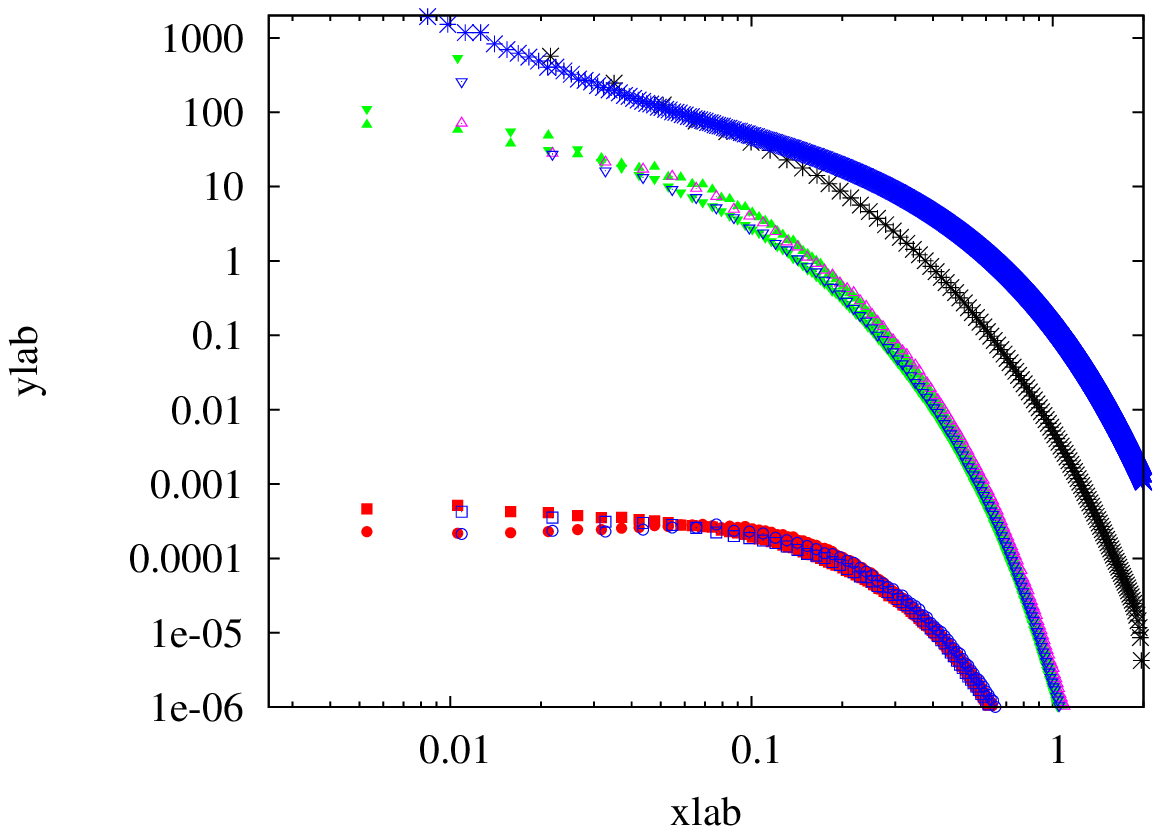}
\hskip -0.9cm
\psfrag{ylab}[][]{$ $}
\psfrag{xlab}[][]{$k^* $}
\includegraphics[width=8.55cm]{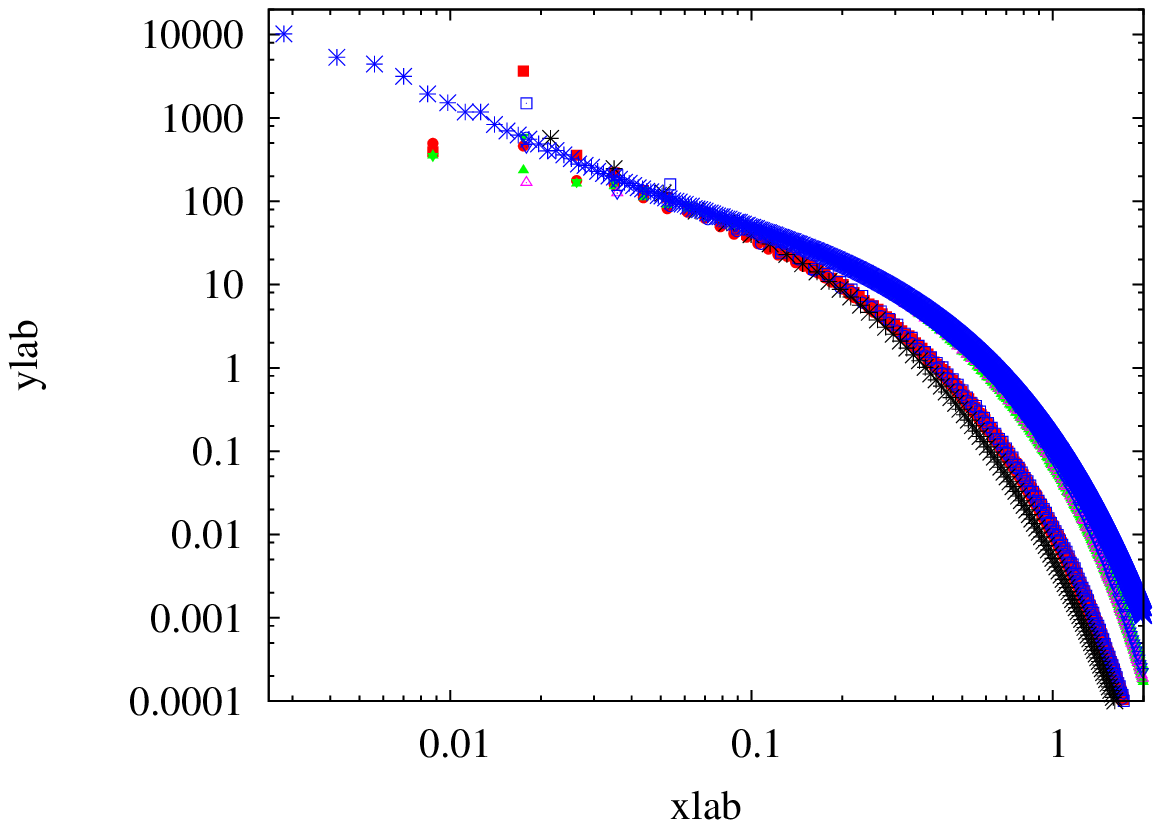}
\vskip -0.1cm
\hskip 4.25cm a) \hskip 6.25cm b) 
\vskip -0.05cm
\caption{Flow at $Re=3000$ and $Pr_S=0.134$: $u_2$ and $\theta$ spectra in  the $x_1$ direction $l=1$
and in the $x_3$ direction at a) $x_2=-0.998$ and b) at $x_2=0$
compared with the transverse spectra of \citet{jimenez_93}
black and with the scalar spectra by 
\citet{donzis_10} blue symbols;
the square are $E^*_{{u_2}1}$  and the circles $E^*_{{u_2}3}$
the down triangles are $E^*_{{\theta}1}$ and the up triangles $E^*_{{\theta}3}$.
Solid symbols are used for $L=8h$, and open symbols for $L=4h$.
}
\label{fig3}
\end{figure}

A proof that the large and the small scales are satisfactorily described by 
the box with $L=4h$ can be demonstrated by showing the $u_2$ and the $\theta$
one-dimensional spectra at the same locations as the visualizations
in figure \ref{fig1} and \ref{fig2}. The calculation of the spectra
was done by a large number of fields saved every time unit.
To point out the  isotropy or anisotropy of the fields at the two different
locations the spectra are given in Kolmogorov units, indicated 
with the superscript $*$. 
The scaled wavenumber is defined as $k^*=k\eta$ 
with $\eta=(1/(Re^2\Omega)^{1/4}$.
$\Omega=<\omega_i^2>$, the scaled
velocity spectra are defined as 
$E^*_{{u_i}l}=E_{{u_i}l}/(\Omega^2\eta^5/Re^2)^{1/3}$,
and the dimensionless temperature spectra are defined as
$E^*_{{\theta}l}=E_{{\theta}l}/(\chi(\Omega/Re)^{-1/3}\eta^{5/3})$
with $\chi=\langle(\frac{\partial \theta}{\partial x_i})^2\rangle$
the rate of temperature dissipation. 
Near the wall the spectra are expected to be far
from isotropic. On the other hand at the channel centerplane 
($x_2=0$) an isotropic condition
should be expected. To investigate whether this is achieved,
the velocity spectra are compared with those obtained by \citet{jimenez_93}
and the temperature spectra with the scalar spectra 
by \citet{donzis_10}.
Figure \ref{fig3}a,  near the wall,
and figure \ref{fig3}b, at the center, show that
the velocity and the temperature spectra do not depend on the
box size, and, that, in particular, at the intermediate and small scales
there is no dependence on the direction. Near the wall $u_2$
is very small and increases for the effect of the buoyancy, 
therefore in  figure~\ref{fig3}a the amplitude of the $u_2$ and 
the $\theta$ spectra are very different. The magnitude of
both the $\theta$ spectra are different from those of
forced isotropic turbulence and are greater than both the
$u_2$ spectra.  From the spectra at several distances
from the wall, not shown for lack of space, it was observed a
rapid tendency to the isotropy. In fact at $x_2=0.5$ the spectra are
similar to those in figure \ref{fig3}b. A remarkable 
similarity between the present and the forced isotropic simulations
spectra is also obtained in this figure. The same differences
between the exponential decay range for the temperature and
the $u_2$ are reproduced, in particular at the end of the inertial range.
This behavior independent on the type of flows indicates 
that the $\theta$ and $u_2$ structures contributing
to the terminal part of the inertial range (the so-called bottle-neck) 
should be very different. 

\begin{figure}
\centering
\hskip -1.5cm
\psfrag{ylab}[][]{$<u_2^2>, \ \ <\theta^2> $}
\psfrag{xlab}[][]{$y/h$}
\includegraphics[width=8.55cm]{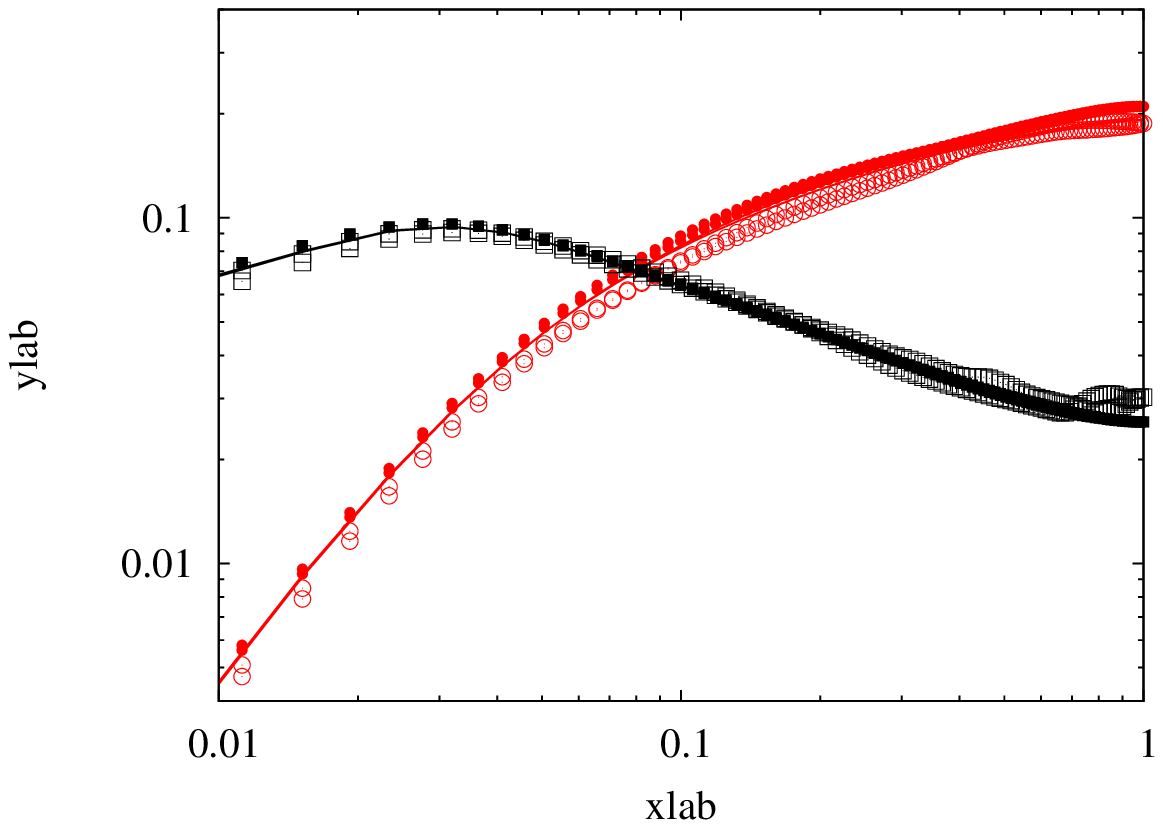}
\hskip -0.4cm
\psfrag{ylab}[][]{$ q,\Omega, \chi$}
\psfrag{xlab}[][]{$y/h$}
\includegraphics[width=8.55cm]{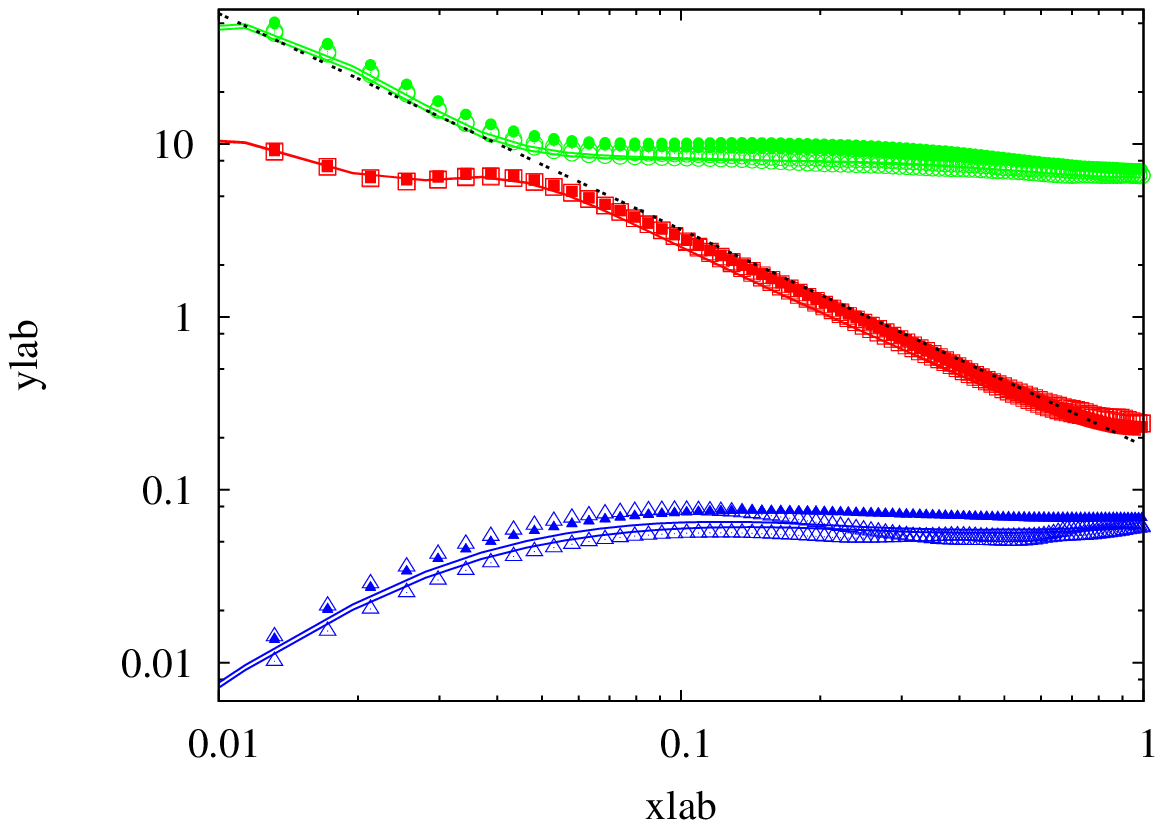}
\vskip -0.1cm
\hskip 4.25cm a) \hskip 6.25cm b) 
\vskip -0.05cm
\caption{Flow at $Re=3000$ and $Pr_S=0.134$. Vertical profiles of: a) $<u_2^2>$ (red) and $ <\theta^2> $ (black);
b) $q$ (blue), $\Omega$ (red) and $ \chi$ (green). 
Open symbols indicate results obtained with 
$L=4h$ $320 \times 256 \times 320$, solid symbols with
$L=8h$ $640 \times 256 \times 640$, and lines with
$L=4h$ $200 \times 128 \times 200$. 
The dashed line denotes $\approx y^{-1.25}$.
The values of the respective Rayleigh numbers are given in table \ref{table1}.
}
\label{fig4}
\end{figure}

To further demonstrate the independence of the box size, the
profiles of the  variance of $u_2$ and $\theta$ were calculated.
In addition to showing that the grid $320 \times 256 \times 320$
is  resolving all the flow scales, the results are compared with those 
obtained with a grid $200 \times 128 \times 200$.
The good resolution can be further appreciated
in figure \ref{fig3} where the spectra extend up to $k^*>2$.
Figure \ref{fig4} shows collapse
of the profiles irrespective
of the grid and the box size.  In this figure the statistics 
are plotted versus the distance from the walls, therefore
the perfect symmetry of the profiles emerges.
In addition the large difference in figure \ref{fig4}a between the
values of $<u_2^2>$ and $ <\theta^2> $ in the near wall region
explains why large differences in the spectra at
$x_2=-0.998$ in figure \ref{fig3}a were observed. 
The reason of the different behavior
between $<\theta^2> $ and $<u_2^2>$ is due to the 
conductivity of the solid that strictly connect the $\theta$
fluctuations inside the solid to those generated by the turbulent
flow motion. In the simulations with isothermal temperature at
$x_2\pm 1$ $<\theta^2>=<u_2^2>=0 $. The profiles in figure \ref{fig4}a
accounts for
the large scales, as well as the profile of
the turbulent kinetic energy $q=<u_i^2>$ in figure \ref{fig4}b.
The satisfactory resolution of the small scales can be appreciated
by the profiles of $\Omega$ and $\chi$ which are necessary to
evaluate the spectra in Kolmogorov units in figure \ref{fig3}.
The relative profiles are shown in figure \ref{fig4}b, which
demonstrate that, also in this case there is an independence
of the profiles from the resolution and from the size of the computational box.
The enstrophy and the rate of temperature dissipation achieve
the highest values near the wall and then have a complete different
behavior, with $\chi$ nearly constant in the entire domain,
while the enstrophy does not largely change near the wall, and
decays approximately as $\Omega \approx y^{-1.25}$
within the box.

Having demonstrated that
for natural convection in the presence of heat conducting smooth walls
at $Re=3000$, when the flow can be considered fully turbulent,
the results do not depend on the lateral size of the box, in
the next section we investigated the influence of the roughness 
shape at lower value of the Reynolds number. 

\subsection{ Roughness influence }
 
{\tiny
\begin{table}[htb]
 \centering
 \begin{tabular*}{1.00\textwidth}
{@{\extracolsep{\fill}}ccccccc}
  \hline
Geometry  &  $Ra$ & $Nu$  & $q_1$ & $q_2$&$q_3$& $T_\theta$ \\
\hline
$SM$ &   0.22664E+07 & 0.11389E+02 & 0.79858E-01 & 0.11134E+00 & 0.44864E-01 & 0.20216E-01\\
$WS$ &  0.14870E+07 &  0.19359E+02 & 0.76241E-01 & 0.11775E+00 & 0.47854E-01 & 0.23993E-01\\
$SC$ &   0.16399E+07 & 0.17266E+02 & 0.59798E-01 & 0.10997E+00 & 0.56664E-01 & 0.24381E-01\\
$LSB$ &   0.18326E+07 & 0.15142E+02 & 0.97332E-01 & 0.12395E+00 & 0.38397E-01 & 0.25379E-01\\
$LTB$ &   0.16993E+07 & 0.15890E+02 & 0.10787E+00 & 0.13016E+00 & 0.33158E-01 & 0.24042E-01\\
$TTB$ &  0.17334E+07 & 0.15264E+02 & 0.53122E-01 & 0.11165E+00 & 0.63454E-01 & 0.21516E-01\\
  \hline
 \end{tabular*}
\caption{Values of some of the global quantities obtained for numerical simulations 
over rough walls at $Re=500$,d $Pr_S=0.134$.
}
\label{table2}
\end{table}
}

The global results for the  six types of surfaces 
mentioned above (see figure~\ref{fig0}) are given 
in table \ref{table2}. The simulations
were performed by fixing the Reynolds number $Re=500$ and 
the solid layer is considered to be made  of copper ($Pr_S=0.134$).
Although $Re$ is fixed, table \ref{table2} indicates a large dependence
of $Ra$ and $Nu$ on the shape of the surface roughness.
In particular, the values of $Nu$ are greater than those obtained by smooth
walls, and  surfaces with three-dimensional elements lead
to higher heat transfer than achieved with two-dimensional 
bars. As should be expected, the $90^0$ change of orientation of
bars with the same shape produces values of $Nu$ not too different.
The slight increase of $Nu$ obtained for the $LTB$
with respect to the $LSB$ surface implies that the  edges  of the
triangular bars improve the heat transfer. The higher efficiency
of the wedges is demonstrated by comparing $Nu$ 
for the $SC$ and $WS$ surfaces. The influence of the orientation
of the two-dimensional elements can be appreciated by comparing the values
of $q_1$ and $q_3$ in table \ref{table2}.   

\begin{figure}[h]
\centering
\hskip -1.5cm
\includegraphics[width=5.95cm]{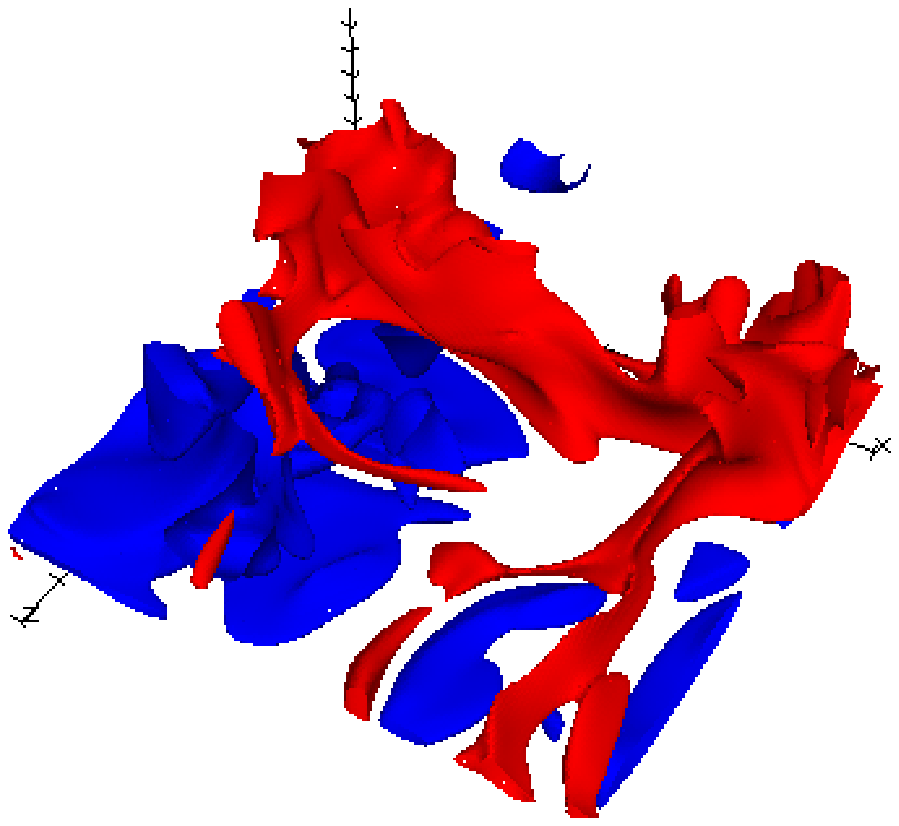}
\hskip -0.9cm
\includegraphics[width=5.95cm]{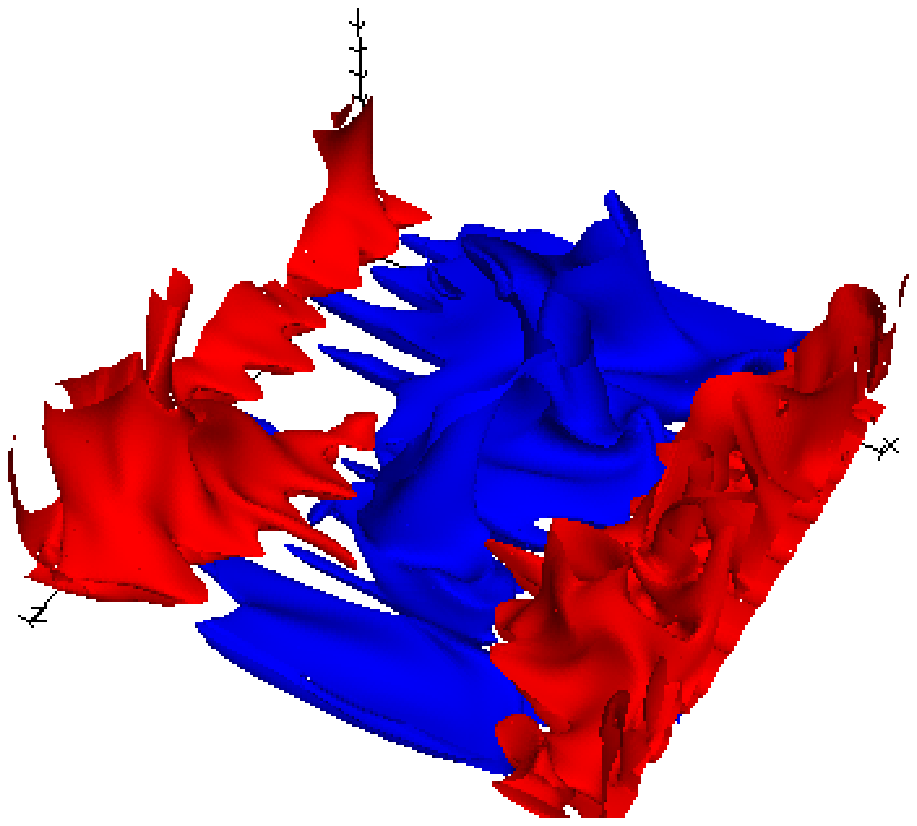}
\hskip -0.9cm
\includegraphics[width=5.95cm]{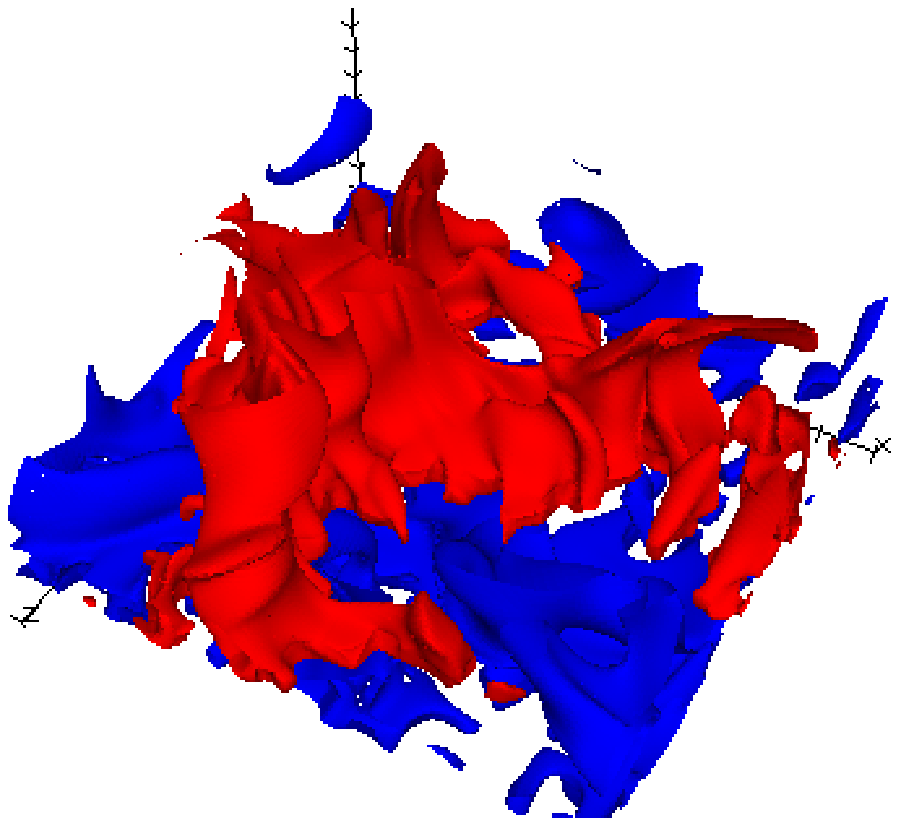}
\vskip -0.1cm
\hskip 1.25cm a) \hskip 5.25cm b) \hskip 5.25cm c) 
\vskip -0.05cm
\hskip -1.5cm
\includegraphics[width=5.95cm]{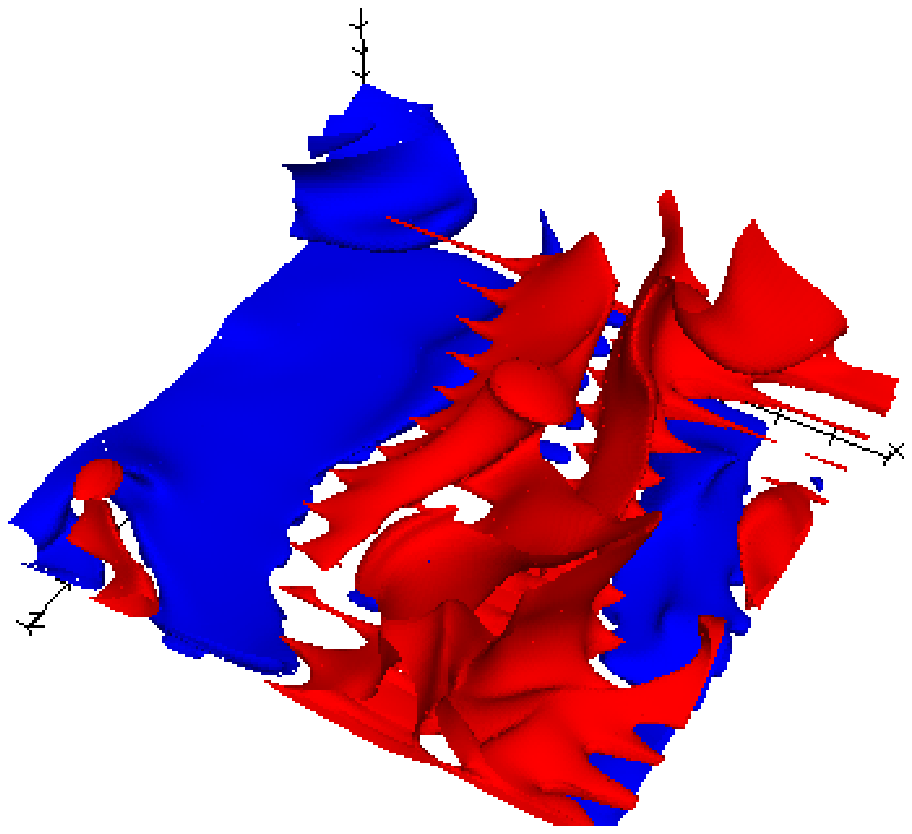}
\hskip -0.9cm
\includegraphics[width=5.95cm]{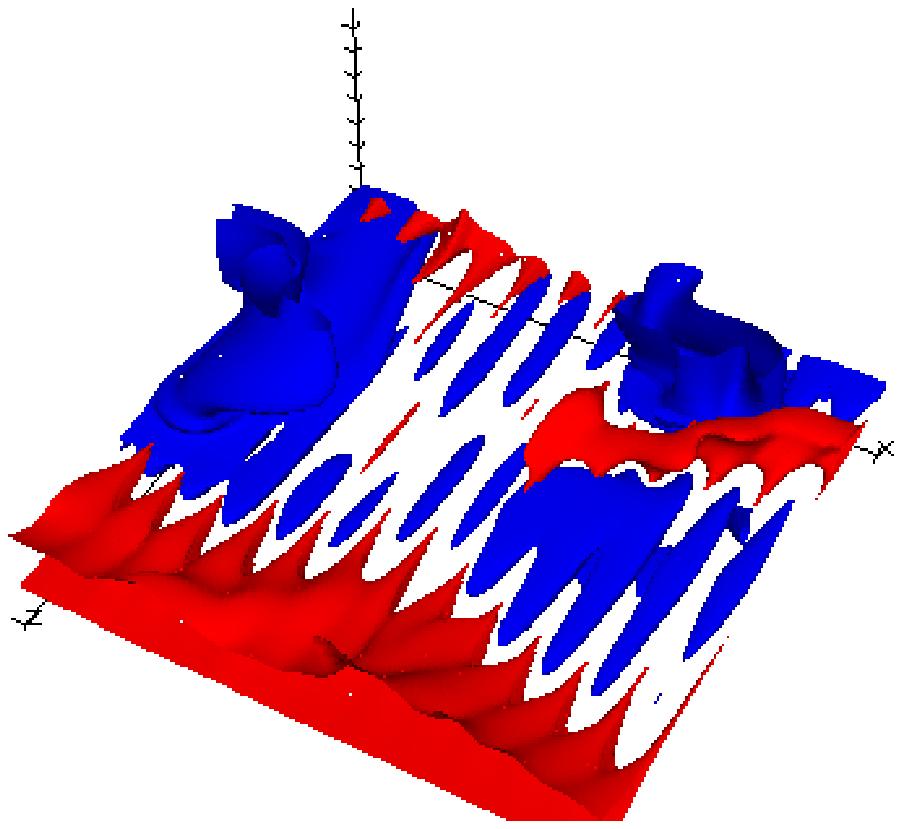}
\hskip -0.9cm
\includegraphics[width=5.95cm]{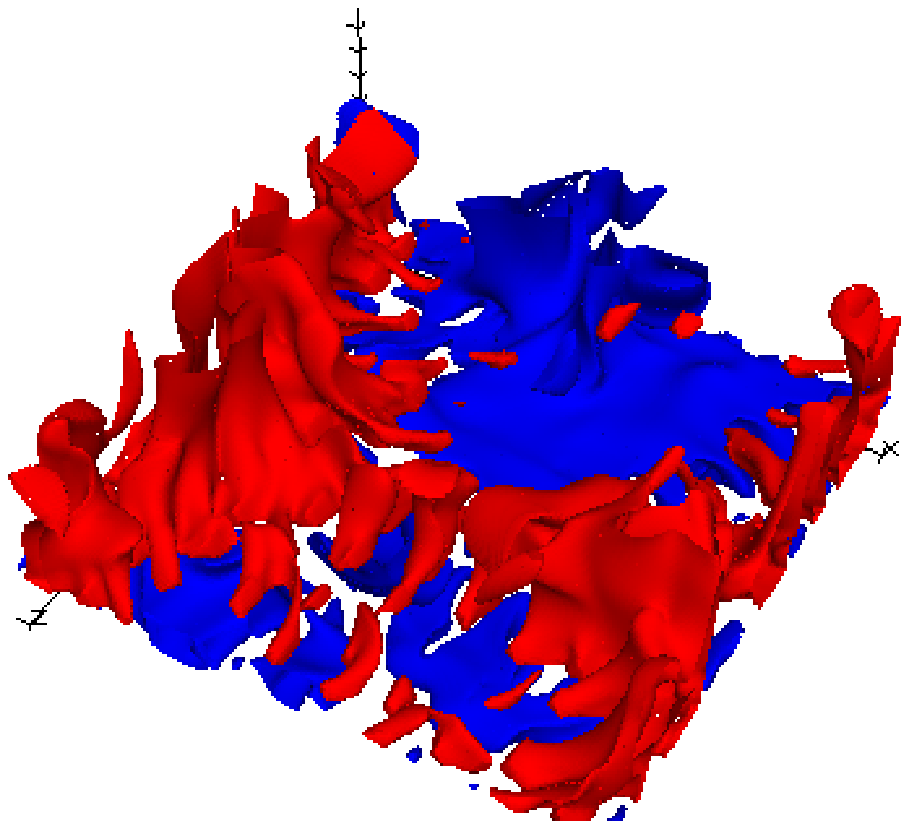}
\vskip -0.1cm
\hskip 1.25cm d) \hskip 5.25cm e) \hskip 5.25cm f) 
\vskip -0.05cm
\caption{Fluctuating $\theta$ surface contours, red $\theta^\prime=+0.1$
blue $\theta^\prime=-0.1$ above the heated surface at $Re=500$ and
$Pr_S=0.134$:
a) smooth $SM$, b) longitudinal square bars $LSB$, c)
three-dimensional staggered cubes $CS$, d) longitudinal triangular
bars $LTB$, e) transverse triangular bars $TTB$, f) three-dimensional
staggered wedges $WS$.
}
\label{fig5}
\end{figure}

The qualitative influence of the shape of roughness on the 
flow structures can be appreciated by surface contours
of $u_2$ and $\theta^\prime$.
Three-dimensional
plots in the region above the rough surfaces are shown at
the values of $Re$ and $Pr_S$ reported in the caption of table \ref{table2}.
The surfaces of $\theta^\prime=\theta-\overline{\theta}=\pm 0.1$ 
in figure \ref{fig5}b-f
for the five rough surfaces,  are compared with
those of the smooth wall given in figure \ref{fig5}a. 
For these visualizations
the averages, indicated by an overbar, 
are performed  in the $x_1$ and $x_3$
homogeneous directions by taking only one field. 
At $Re=500$ there is a marginal influence of the small scale turbulence on the 
large scale structures that are mainly related to the shape of the surface. 

\noindent In figure \ref{fig5}a the rollers for the $SM$ surface,
previously described at a higher
Reynolds number, are oriented in the $x_1$ direction.
Some  unstable small structures are visible, which at this
value of $Re$ have a size comparable to that of the rollers.
Longitudinal square bars $LSB$, in figure \ref{fig5}b 
promote a longitudinal motion
within the bars with the effect of orienting the rollers 
in the $x_3$ direction. This effect is corroborated by 
the orientation induced by the triangular longitudinal bars $LTB$
in figure \ref{fig5}d. The  triangular shape 
of the elements promotes a greater penetration of the 
hot rollers in the space between the bars, leading to a greater value
of $q_1$ for $LTB$ than that for $LSB$ in table \ref{table2}.
\begin{figure}[h!]
\centering
\hskip -1.5cm
\includegraphics[width=5.95cm]{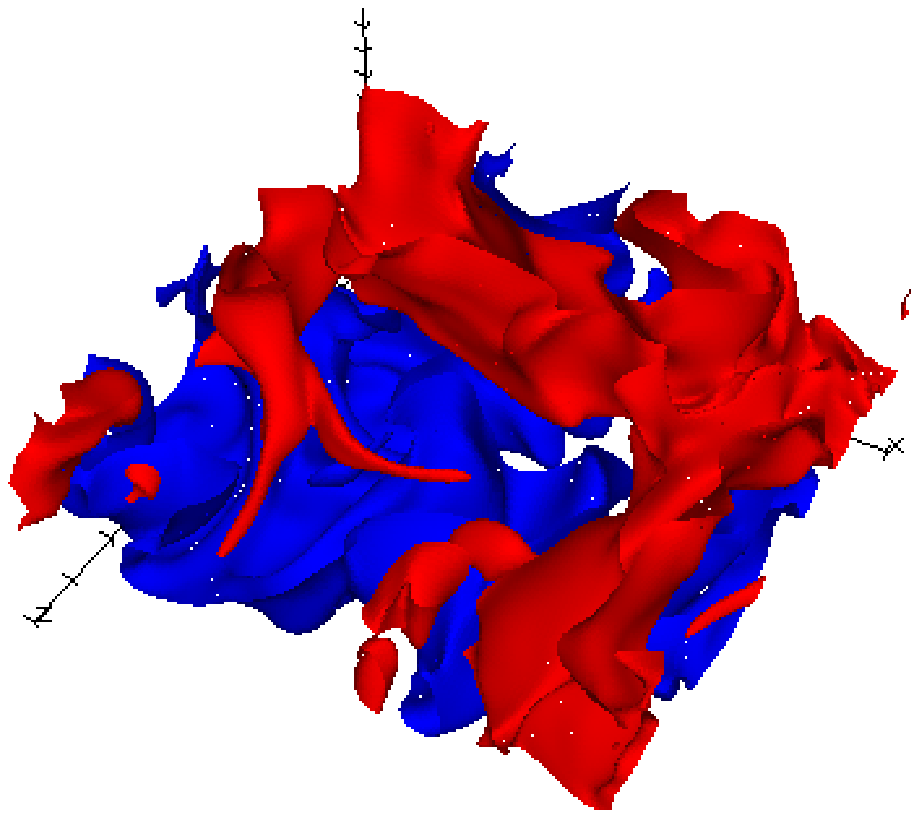}
\hskip -0.9cm
\includegraphics[width=5.95cm]{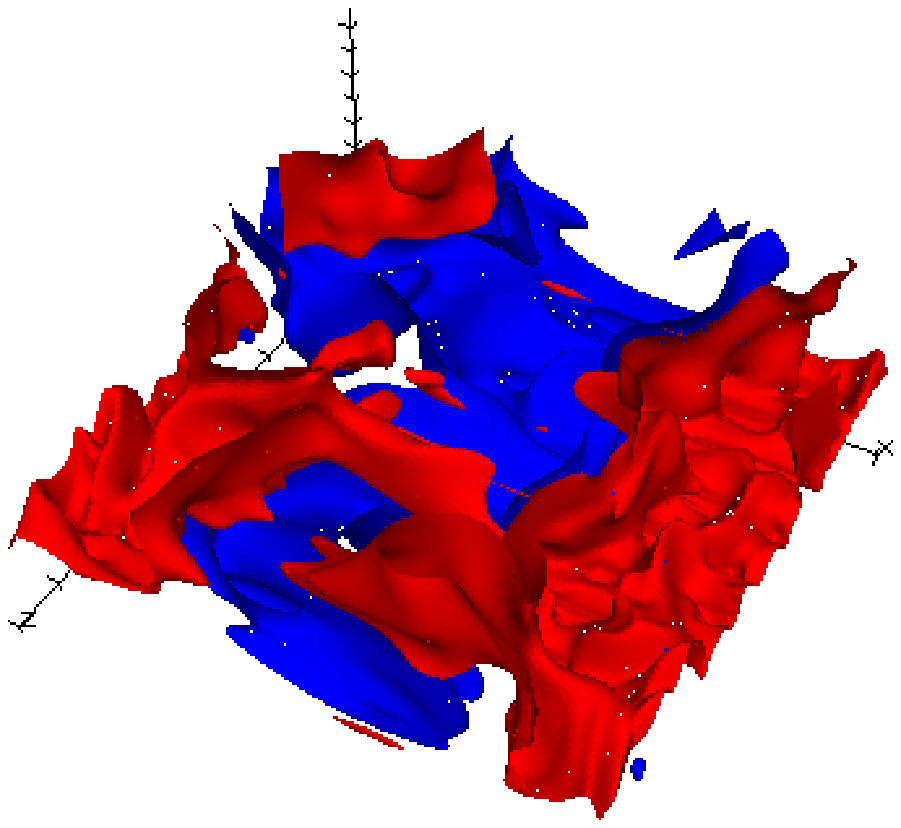}
\hskip -0.9cm
\includegraphics[width=5.95cm]{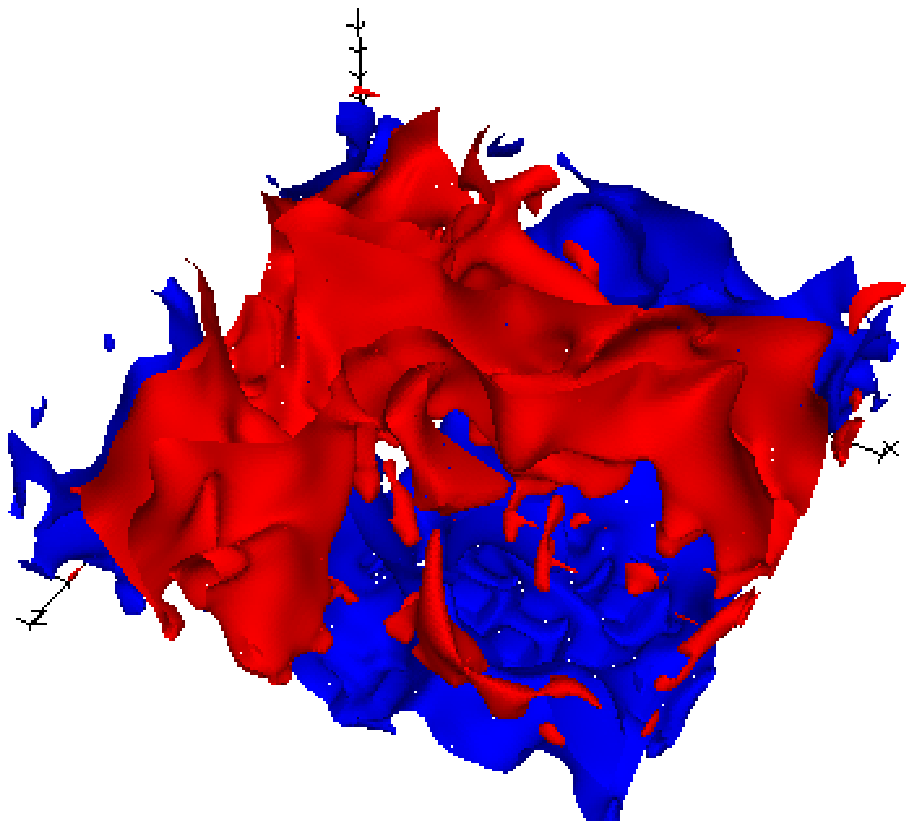}
\vskip -0.1cm
\hskip 1.25cm a) \hskip 5.25cm b) \hskip 5.25cm c) 
\vskip -0.05cm
\hskip -1.5cm
\includegraphics[width=5.95cm]{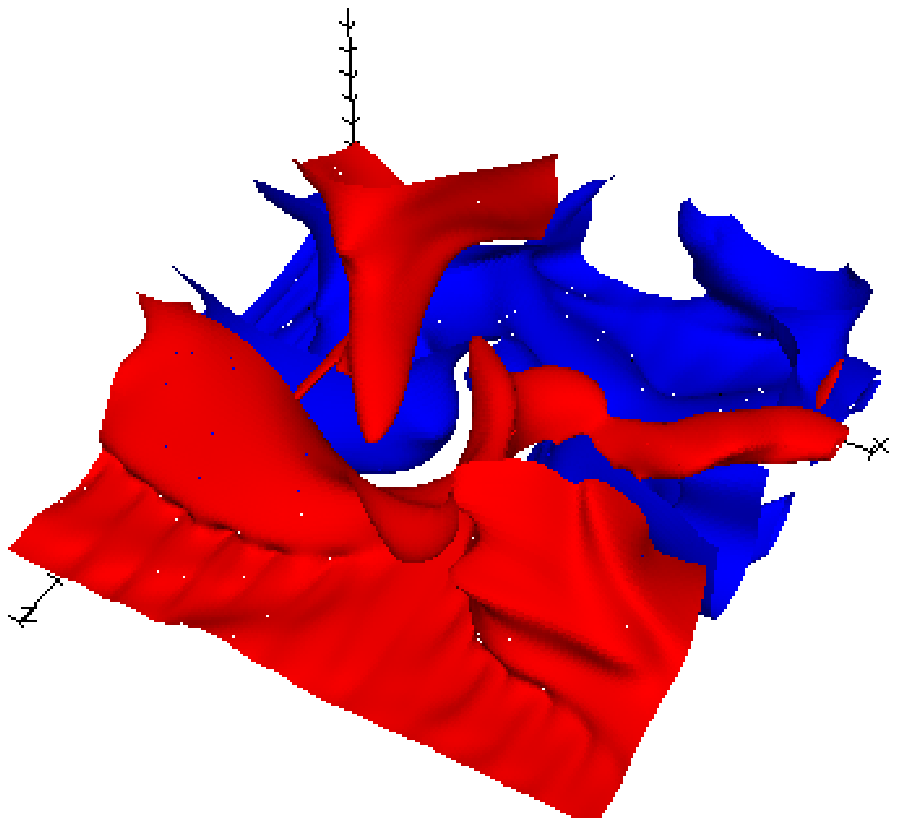}
\hskip -0.9cm
\includegraphics[width=5.95cm]{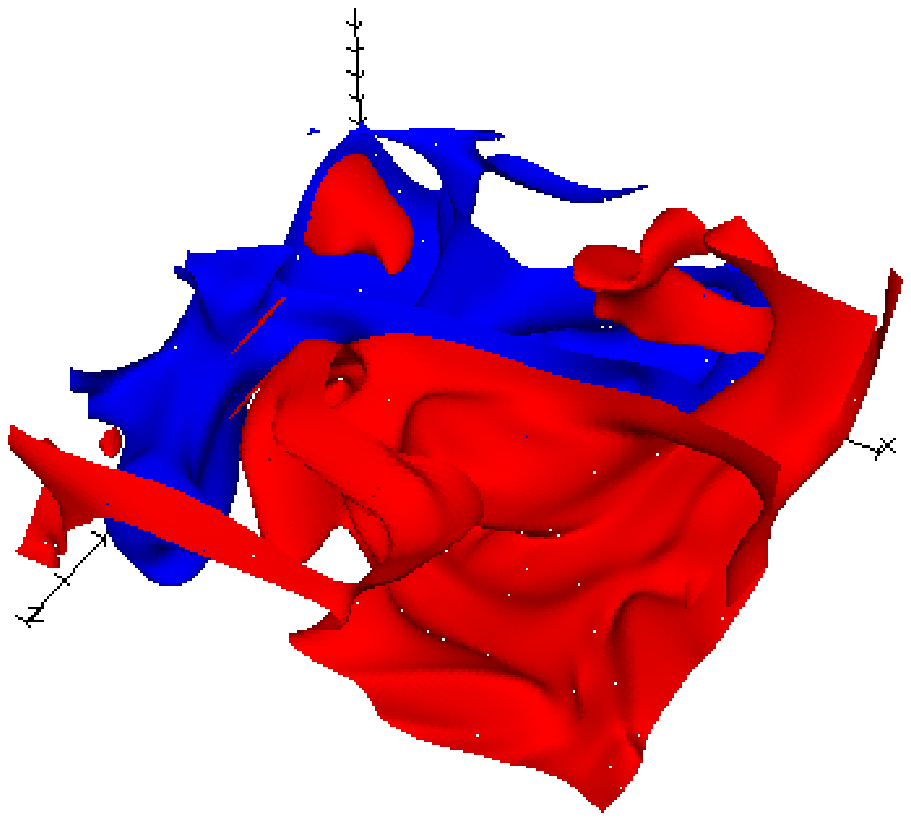}
\hskip -0.9cm
\includegraphics[width=5.95cm]{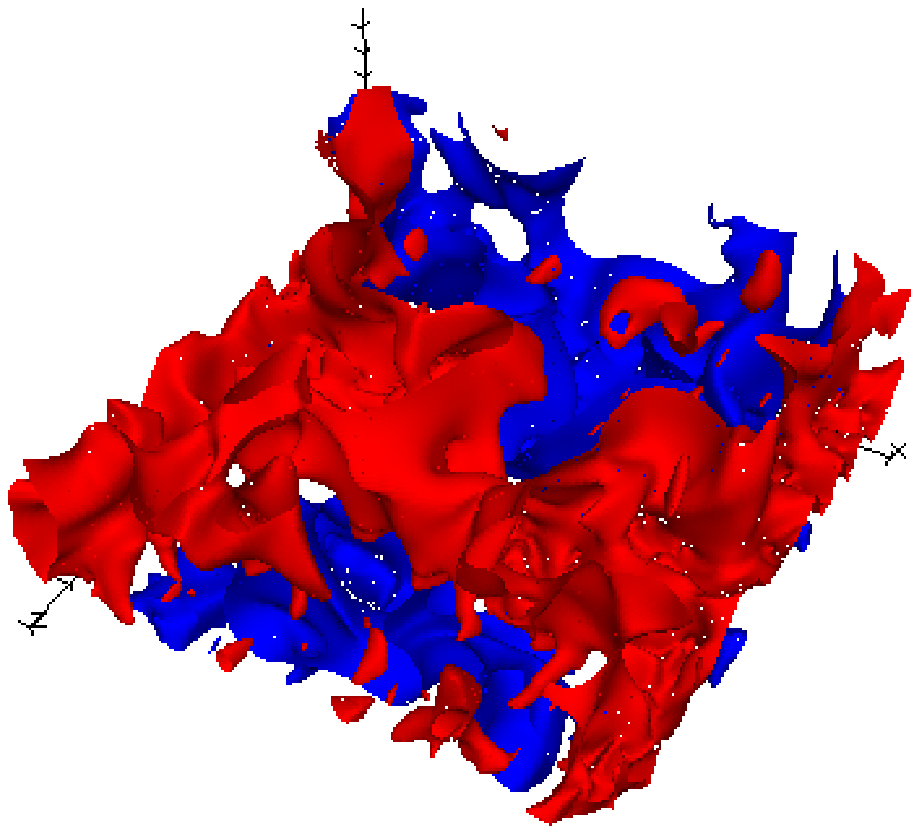}
\vskip -0.1cm
\hskip 1.25cm d) \hskip 5.25cm e) \hskip 5.25cm f) 
\vskip -0.05cm
\caption{Fluctuating $u_2$ surface contours, red $u_2^\prime=+0.1$
blue $u_2^\prime=-0.1$ above the heated surface at $Re=500$ and
$Pr_S=0.134$:
a) smooth $SM$, b) longitudinal square bars $LSB$, c)
three-dimensional staggered cubes $CS$, d) longitudinal triangular
bars $LTB$, e) transverse triangular bars $TTB$, f) three-dimensional
staggered wedges $WS$.
}
\label{fig6}
\end{figure}
\noindent These two surfaces produce values of $q_1$ greater than those
of the other surfaces (see table \ref{table2}).
The influence of the bars orientation on the shape of
the structures is depicted by the $90$ degree change of the orientation
from $LTB$ (figure \ref{fig5}b).
to $TTB$ (figure \ref{fig5}e).
Depending on the time chosen for the visualizations of the
three-dimensional elements, both for cubes 
$CS$ (figure \ref{fig5}c)
and wedges $WS$ (figure \ref{fig5}f)
the rollers can be oriented in different directions.

\begin{figure}
\centering
\hskip -1.5cm
\psfrag{ylab}[][]{$<u_2^2>^{1/2} $}
\psfrag{xlab}[][]{$x_2 $}
\includegraphics[width=5.05cm]{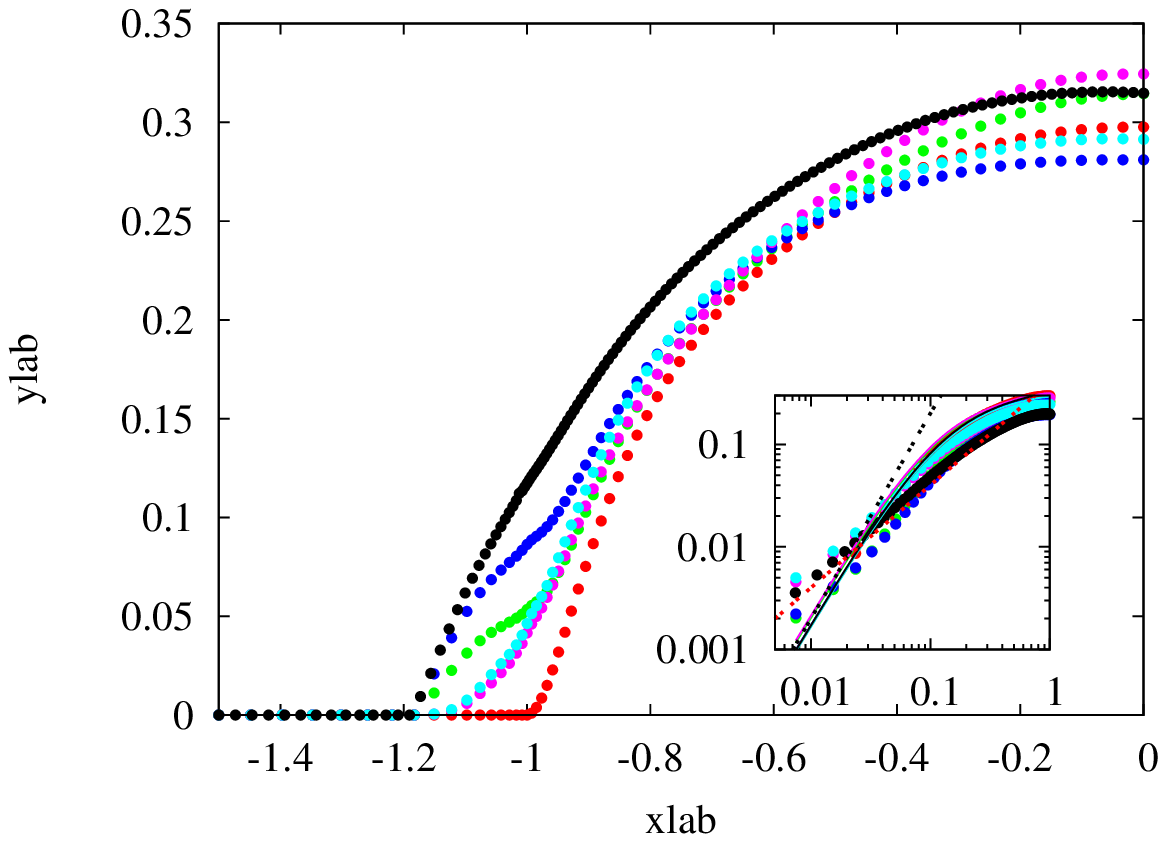}
\hskip -0.0cm
\psfrag{ylab}[][]{$ <\theta^{\prime 2}>^{1/2}$}
\psfrag{xlab}[][]{$x_2 $}
\includegraphics[width=5.05cm]{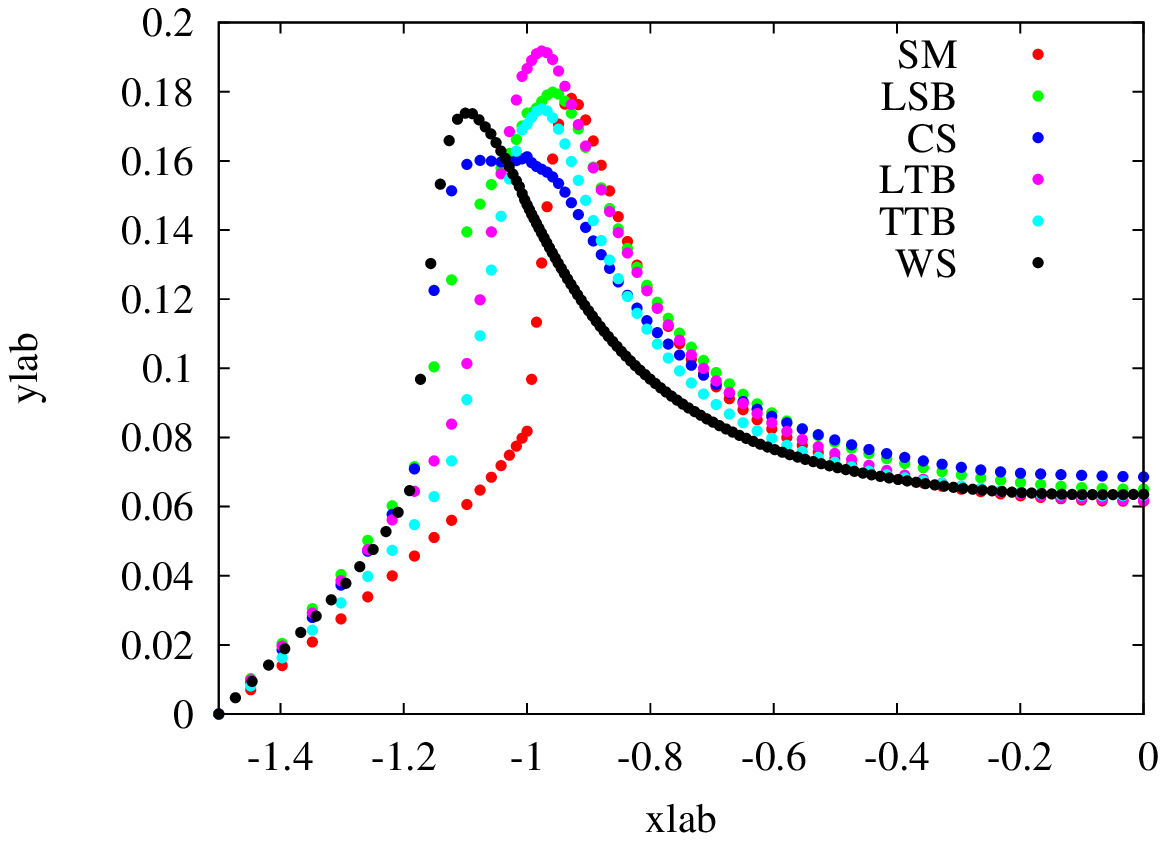}
\psfrag{ylab}[][]{$ C_{v\theta}$}
\psfrag{xlab}[][]{$x_2 $}
\includegraphics[width=5.05cm]{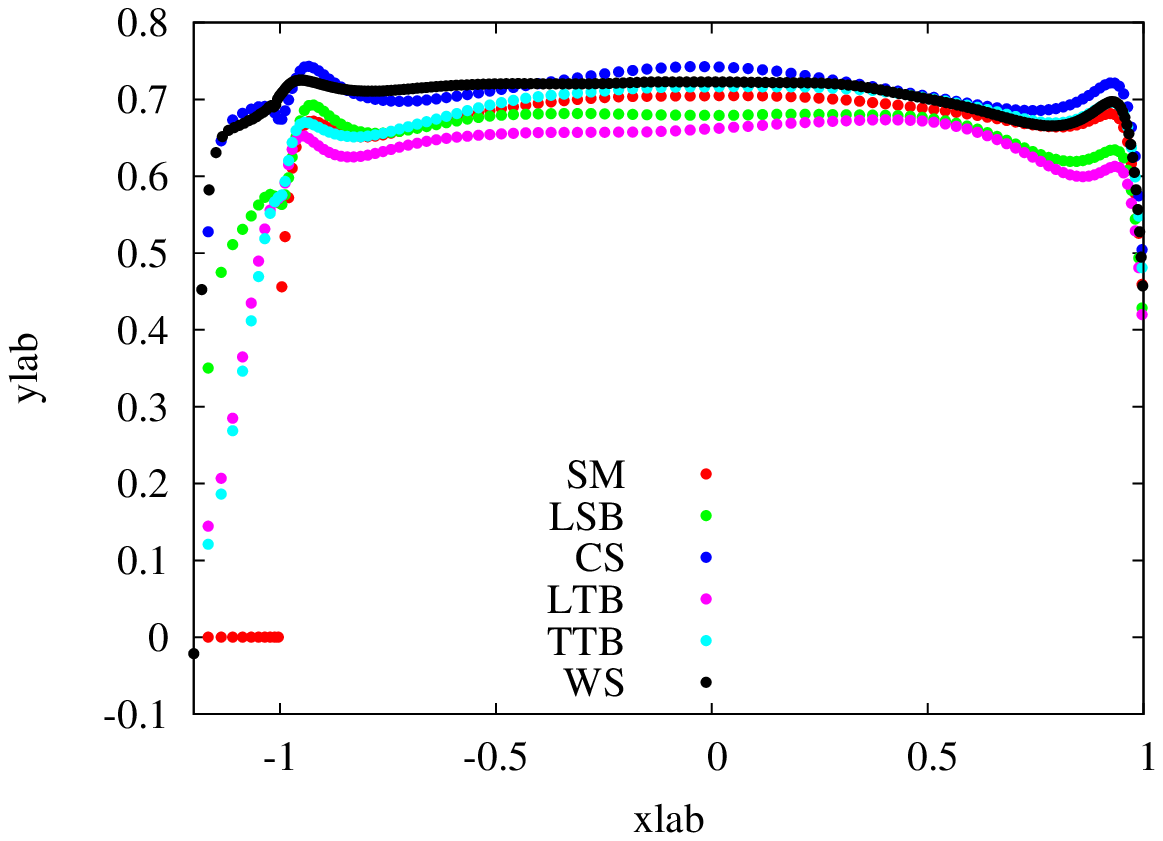}
\vskip -0.1cm
\hskip 1.25cm a) \hskip 5.25cm b) \hskip 5.25cm c)
\vskip -0.05cm
\caption{Profiles of a) vertical velocity variance, b) temperature variance,
and c) $v\theta$ correlation coefficient,
for different surfaces as indicated by the insets in panels b) and c).
}
\label{fig7}
\end{figure}

The surface contours of $u_2$ given in figure \ref{fig6} lead to the
same conclusions concerning the orientation of the structures. In addition,
comparing figure \ref{fig5} and \ref{fig6},
the qualitative conclusion of a higher correlation between
$u_2$ and $\theta^{\prime}$ for the three-dimensional surfaces
can be drawn. It seems also that for the other surfaces
these two quantities are less correlated. A more quantitative
result is obtained by evaluating the correlation coefficient
$C_{v\theta}=<u_2\theta^\prime>/\sqrt{<u_2^2><\theta^\prime 2>}$
evaluated over a large number of fields, which is plotted in 
figure~\ref{fig7}c. The values of
$C_{v\theta}$ ($C_{v\theta}=0.4$) are  low at the plane of the crest
for $SM$ and for the geometries with two-dimensional bars.
On the other hand $C_{v\theta}$ is more uniform in the near-wall
region for walls with three-dimensional roughness,
implying that the velocity fluctuations begin to be linked to
the thermal fluctuations in the region within the
solid elements. 
The $C_{v\theta}$ correlations grow in the region above the
surface and reach their maximum at the location where the 
production of $<\theta^{\prime 2}>$ 
($<u_2\theta^\prime>\frac{\partial <\theta>}{\partial x_2}$
has a maximum. This point is located within the thermal boundary
layer, which is however only well defined for smooth surfaces.

Profiles of the vertical velocity variance $v_2=<u_2^2>^{1/2}$ and 
of the temperature variance
$T^\prime=<\theta^{\prime 2}>^{1/2}$
obtained by averaging in the homogeneous direction and in time 
show a different trend near rough surfaces. The $v_2^2$ start from
different  values at the plane of the crest (figure \ref{fig7}a), 
depending on the shape of the hot surfaces, and reach the highest value at the
center (figure \ref{fig7}a). As expected, three-dimensional roughness
produces stronger values of $v_2$ at the plane of the crests
than those generated by two-dimensional roughness.
The main effect of the $u_2$ disturbances at the plane of
the crest is to promote a growth of $v_2$ proportional
to $y$ ($y=1-|x_2|$ for $|x_2|<1$), namely the distance from the surface.
In the near-wall region of the smooth surfaces instead
it has been observed that $v_2\approx y^2$, as shown in the inset
of figure \ref{fig7}a, where the lines are evaluated
in the region near the smooth upper wall.
The temperature variances $T^\prime$ depicted in figure \ref{fig7}b 
have a rather complex behavior
inside the roughness layer and near the plane
of the crests ($x_2=-1$). The differences among the values of $T^\prime$ for
$x_2<-1.2$ implies that the flow within the elements and
the shape of the solid elements do affect the thermal field
in the solid layer on which the elements
are mounted. 
Due to the enhancement of the lateral heat
transfer within the three-dimensional elements the
maximum of $T^\prime$ is  typically below $x_2=-1$. For the two-dimensional
surfaces the maximum, above the plane of the crests, moves 
closer to $x_2=-1$ with respect to that of $SM$ located
at the location of the maximum production of  $<\theta^{\prime 2}>$.
This behavior, as mentioned before discussing figure \ref{fig7}a,
implies that in the presence of rough surfaces the definition  of 
a thermal boundary layer thickness is no longer appropriate. 
Figure \ref{fig7}b in addition shows that nearly the same value 
of $T^\prime$ at the center is reached for all geometries.

The strong correlation between $\theta^\prime$ and $u_2$
producing the prevalence of $Q_T$ in the interior of the
fluid region can be understood in more detail by looking
at figure \ref{fig5} and figure \ref{fig6}, in which 
a strong correspondence of red and blue regions
of $u_2$ and $\theta^\prime$ can be appreciated, 
meaning that the turbulent
heat transfer is associated with hot ascending and cold descending
plumes. 
In addition it appears that $u_2$ surfaces
reproduce rather well the formation 
and orientation of the large-scale rollers.
Qualitatively the strong correlation between $u_2$ and $\theta^{\prime }$
more uniformly distributed near the three-dimensional
surfaces and extending deeply in the flow helps to understand
why the heat transfer is enhanced. The quantitative 
measure is obtained by the values of the $Nu$ numbers
in table \ref{table2}. The present simulations
have demonstrated that the highest heat transfer obtained by three-dimensional  
wedges is due the formation of strong plumes at the cusp of the solid elements.
Therefore it is worth investigating in
greater detail the differences between the statistics and the structures
generated by the $WS$ surface with respect to those by the
smooth wall when $Re$ and $Pr_S$ are changed.

\subsection{ Transitional regime }

\subsubsection{Time evolution and visualizations }

To investigate the transitional  regime simulations at low
and intermediate values of the Reynolds number should be performed. 
At low $Re$
the numerical procedure requires implicit treatment of the viscous terms 
to avoid the time step restriction  $\Delta t/\Delta x Re <2$.
The increase of CPU due to the inversion of tridiagonal
matrices and to the enhancement of  the data transfer among
the processor is not so penalising, at low $Re$, because 
the small velocity and temperature  gradients require less resolution.

\begin{figure}
\centering
\hskip -1.0cm
\psfrag{ylab}[][]{$ q_2$}
\psfrag{xlab}[][]{$t $}
\includegraphics[width=7.75cm]{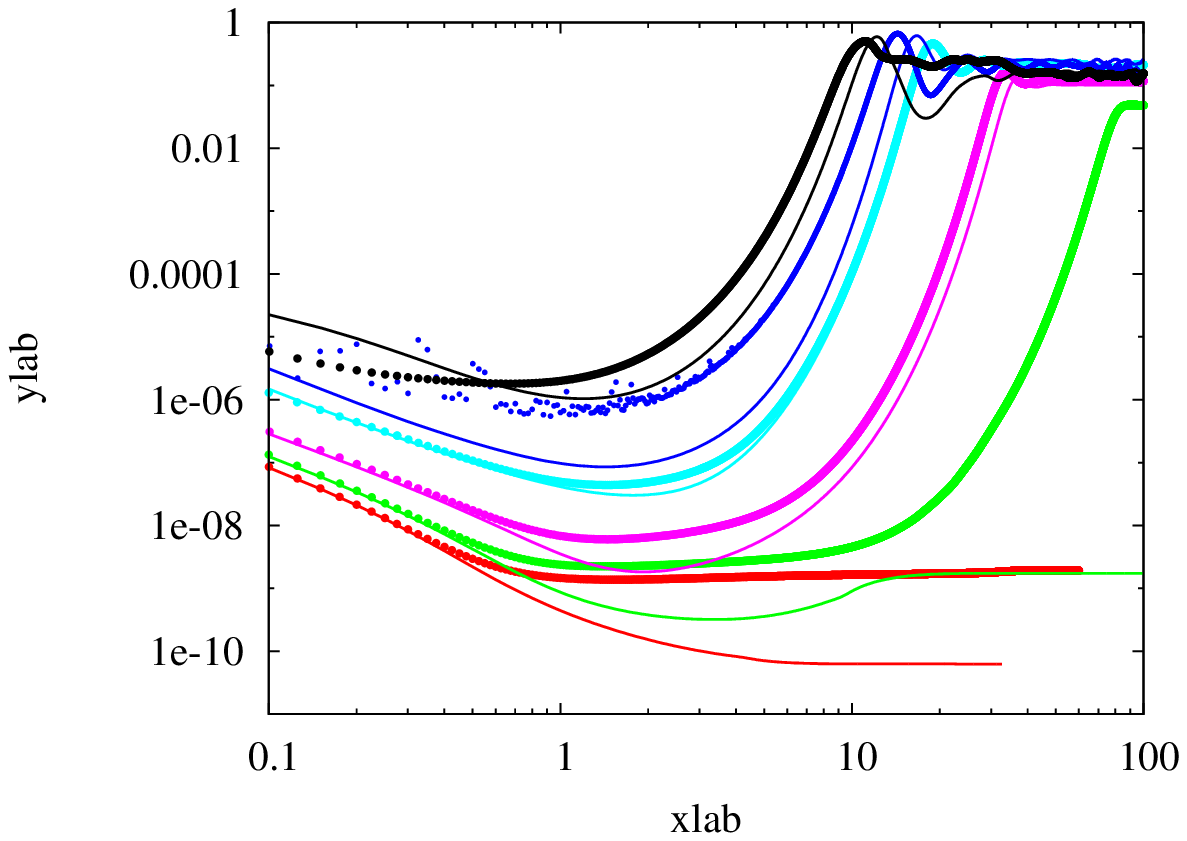}
\hskip -0.5cm
\psfrag{ylab}[][]{$ q_3 $}
\psfrag{xlab}[][]{$t $}
\includegraphics[width=7.75cm]{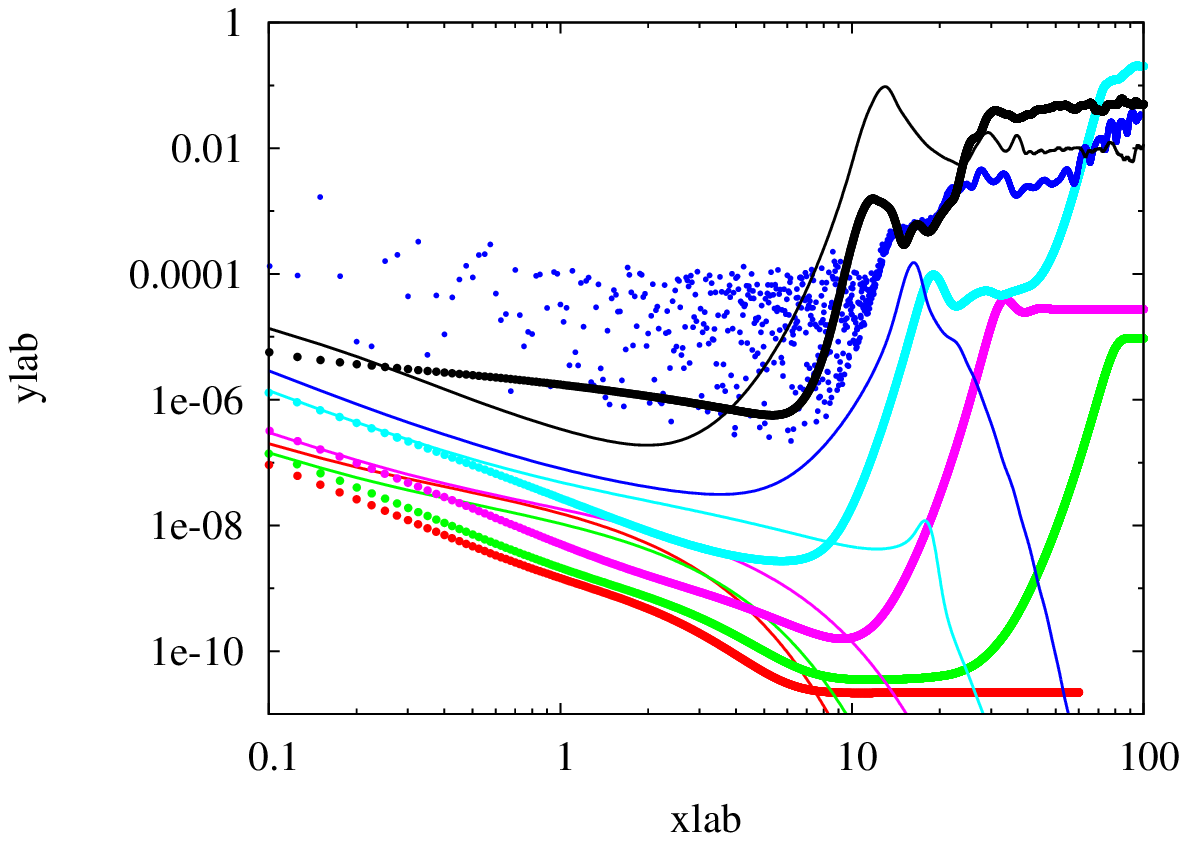}
\vskip -0.55cm
\hskip 5.25cm a) \hskip 6.5cm b)
\caption{Time history of integrated squared velocity disturbances: a) $q_2$ and
b) $q_3$ at red $Re=10$,
green $Re=12.5$,
magenta $Re=20$,
cyan $Re=50$,
blue $Re=75$,
and black $Re=250$;
symbols $WS$, lines $SM$.
}
\label{fig8}
\end{figure}

The experiments of \citet{silveston_58} confirmed
the theoretical value $Ra_c=1707$, 
of the transitional Rayleigh number for smooth walls.
In addition, the experiments showed that 
$Nu$ was growing as $Ra^n$, with
large variations of $n$ for $1078<Ra<10^6$. To investigate
whether the present numerical simulations with conducting
solid walls reproduce these observations, at $t=0$
a uniform temperature
distribution along the $x_1$ and $x_3$ directions is imposed in the fluid layer
with $\frac{\partial \theta}{\partial x_{2}}=1$,
and the flow instability 
is triggered by assigning 
random velocity disturbances.
The time evolution of the integrated velocity disturbances
$q_i=\int \overline{u_i^2}^{1/2}dx_2$ are analysed in figure~\ref{fig8}.
Initially, the velocity disturbances
decrease due to viscous energy dissipation
at the small scales. The large viscosity
at low Reynolds number,
also dissipates energy at the large scales leading to a laminar state
characterised by a linear temperature profile,
irrespective of the shape  of the surface. For smooth 
conducting surfaces at $Pr_S=0.0134$ (highly conductive material)
the temperature at the walls are found to be $\pm 0.993$. For the $WS$ surface 
we find $\theta_R=0.84$ at the plane of the crest,
and $\theta_S=-0.94$ at the opposite smooth wall.
To understand why $|\theta_R| < |\theta_S|$,
for $Ra<Ra_c$, it should be recalled that  inside the solid 
elements the temperature decreases less than 
in the fluid regions surrounding the solid elements, as $Pr_F>Pr_S$. 
The different decrease of temperature in the layer with the
solid elements produces undulations of $\theta$ in the fluid layer
near the plane of the crests, whose amplitude 
rapidly decreases moving towards the centre.
Regarding the flow structures at $Ra < Ra_c$, the contours 
of the velocity components $u_1$ and $u_2$ show at  a certain
time the formation of two convective cells whose
strength decreases in time in the case of smooth walls.
This can be appreciated
by the time evolution of $q_2$ in figure \ref{fig8}a.
For $WS$ the motion within the rough elements 
does not largely affect  the large-scale cells. 
The three-dimensional velocity and temperature disturbances 
generated by the $WS$ surfaces are not strong enough to
trigger the spanwise motion on the large scale cells.
The flow structures occurring in the initial time evolution for
$Ra< Ra_c$ are not presented, being
similar to those at $Re=20$, discussed later on.

Figure \ref{fig8} shows the time history of the $q_i$
($q_1$ is not plotted being
similar to that of $q_3$).
The discussion about the  large values found for the correlation coefficient
$C_{v\theta}$ suggests to get an exponential growth 
of $q_\theta$ close to that of $q_2$.
Indeed this has been observed
with  a time behavior for $q_\theta$ similar to that in figure \ref{fig8}a.
The difference is in the initial evolution because of the absence
of $\theta$ fluctuations at $t=0$.
At any $Re$ (the corresponding values of $Ra$, at the end
of the simulations, are given in the figure caption), the decrease
in the short initial period,  occurs for the 
dissipation at different rates of the different scales
contributing to  $q_i$.
At $Re=10$, $q_2$ in figure \ref{fig8}a
reaches a constant level which depends on the shape of the surface.
On the other hand, at $Re=12.5$, while 
$q_2$ tends to a constant for $SM$, it 
grows exponentially for $WS$.
This is a first indication
that the critical Rayleigh number for $WS$ is smaller than
than for smooth wall. In figure \ref{fig8}b it is seen that 
at $Re<50$, for both geometries $q_3$ does not reach values close to $q_2$,
which is the condition supporting the formation of large amplitude  
three-dimensional instabilities.  For the $WS$ geometry at
$Re=12.5$ (green symbols) the value $q_3=0.00001$ at $t=100$ 
accounts for the three-dimensional disturbances within
the roughness, which affect a thin layer of fluid near the plane of
the crests. 
\begin{figure}
\centering
\hskip -1.0cm
\includegraphics[width=4.05cm]{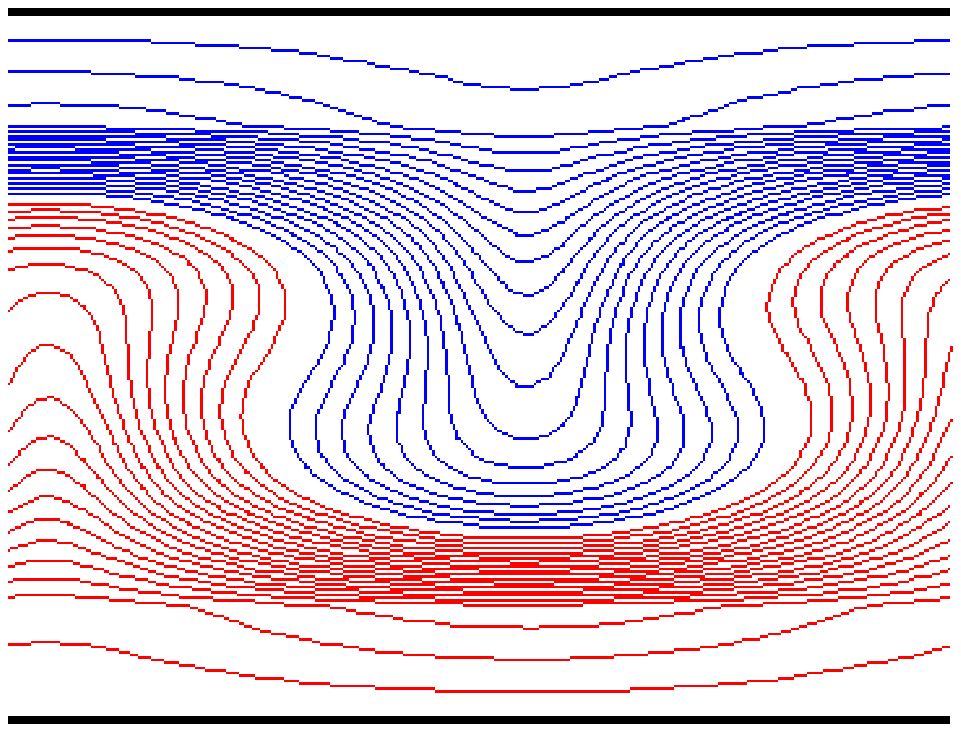}
\hskip -0.5cm
\includegraphics[width=4.05cm]{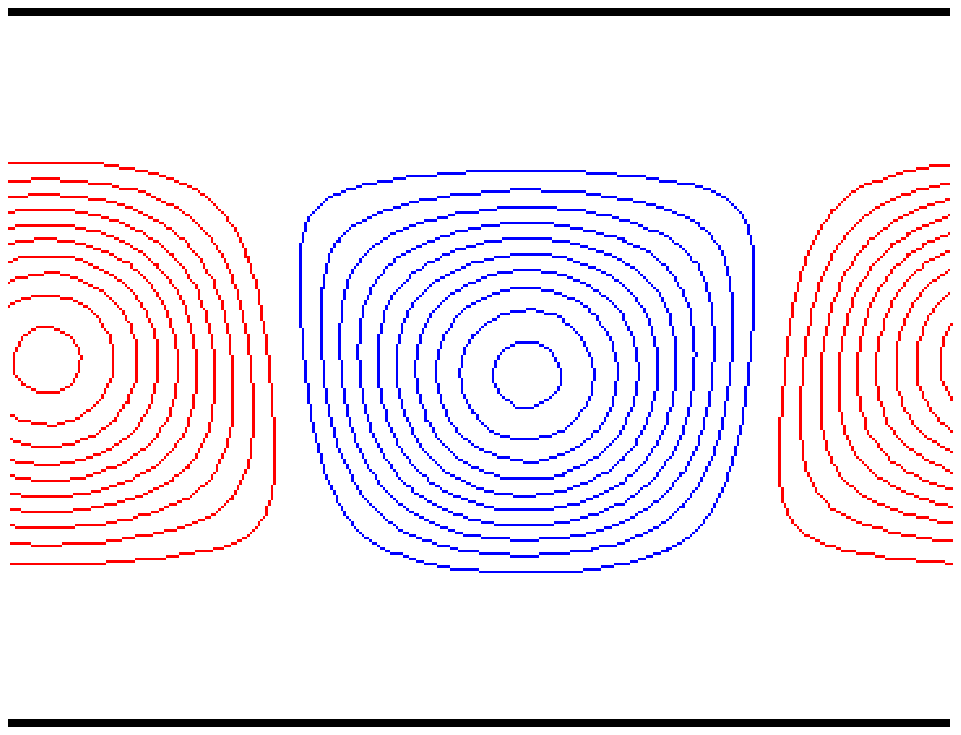}
\hskip -0.5cm
\includegraphics[width=4.05cm]{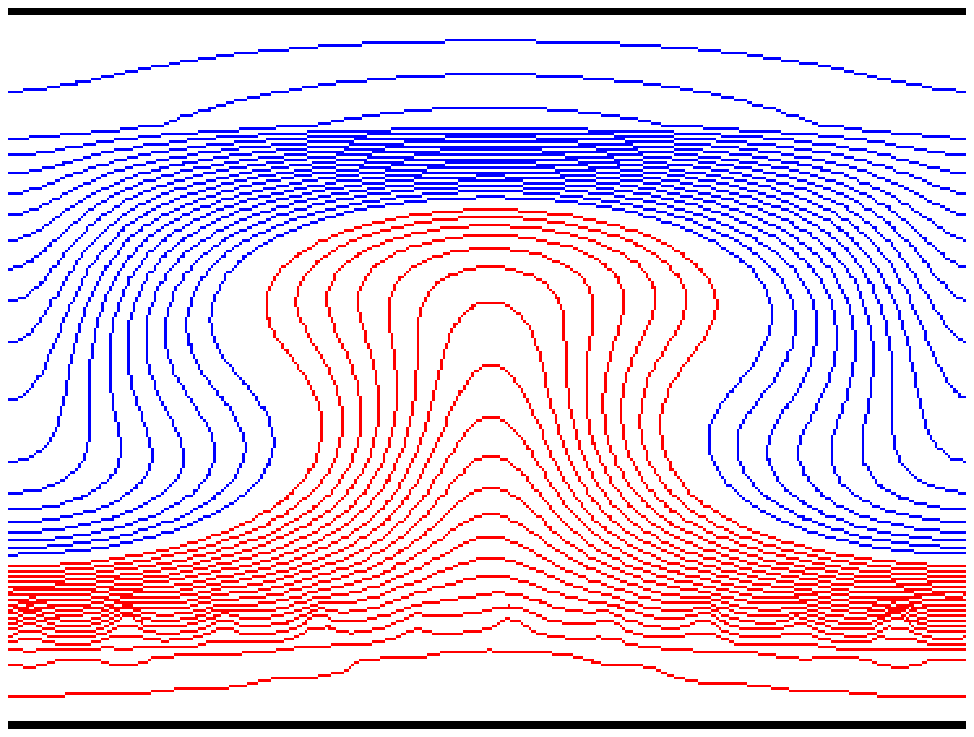}
\hskip -0.5cm
\includegraphics[width=4.05cm]{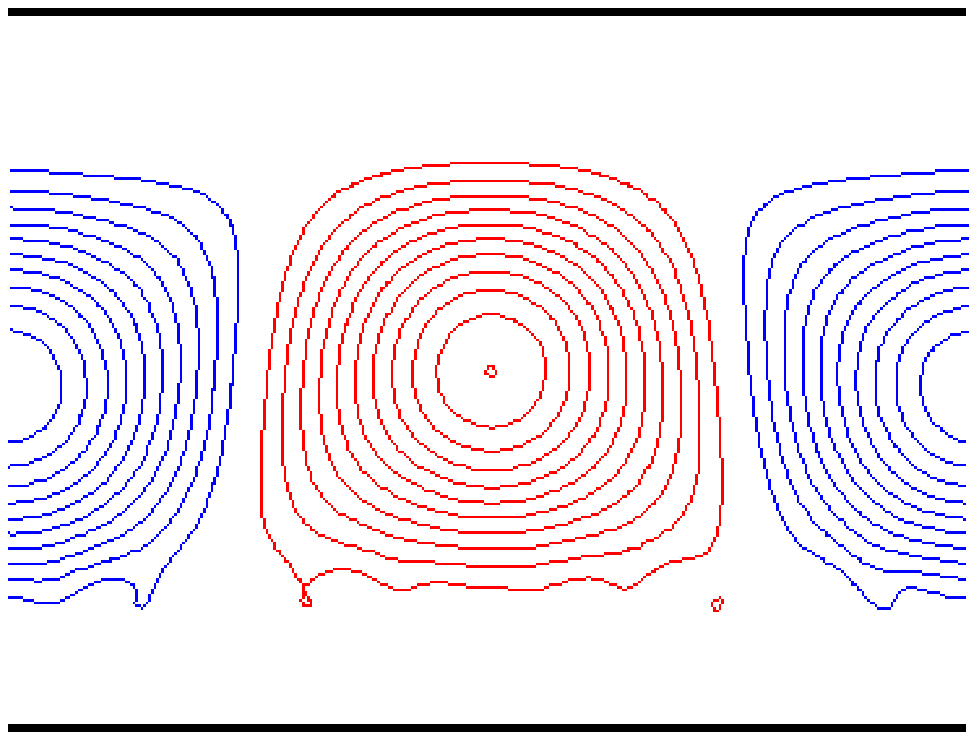}
\vskip -0.1cm
\hskip 1.25cm a) \hskip 2.25cm b) \hskip 2.25cm c) \hskip 2.25cm d)
\vskip -0.05cm
\caption{
Flow at $Re=20$, $Pr_S=0.134$:
Contour plots in a $x_1-x_2$ plane at
$x_3=2$ of $u_2$ (b, d) and of $\theta$ (a, c),
for smooth walls (a, b) and for the $WS$ geometry (c, d).
Contours are given in intervals $\delta=0.05$, blue negative,
red positive. 
}
\label{fig9}
\end{figure}
\noindent At $Re=20$ both flows have a  similar growth of $q_2$ 
(magenta in figure \ref{fig8}a), and the corresponding
flow structures at $t=100$ are depicted in figure \ref{fig9}  
trough $\theta$ and $u_2$ contour
plots in a $x_1-x_2$ plane at $x_3=2$. The 
temperature distribution  inside the solid are better visualized at
$Pr_S=0.134$ than at $Pr_S=0.0134$, where the contours are clustered
near the plane of the crests.  The $u_2$ contours in figure \ref{fig9}b ($SM$)
and figure \ref{fig9}d ($WS$) are similar except for 
a spatial shift.
In the latter case the small disturbances
emerging from the interior of the rough surface are barely
appreciated, and do not affect the contours in the central part of
the domain. The strong $\theta$ undulations inside the roughness 
(figure \ref{fig9}c) do not protrude into the flow, in fact,
far from the plane of the crests the contours of $\theta$
in figure \ref{fig9}a are similar to those in figure \ref{fig9}c.

In figure \ref{fig8}a 
the time history of $q_2$ at $Re=20$ (magenta) and $Re=50$ 
(cyan) for the $SM$ flow are very similar, however
the visualizations at $Re=50$ (not shown)
show two vertically oriented cells 
rather than a single cell.
The same number of cells also appear
for $WS$, with the $u_2$ and $\theta$ disturbances due 
to the three-dimensional surface, 
as at $Re=20$, localised in a thin layer of fluid
near the plane of the crest. This was observed at
$t=30$, at the end of the exponential growth. 
\begin{figure}
\centering
\hskip -1.0cm
\includegraphics[width=4.05cm]{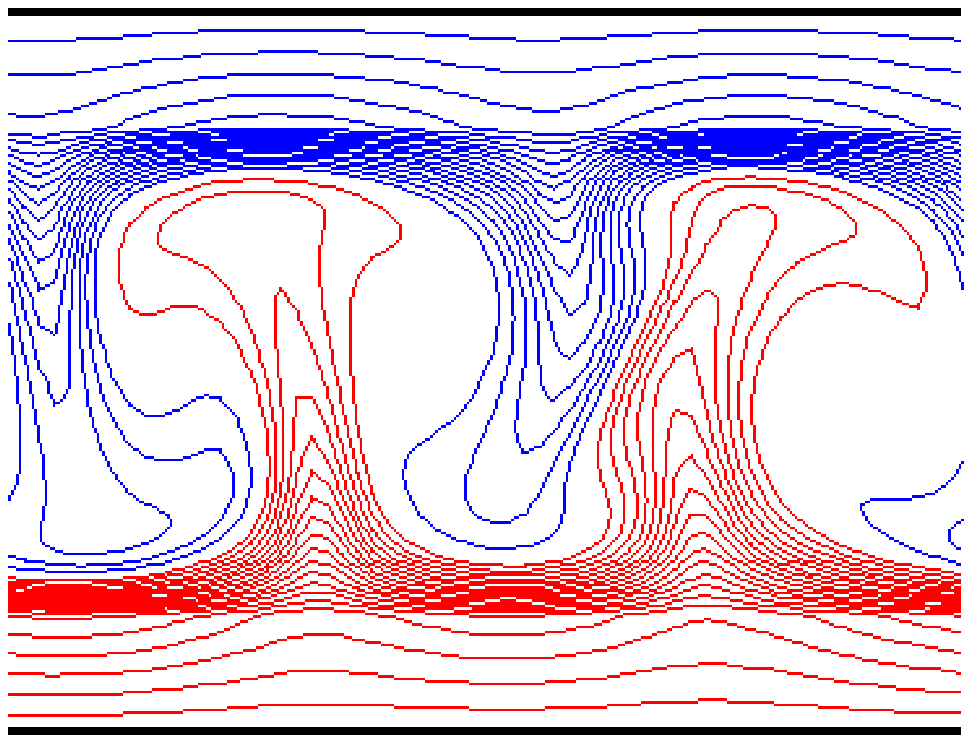}
\hskip -0.5cm
\includegraphics[width=4.05cm]{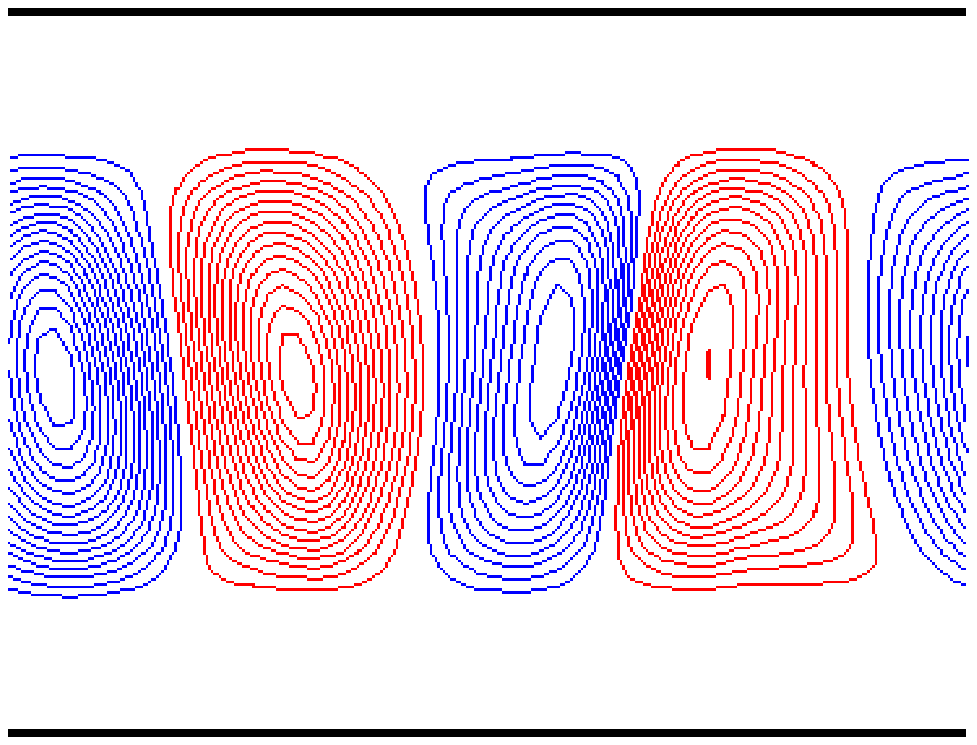}
\hskip -0.5cm
\includegraphics[width=4.05cm]{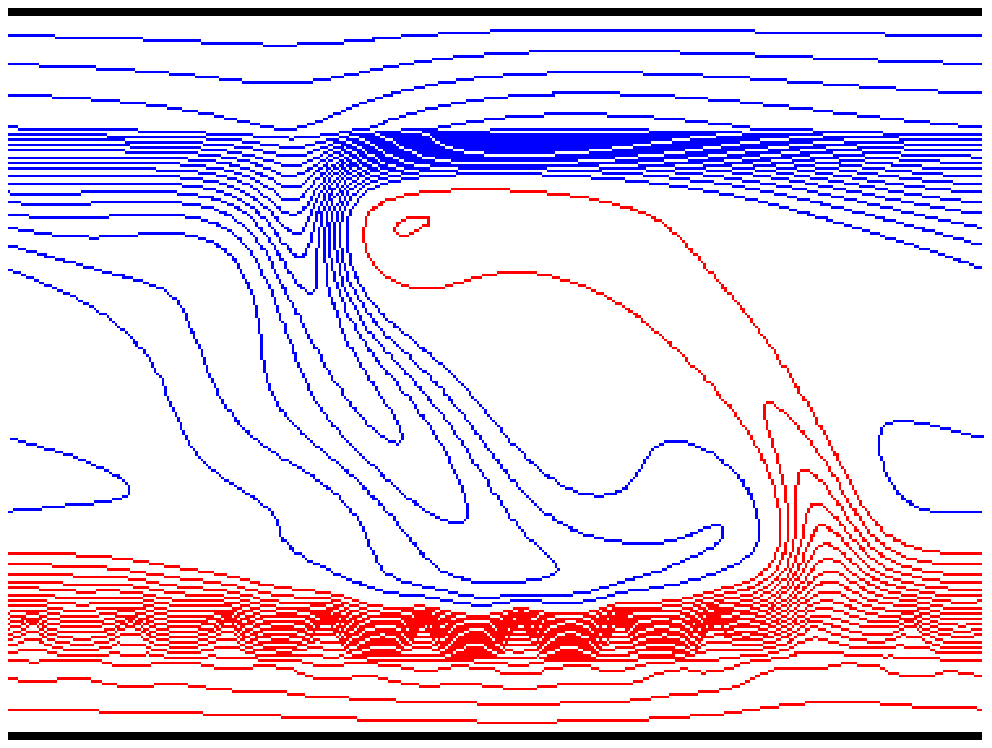}
\hskip -0.5cm
\includegraphics[width=4.05cm]{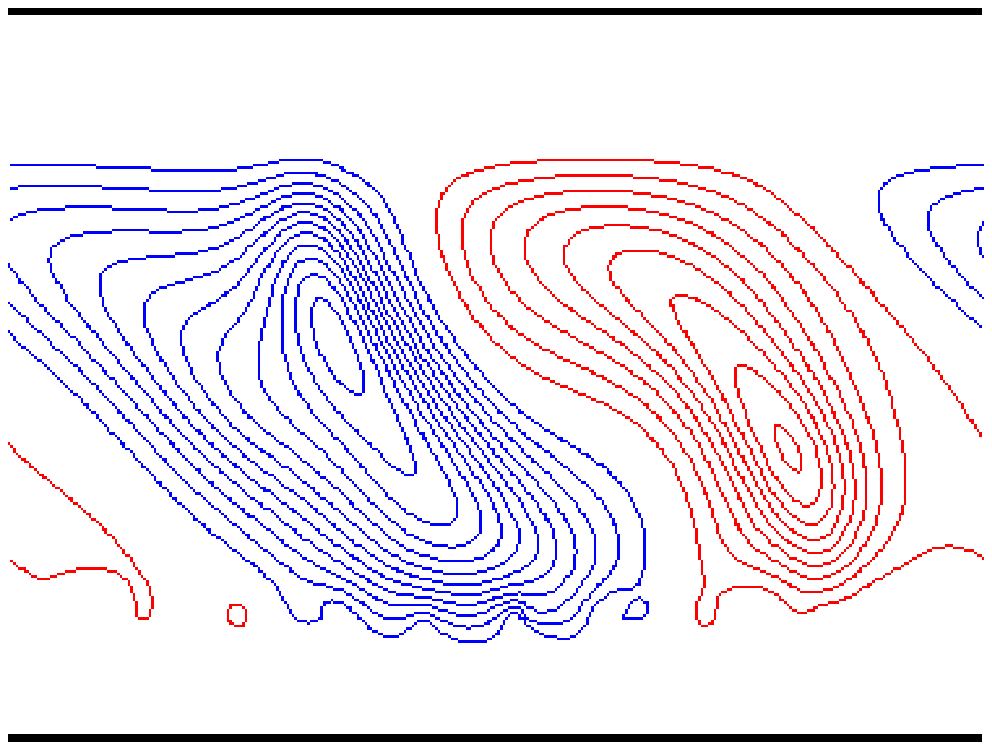}
\vskip -0.1cm
\hskip 1.25cm a) \hskip 2.25cm b) \hskip 2.25cm c) \hskip 2.25cm d)
\vskip -0.05cm
\caption{
Flow at $Re=75$, $Pr_S=0.134$:
Contour plots in a $x_1-x_2$ plane at
$x_3=2$ of $u_2$ (b, d) and of $\theta$ (a, c),
for smooth walls (a, b) and for the $WS$ geometry (c, d).
Contours are given in intervals $\delta=0.05$, blue negative,
red positive. 
}
\label{fig10}
\end{figure}
\begin{figure}
\centering
\hskip -1.0cm
\includegraphics[width=4.05cm]{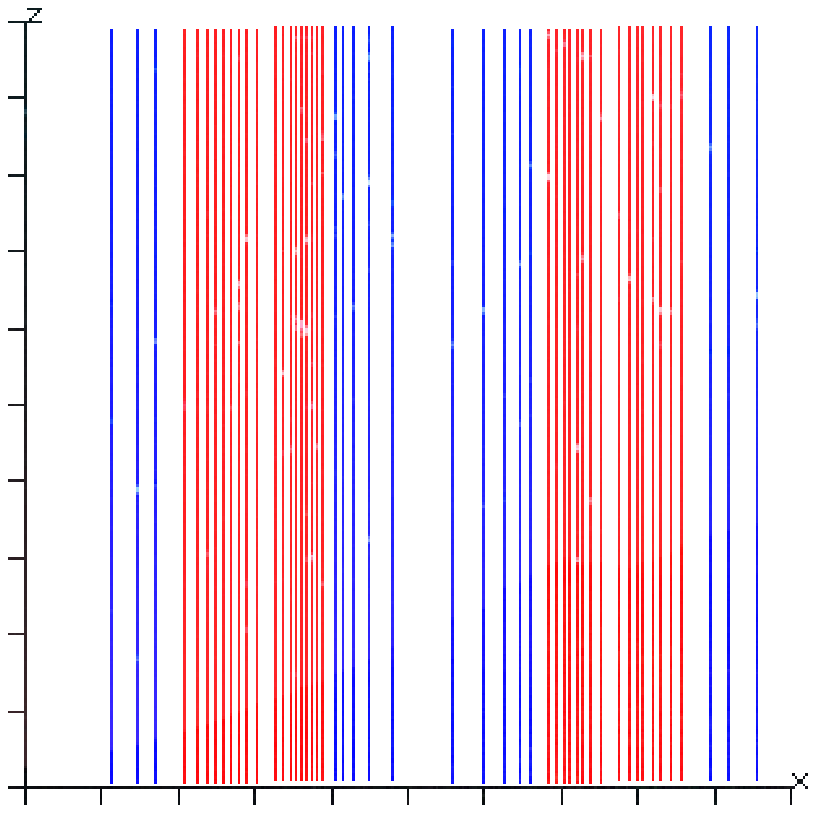}
\hskip -0.5cm
\includegraphics[width=4.05cm]{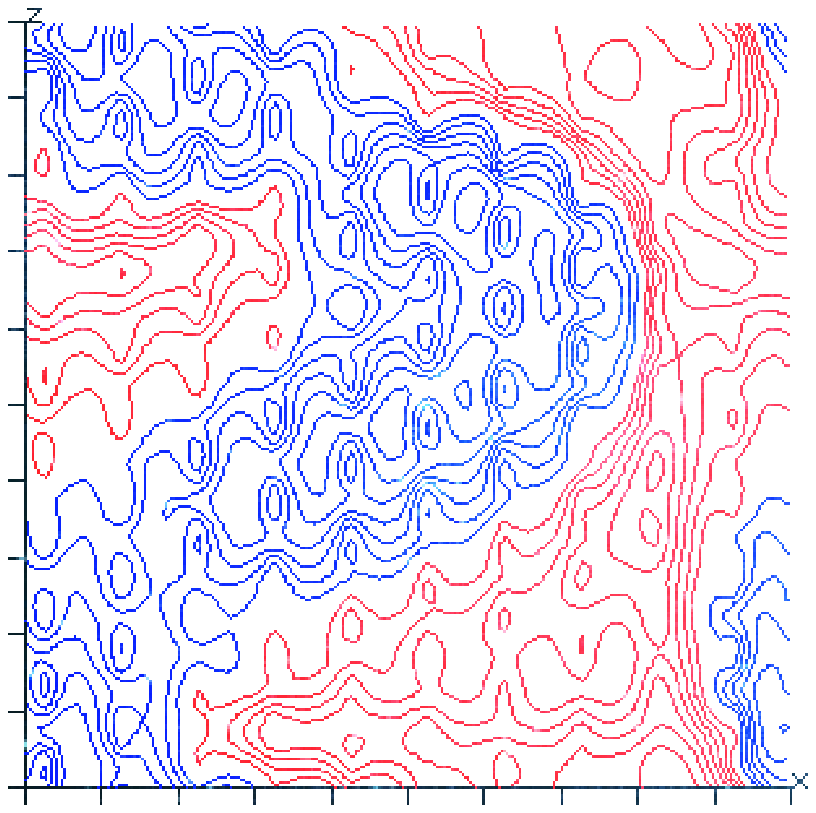}
\hskip -0.5cm
\includegraphics[width=4.05cm]{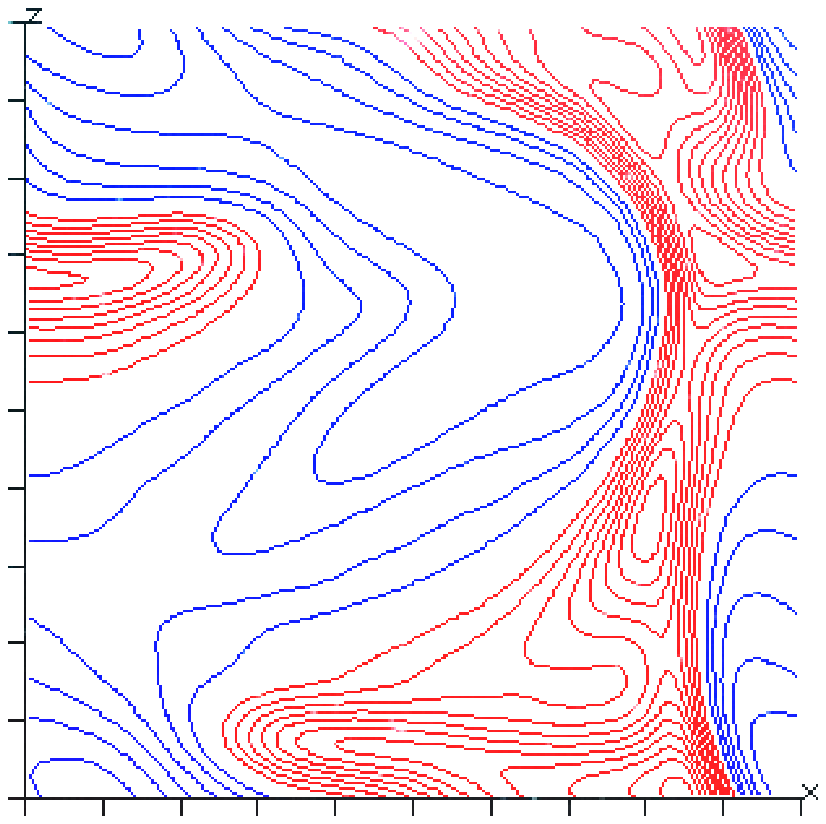}
\hskip -0.5cm
\includegraphics[width=4.05cm]{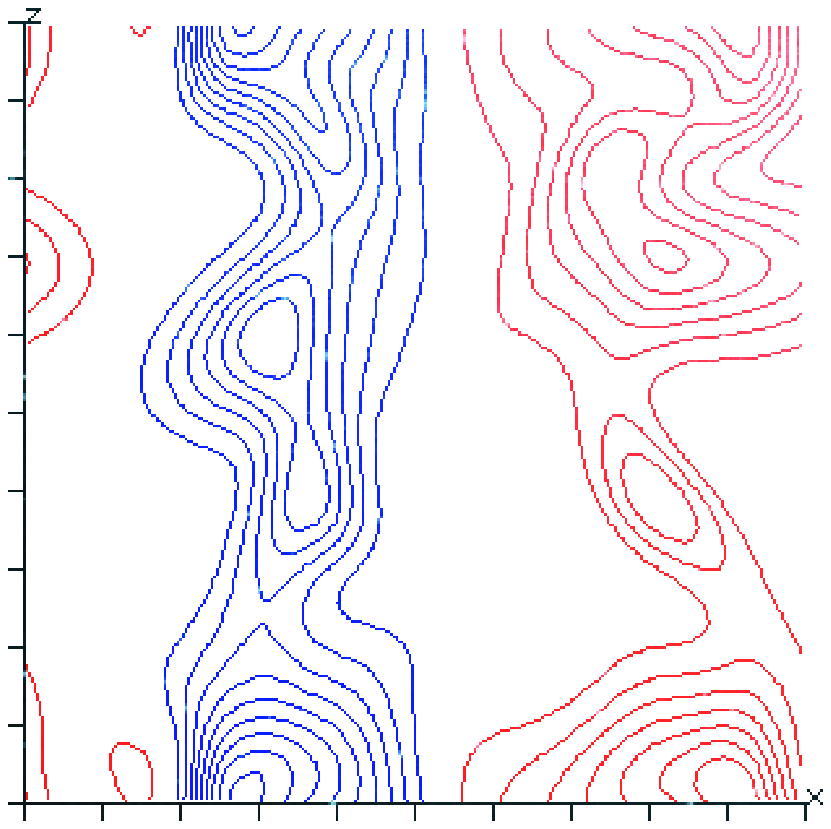}
\vskip -0.1cm
\hskip 1.25cm a) \hskip 2.25cm b) \hskip 2.25cm c) \hskip 2.25cm d)
\vskip -0.05cm
\caption{
Flow at $Re=75$, $Pr_S=0.134$:
Contour plots of $\theta^\prime$ in a $x_1-x_3$ plane at
$x_2=-0.76$ for $SM$ (a) and $WS$ (b),
at $x_2=-0.99$ for $WS$ (c), and
at $x_2=0$ for $WS$.
Contours are given in intervals $\delta=0.05$, blue negative,
red positive. 
}
\label{fig11}
\end{figure}
\noindent The undulations of the $q_i$, after the exponential growth, 
in particular in figure \ref{fig8} are related to the unsteadiness 
of the structures.
In this figure the second growth 
for $q_3$ at $Re=50$ for $WS$ (cyan symbols) is due 
to the formation of disturbances
oriented in the spanwise direction. In figure \ref{fig8}b 
there is a similarity between the evolution of $q_3$ 
for $Re=75$ (blue symbols) and for $Re=50$; the difference is
the time when the fast growth is initiated. 
The higher amplitudes of $q_3$ for $Re=75$
allow better  visualizations. 
Figure \ref{fig10}a and figure \ref{fig10}b indeed shows, for smooth walls,
the formation of two slightly
inclined cells. The cells oscillate from left to right in time 
producing the small undulations of the blue line in 
figure \ref{fig8}a. For this flow the
spanwise disturbances (indicated by $q_3$) grow for a short time 
($8<t<16$) to reach a small value ($q_3=0.0001$), 
and then sharply decrease 
(blue line in figure \ref{fig6}b). Differently, for the $WS$ flow
the spanwise velocity disturbances grow continuously in time, 
with resulting change in the shape of the main rollers. 
Figures \ref{fig10}c and figure \ref{fig10}d indeed depict the formation of strong
inclined thermal plumes. In these
circumstances the flow has a chaotic behavior 
as emerges from the oscillations in figure \ref{fig8}b,
therefore the spanwise disturbances do not have a well defined wavenumber.

Contour plots of $\theta^\prime$ in wall-parallel $x_1-x_3$ planes 
at different $x_2$ locations,
better emphasize the shape of the spanwise structures.
Indeed the parallel straight lines 
in figure \ref{fig11}a evaluated at $x_2=-0.76$ 
demonstrate the absence of thin temperature plumes with ordered or disordered
for the $SM$ geometry. 
The visualization at $x_2=0$   
(not reported) is similar, with the lines shifted to the left
or to the right, indicating that the thermal plumes in
figure \ref{fig10}a oscillate in the $x_1$ direction. On the other
hand, the velocity and thermal disturbances
generated  by the three-dimensional surface 
affect the shape of the plumes in a complicated way,
as shown in figure \ref{fig11}(b,c,d).  
The interaction between the disturbance of each wedge,
at $x_2=-0.99$ (figure \ref{fig11}b) causes the bending of 
the large-scale rollers, and, in certain regions
strong temperature gradients are created,
with local increase of the heat flux. 
At $x_2=-0.76$ (figure \ref{fig11}c) the effect
of the single solid element disappears but the shape of
the rollers, and the sharp $\theta^\prime$ gradients
are still visible.  At $x_2=0$ (figure \ref{fig11}d) the rollers 
have wave-like disturbances aligned along the $x_3$ direction. 
Near the smooth top  surface, at $x_2=0.76$
contours similar to those in figure \ref{fig11}c 
are obtained with blue and red interchanged, therefore this picture 
is not shown.

\begin{figure}
\centering
\hskip -1.0cm
\includegraphics[width=4.05cm]{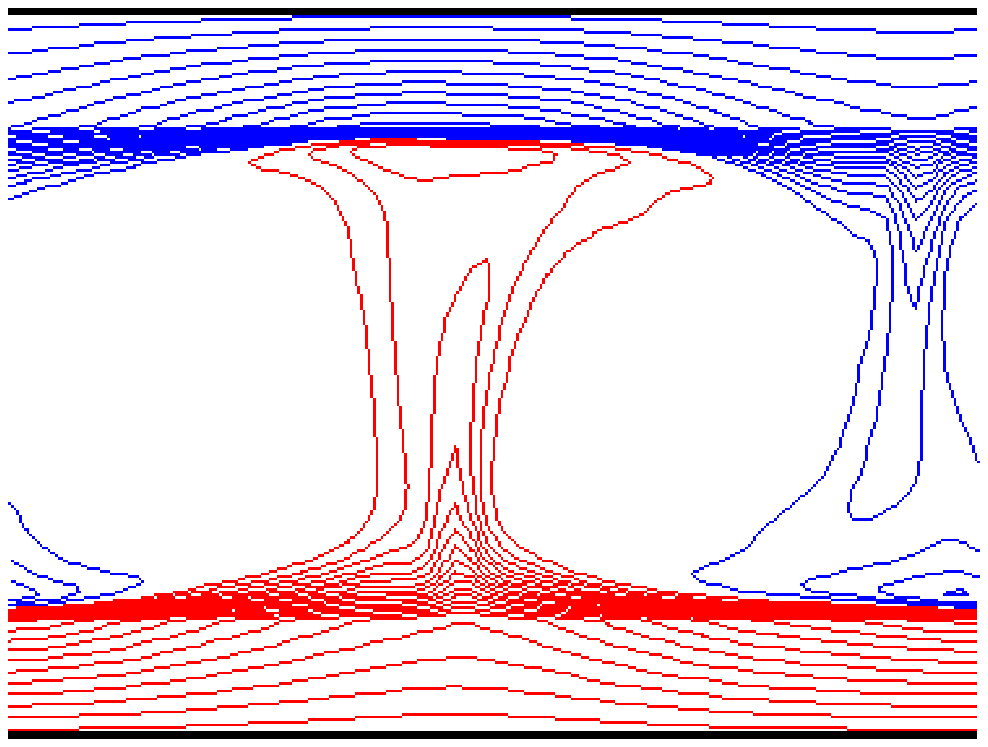}
\hskip -0.5cm
\includegraphics[width=4.05cm]{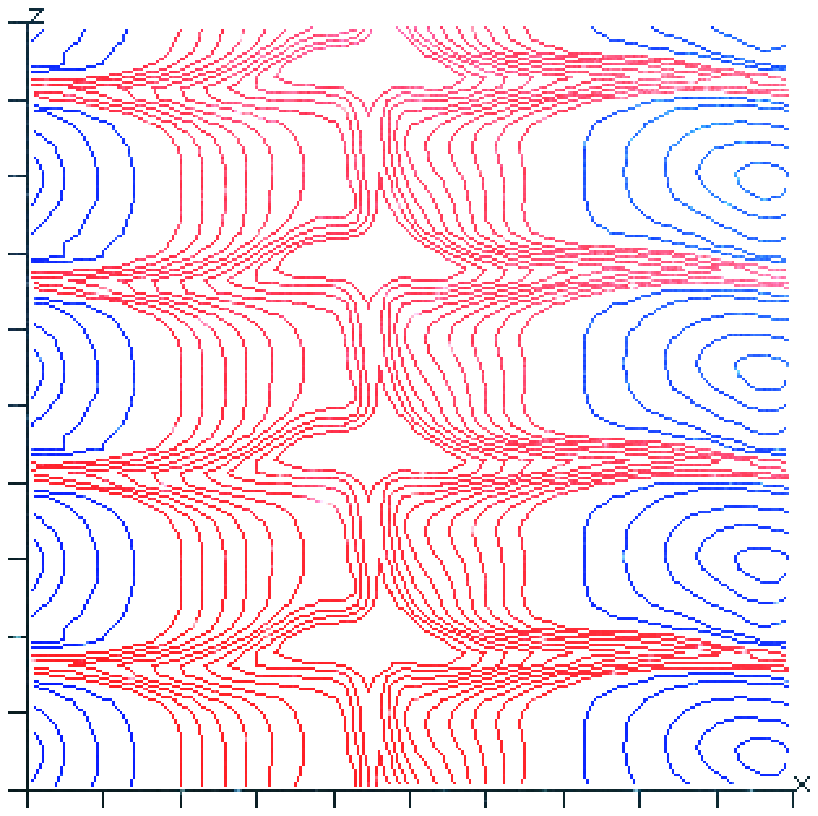}
\hskip -0.5cm
\includegraphics[width=4.05cm]{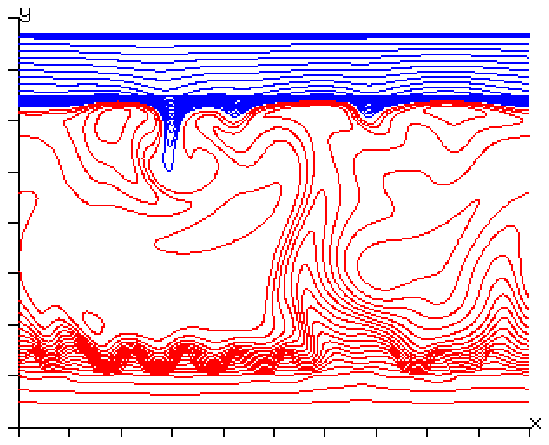}
\hskip -0.5cm
\includegraphics[width=4.05cm]{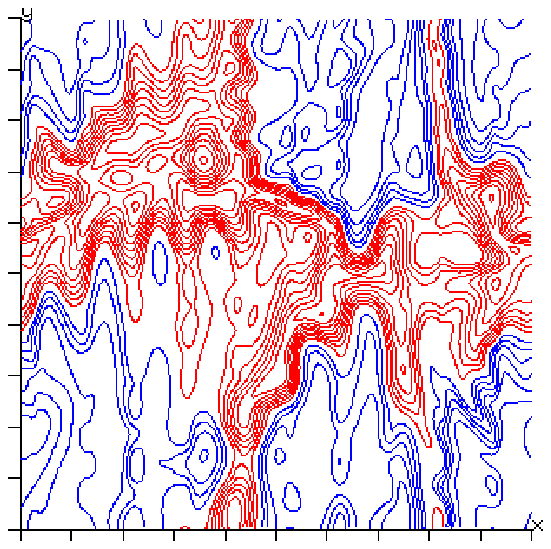}
\vskip -0.1cm
\hskip 0.25cm a) \hskip 2.25cm b) \hskip 2.25cm c) \hskip 2.25cm d)
\vskip -0.05cm
\caption{
Flow at $Re=250$, $Pr_S=0.134$:
Contour plots of $\theta^\prime$ 
in a $x_1-x_2$ plane at
$x_3=2$ for $SM$ (a) and $WS$ (c), and
in a $x_1-x_3$ plane at
$x_2=-0.976$ for $SM$ (b) and $WS$ (d).
Contours are given in intervals $\delta=0.05$, blue negative,
red positive. 
}
\label{fig12}
\end{figure}

At $Re=250$ the $SM$ and the $WS$ surfaces produce only one cell 
with high temperature gradients near the walls, with that 
for $SM$, in figure \ref{fig12}a,  better defined than
than for $WS$ in figure \ref{fig12}c. 
The visualizations in the $x_1-x_2$ planes, for the $SM$ surface 
do not allow to make conjectures on  the formation of spanwise disturbances.
The formation, growth and persistence of these disturbances
is the key process
leading  to have a fully turbulent regime when the Reynolds number increases.
Indeed the appearance of the spanwise disturbances at $Re=250$ for 
$SM$ could be conjectured based on the time evolution of $q_3$
in figure \ref{fig8}b. 
In fact the black line for $SM$ at $t>30$ was 
tending to a constant value smaller than for $WS$. To better see the shape
of the spanwise disturbances the contours of $\theta$ in $x_1-x_3$
planes are presented at $x_2=-0.76$ in figure \ref{fig12}b for
$SM$ and in figure \ref{fig12}d for $WS$. 
For $SM$ a short wave instability in the $x_3$ direction
is visible, whereas for $WS$ a more complex thermal structure than
previously discussed at $Re=75$ (figure \ref{fig11}c), can be observed at 
$x_2=-0.76$. As in figure \ref{fig11}c, at this distance
from the wall, the imprinting of the solid elements is barely appreciated.

As previously shown at 
$Re=500$ in figure \ref{fig5} and in figure \ref{fig6}
the shape of  $\theta^\prime$ surfaces (smoother than that of $u_2$) does
not largely differ from that of $u_2$,
and with a correspondence of the negative and of the positive values.
At $Re=500$ the surfaces were corrugated implying that
flows with a wide range of scales were generated, and
the spectra were highlighting the
energy and thermal distribution among the scales.
\begin{figure}[ht]
\centering
\hskip -1.0cm
\psfrag{ylab}[][]{$ Q_T,Q_F  $}
\psfrag{xlab}[][]{$x_2 $}
\includegraphics[width=7.25cm]{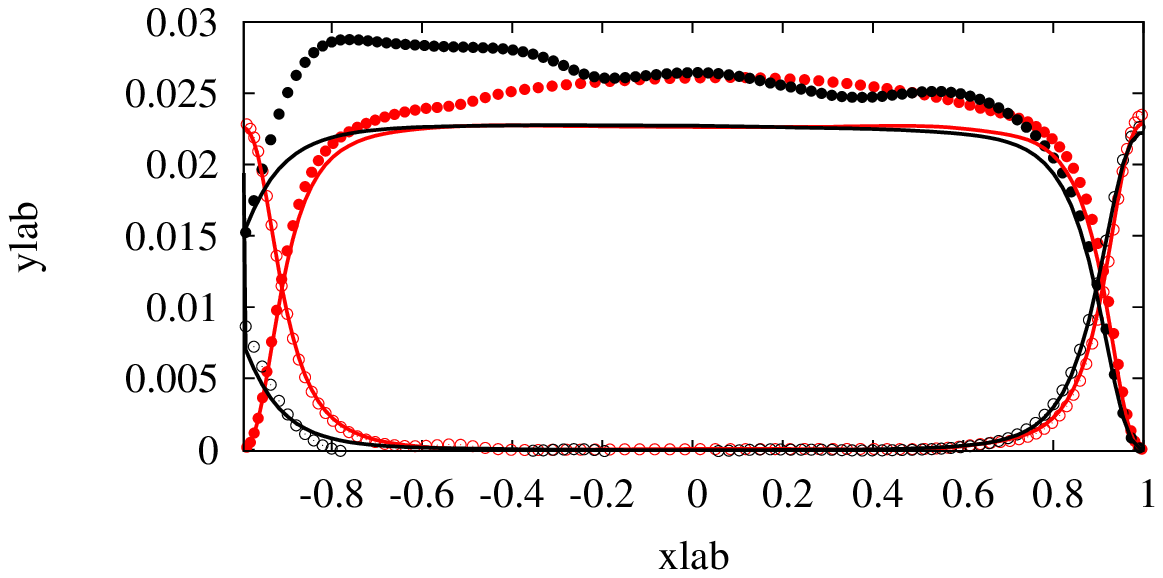}
\hskip -0.5cm
\psfrag{ylab}[][]{$ Nu_T,Nu_F  $}
\psfrag{xlab}[][]{$x_2 $}
\includegraphics[width=7.25cm]{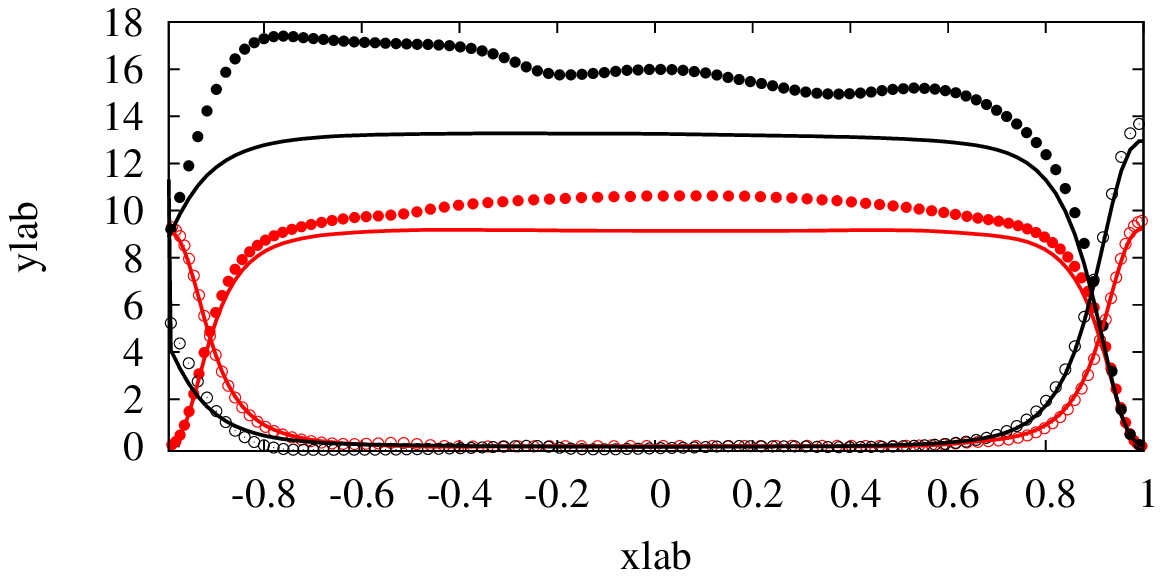}
\vskip -0.1cm
\hskip 0.25cm a) \hskip 6.25cm b) 
\vskip -0.1cm
\hskip -1.0cm
\includegraphics[width=4.25cm]{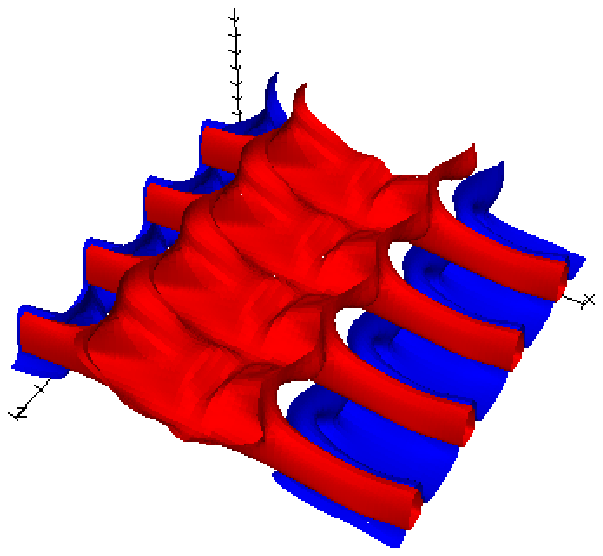}
\hskip -0.5cm
\includegraphics[width=4.25cm]{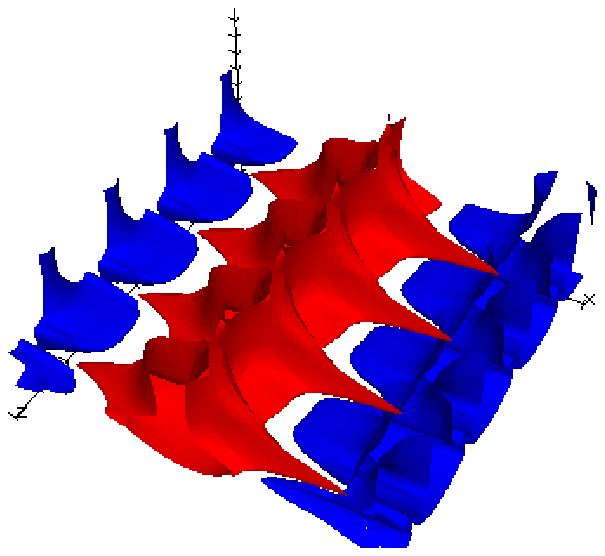}
\hskip -0.5cm
\includegraphics[width=4.25cm]{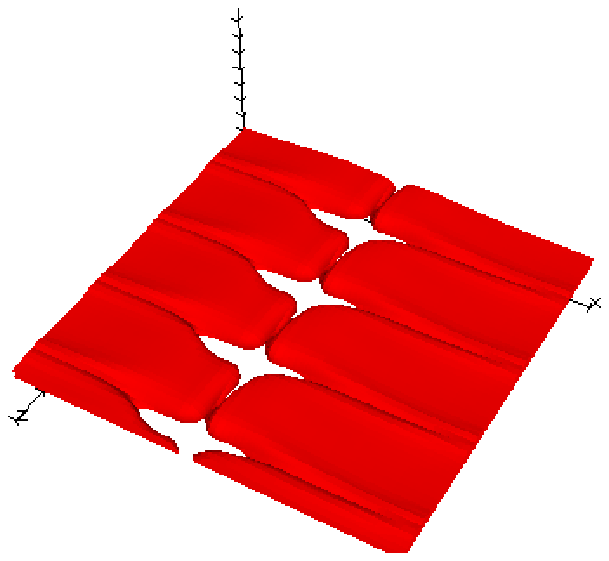}
\hskip -0.5cm
\includegraphics[width=4.25cm]{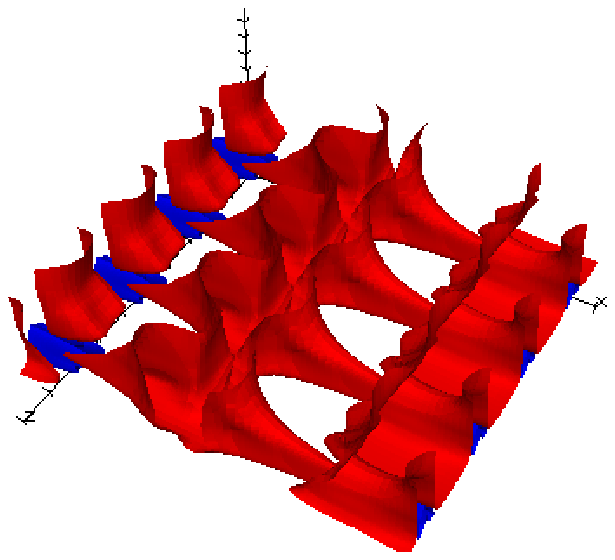}
\vskip -0.1cm
\hskip 0.25cm c) \hskip 2.25cm d) \hskip 2.25cm e) \hskip 2.25cm f)
\vskip -0.1cm
\hskip -1.0cm
\includegraphics[width=4.25cm]{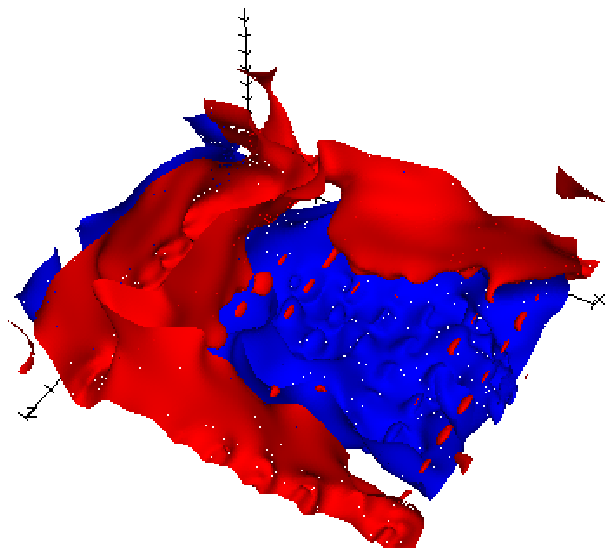}
\hskip -0.5cm
\includegraphics[width=4.25cm]{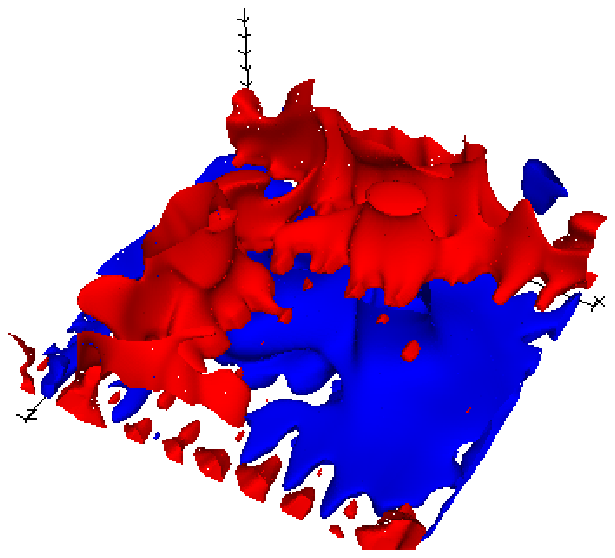}
\hskip -0.5cm
\includegraphics[width=4.25cm]{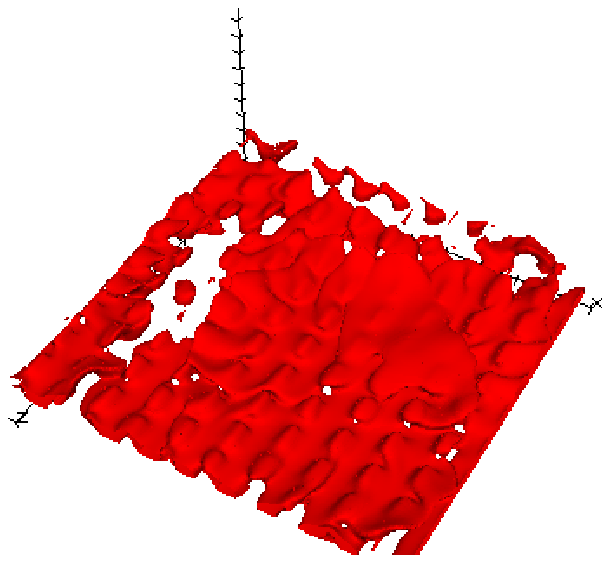}
\hskip -0.5cm
\includegraphics[width=4.25cm]{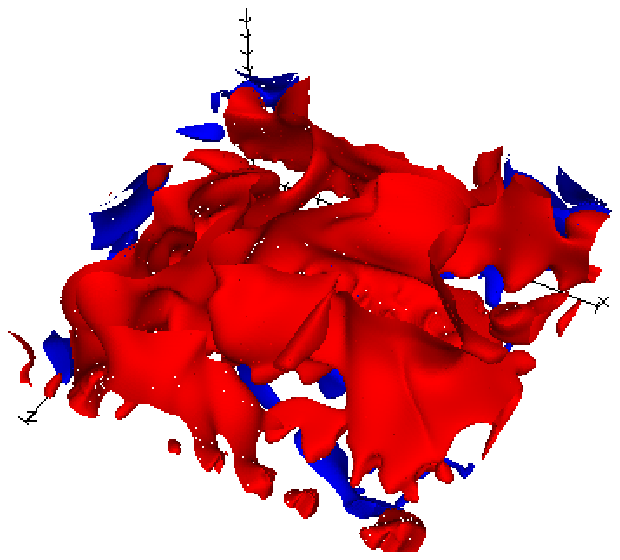}
\vskip -0.1cm
\hskip 0.25cm g) \hskip 2.25cm h) \hskip 2.25cm i) \hskip 2.25cm j)
\caption{Flow at $Re=250$, $Pr_S=0.134$: profiles of the heat fluxes $Q_T$ and $Q_F$ (a) and of their
contribution to the Nusselt number (b).
The solid symbols denote statistics taken by averaging over 
one realisation, in the text indicated by $\overline { }$;
solid lines denote time averaged. Red stands for $SM$, and black for $WS$.
Visualizations through surface contours are shown in panels b-f, 
for $u_2$ (c,g),
$\theta^\prime$ (d,h),
$\frac{1}{Re}\frac{\partial {\theta}}{\partial x_2}$ (e,i),
${\theta^\prime u_2}$ (f,j).
Data are shown for $SM$ in (c,d,e,f) and for
$WS$ in (g,h,i,j).
Blue and red contours correspond to negative red positive values, respectively,  
with value $\pm 0.1$ (c,d,g,h) and
$\pm 0.01$ (e,f,i,j).
}
\label{fig13}
\end{figure}
\noindent From the spectra, not shown here for lack of space,
it was possible to see that at
this Reynolds number the influence of the shape
of the surface was barely detected at the location
of maximum production of $<\theta^{\prime 2}>$.
In the spectra at $Re=250$ at the centerline there
are differences between the $SM$ and the $WS$, that did not
appear at $Re=500$.  At the higher Reynolds number the 
one-dimensional spectra of $u_2$ and $\theta$ in $x_1$ coincided
with those in $x_3$ at the center. On the other hand, at
$Re=250$ in the spectra along $x_3$ a clear signature
of discrete wavenumber was detected. This occurrence
suggests that the well organised structures in
figure \ref{fig12}b extend up to the center of the
box. Therefore to understand in greater detail, at this Reynolds number, 
the heat transfer mechanism due to well organised structures for
$SM$, and to a wider range of scales for $WS$  a global picture
of the quantities contributing 
to  the total heat flux ($Q_T$ and $Q_F$) are shown.

The visualizations are evaluated with a single field,
which approximates within a certain measure the averaged
heat transfer. The profiles of the two contributions 
in figure \ref{fig13}a prove that for $SM$
the approximation is better than that for 
$WS$ surfaces, implying a stronger dynamics of the
turbulent component $\overline {Q_T}$ generated by the flow motion
within the roughness elements. This difference in fact is
not observed  in the
profiles of $\overline {Q_T}$ and $\overline {Q_F}$ near the top surface.
This figure emphasizes that the averaged heat fluxes
for the two surfaces  do not largely
differ, with the exception of the thin region near the
plane of the crests. On the other hand when the respective
contributions to the Nusselt number are calculated
($\overline {Nu_T}=\overline {Q_T}/Q_K$ and $Nu_F=\overline {Q_F}/Q_K$
with $Q_K= \frac{\overline {\theta_R}-\overline {\theta_S}}{Re}$)
a large difference between the
$SM$ and the $WS$ surfaces appears in the central region in
figure \ref{fig13}b. This is due to the large
differences in the values of $Q_K$
for the $SM$ and $WS$ surfaces. From these two figures
and in particular from the behavior in the 
near-wall region of $\overline {Q_T}$ and $\overline {Q_F}$ it turns out
that large differences in the distribution of the two
fluxes between the $SM$ and $WS$ surfaces should occur. 
To explain why this happens, the surface
contours of $u_2$, $\theta^\prime$, $u_2\theta^\prime$ and
$\frac{1}{Re}\frac{\partial {\theta}}{\partial x_2}$ 
in figure \ref{fig13} are given. 
The $\theta^\prime$ and $u_2$ surfaces  in
figure \ref{fig13}c and figure \ref{fig13}d 
show that the corrugation in the presence of the smooth surface is
regular and it consist on four thermal plumes due to the 
exponential growth of the secondary instability of the $u_3$ disturbances. 
These disturbances extend in a large region above the wall. 
The rather good correspondence between red and blue regions of $u_2$ and
$\theta^\prime$ cause the formation of  large
regions of positive turbulent heat flux$\theta^\prime u_2$   depicted in 
figure \ref{fig13}f. The blue regions 
have a limited extension. Figure \ref{fig13}e shows that 
the conductive heat flux 
$\frac{1}{Re}\frac{\partial {\theta}}{\partial x_2}$ 
is localised near the smooth wall, with reduced effect of the
spanwise disturbances.
The comparison between the profiles of the averaged and the 
instantaneous contributions emphasizes that at this value of $Re$
in the presence of smooth surfaces the dynamics is rather slow. 
The visualizations of the same quantities for the $WS$
surfaces, on the other hand, show a thicker layer where the disturbances
emanating from the roughness affect the turbulent heat flux.
This is particularly evident by the wider
and taller red regions in figure \ref{fig13}j with respect to those in
figure \ref{fig13}f. The dynamics of this thick layer produces at this
time an overshoot of $\overline {Q_T}$ with respect to the averaged
$Q_T$ linked to a high correlation between $u_2$ and $\theta^\prime$.
It has been found that at a previous time there was an undershoot
caused by  a reduction of the $\theta^\prime u_2$ correlation, in fact 
a reduced dynamics was observed in the profiles
of $\overline {u_2^2}$ and of $\overline {\theta^{\prime 2}}$.
The imprinting of the 
roughness elements is evident in figure \ref{fig13}i, and
not on the distribution of $u_2$ in figure \ref{fig13}g and
of $\theta^\prime$ in figure \ref{fig13}h.

\subsubsection{Rayleigh dependence of the heat fluxes and rms profiles }
 
\begin{figure}
\centering
\hskip -1.0cm
\psfrag{ylab}[][]{$ Nu  $}
\psfrag{xlab}[][]{$Ra $}
\includegraphics[width=11.75cm]{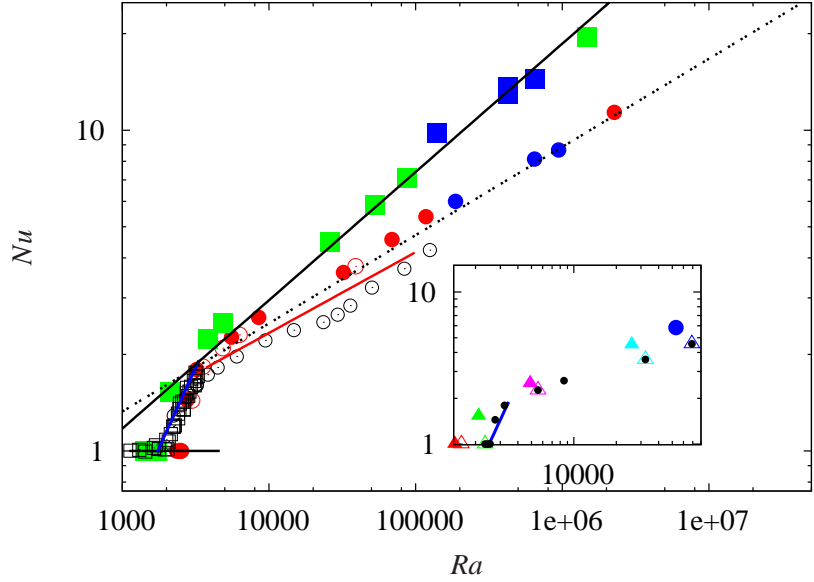}
\caption{Nusselt number versus Rayleigh number for 
$SM$ at $Pr_S=0.134$ (open red circles),
$Pr_S=0.0134$ (solid red circles), 
$SM$ for $Re=250$ at $Pr_S=0.0134$, $Pr_S=0.0134$ and $Pr_S=1.34$ (blue solid circles),
$WS$ at $Pr_S=0.134$ (green solid squares),
$WS$ for $Re=250$ at $Pr_S=0.0134$,  $Pr_S=0.0134$  and $Pr_S=1.34$ (blue solid squares).
The black open circles indicate experimental 
data by Silveston~\citep[p. 69]{chandrasekhar_61}.
The solid black line is $0.0074 Ra^{0.4}$, the dashed
black line is $0.21 Ra^{0.275}$,
the red line is $0.229 Ra^{0.252}$. 
In the inset
the solid triangles are for $WS$ and the open triangles for $SM$
the blue line is $0.0004 Ra^2$.
}
\label{fig14}
\end{figure}

\begin{figure}[ht]
\centering
\hskip -1.0cm
\psfrag{ylab}[][]{$ Nu_F, Nu_T  $}
\psfrag{xlab}[][]{$  $}
\includegraphics[width=7.000cm]{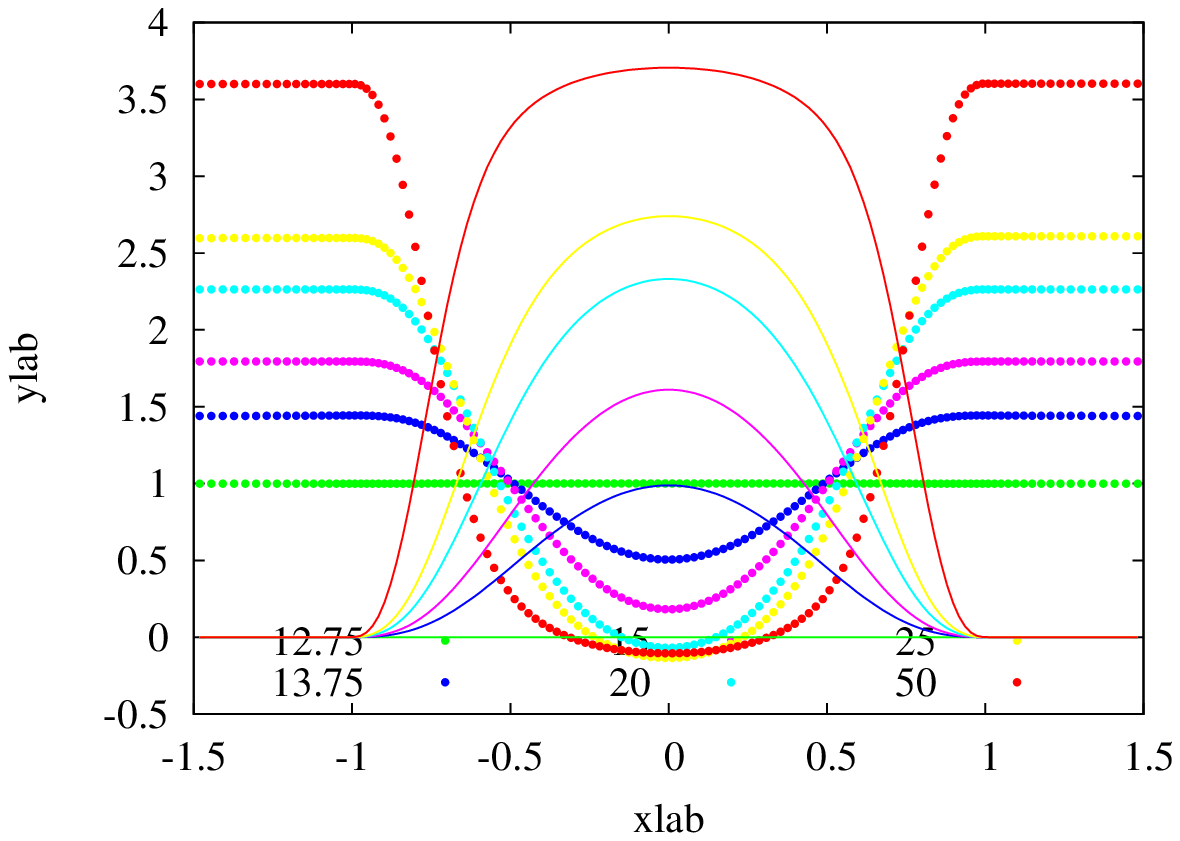}
\hskip -0.2cm
\psfrag{ylab}[][]{$   $}
\psfrag{xlab}[][]{$  $}
\includegraphics[width=7.000cm]{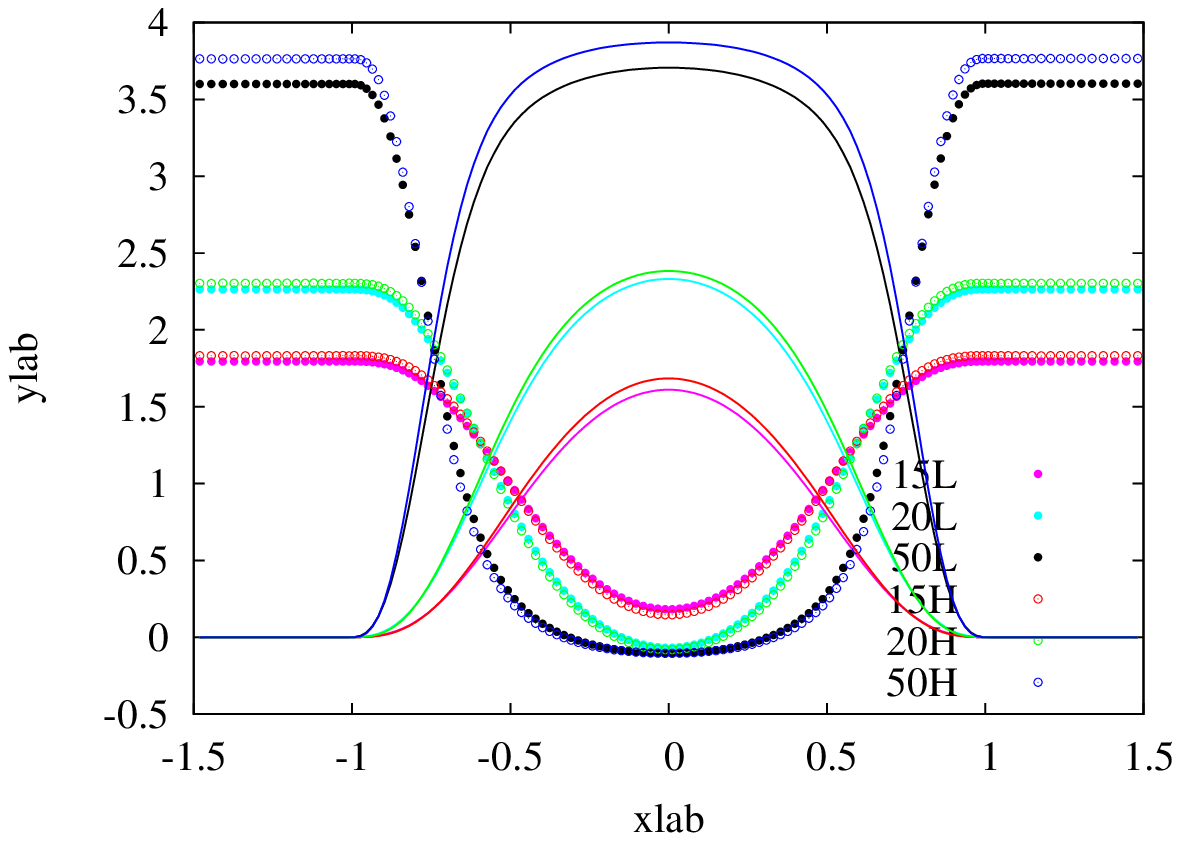}
\vskip -0.55cm
\hskip 5.25cm a) \hskip 6.5cm b)
\vskip -0.25cm
\hskip -1.0cm
\psfrag{ylab}[][]{$ Q_F, Q_T  $}
\psfrag{xlab}[][]{$x_2 $}
\includegraphics[width=7.000cm]{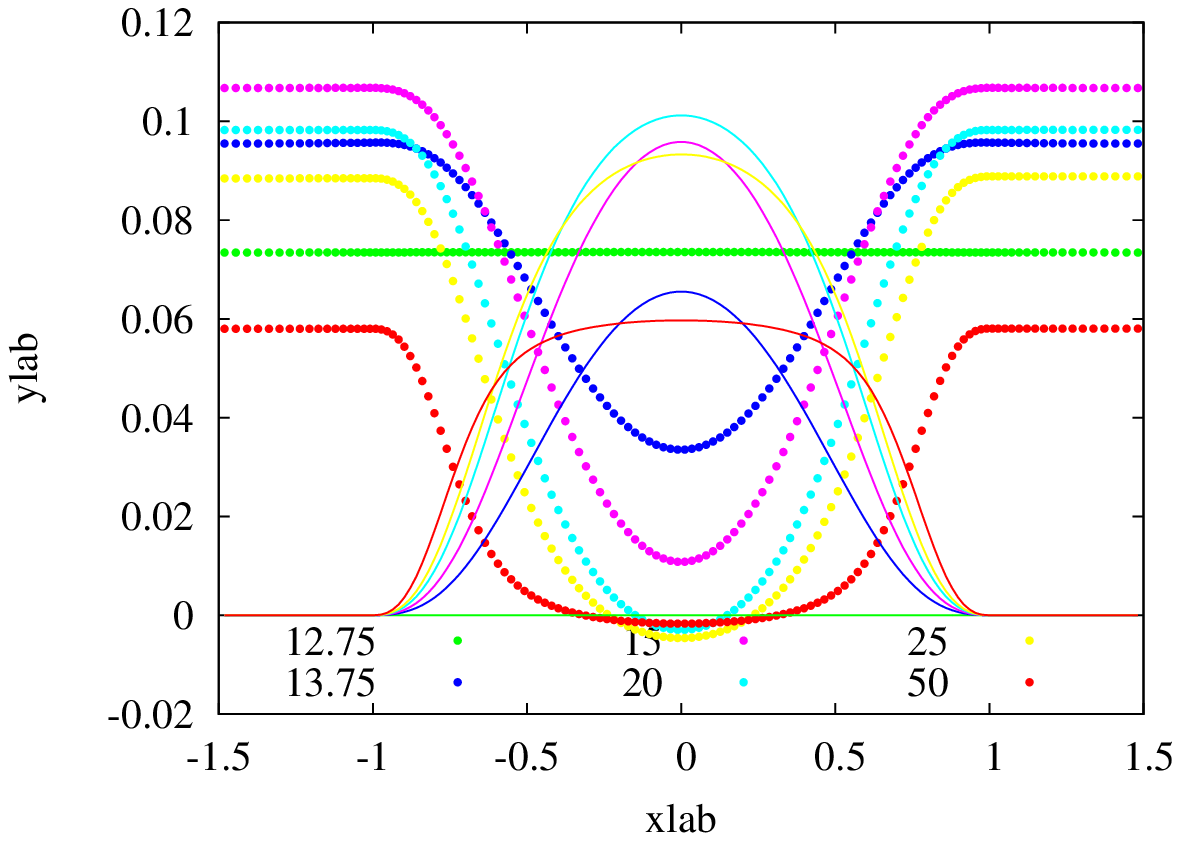}
\hskip -0.2cm
\psfrag{ylab}[][]{$   $}
\psfrag{xlab}[][]{$x_2 $}
\includegraphics[width=7.000cm]{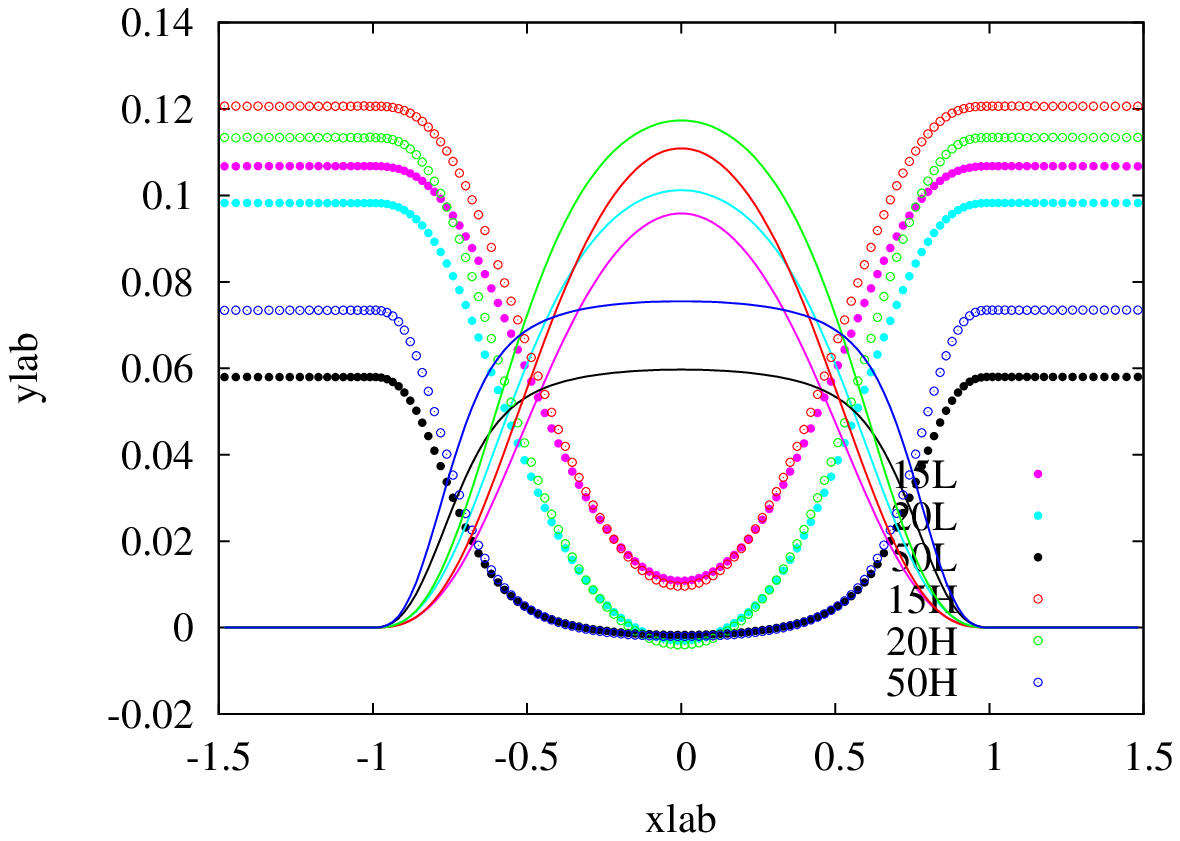}
\vskip -0.55cm
\hskip 5.25cm c) \hskip 6.5cm d)
\caption{Results for $SM$ surface: a) profiles of conductive 
($Nu_F$, solid symbols) and turbulent ($Nu_T$, lines with same color) 
contribution to Nusselt number at low $Re$ for 
low-conducting walls with $Pr_S=0.134$.
b) same data compared with data at $Pr_S=0.0134$ 
(open symbols for $Nu_F$, lines with same color
for $Nu_T$).
Panels c) and d) provide the same information for $Q_F$, $Q_T$.
The values of $Re$ are given in the inset 
($L$ and $H$ stand for $Pr_S=0.134$ and $Pr_S=0.0134$, respectively.
}
\label{fig15}
\end{figure}

From the time evolution of the global rms in figure \ref{fig8} and from  the 
visualizations it is possible to understand the different stages
of natural convection flows starting from a laminar status  by increasing  the
Reynolds number. In the presence of smooth walls
a first transition generates rollers without 
spanwise disturbances, with one or two cells depending
on the $Re$ number. The structures with two cells are less
stable and may merge. A successive transition leads to the
growth of short wavelength disturbances in the spanwise
directions which generate thin layers with high temperature gradients,
the precursors of the intense plumes characteristic
of the turbulent regime. The spanwise disturbances facilitate the
formation  of one cell. 
In presence of rough walls with three-dimensional
elements the growth of the spanwise disturbances is enhanced, and 
therefore the transition to a turbulent regime is faster than 
in the presence of smooth walls. 
The global result of the growth of the instabilities described in
figure \ref{fig8} produces large variations in the values
of the Nusselt number at increasing 
Rayleigh number. In figure \ref{fig14}
several regimes can be detected. In the first regime, for $Ra<Ra_c$, the value
$Nu=1$ is achieved for any type of surface.  For the $SM$ geometry,
at $Ra_c<Ra<3000$,  $Nu$ grows with $Ra$ faster than the linear
trend reported by Silveston~\citep[p. 69]{chandrasekhar_61}. 
The inset in figure \ref{fig14} shows that in 
this range of Rayleigh numbers
the $SM$ and $WS$ data differ, however both show a rapid growth.
The present simulations predict for $SM$ a critical Rayleigh number 
$Ra_c=2500$, and for $WS$ $Ra_c=1800$.
For smooth walls the value of $Ra_c$ differs from the theoretical
value ($Ra=1700$),  as we have assumed a finite width
of the computational box $L_1=L_3=4$. Having two cells in  this box
the most unstable mode is given by $a=2$. The  
theoretical value of $Ra_c$ was obtained for $a=3.117$.
Since the growth depends on the value of $a$, it can be understood
why the two values differ. At higher values of $Ra$  
a  reduction of the growth of $Nu$ versus $Ra$,
for $SM$ and for $WS$, can be appreciated in figure \ref{fig14}
($4000<Ra<10000$). A $Nu\approx Ra^n$ power-law well fits the data
with $n=0.4$ for $WS$ and by $n=0.25$ for $SM$ at higher values of $Ra$. 
From the previous discussion
on the $q_i$ temporal growth (figure \ref{fig8}) 
and on the flow structures reported in the following figures,
it is possible to infer why there is
the  change in slopes around $Ra=4000$ for both surfaces.
To connect the results in figure \ref{fig8} and the
$Nu(Ra)$ relationship, the same colours of figure \ref{fig8} are used 
for the open ($SM$) and for the closed ($WS$) triangles  
in the inset of figure \ref{fig14}.
The visualization in figure \ref{fig9}b
at $Re=20$ ($Ra=5556$) show
a relative strong convective cell advecting $\theta$ and producing 
ascending hot and the descending cold layers. 
This results (see figure~\ref{fig15}a)
in steep mean temperature gradients near the two smooth walls 
and negative $<\theta>$ 
gradients at the center. Negative values of the temperature
gradients at $x_2=0$ do not occur at $Re=13.75$, ($Ra=2758$)
and at $Re=15$ ($Ra=3213$). At these values of $Ra$ the inset of 
figure \ref{fig14} shows fast growth of $Nu$ ($Nu\approx Ra^2$). 
Figure \ref{fig15}a emphasizes that for $Re<20$ the growth 
of $Nu$ is mainly due to 
the contribution of $Nu_F$. In fact, in this figure 
the $Nu_T$ contribution at the centre, due to the convective cell, 
does not overcome $Nu=Nu_S$. On the other hand, starting from 
$Re=20$ the contribution of $Nu_F$ at the center are
negative and therefore $Nu_T>Nu_S$. This explains
the change of the growth rate in the $Nu(Ra)$ relationship.
Figure \ref{fig15}a in addition emphasize that at $Re=25$ ($Ra=8513$)
the negative $Nu_F$ at the centre increases with respect to that at
$Re=20$, and that a further increase of 
the Reynolds number $Re=50$ ($Ra=32215$) reduces the amplitude of the
peak, with a widening of the region with predominance of $Nu_T$. 
\noindent Usually the Nusselt number and the heat flux are
function of the fluid Prandtl and Reynolds numbers, 
($Nu_T=Q_T/Q_K$, with $Q_K= \frac{\Delta T }{RePr_F}$)
because of  the scaling of the temperature with the temperature 
difference between the hot and cold walls.
In the present simulations, which also account for the heat transfer
in the solid, as previously mentioned the $Q_K$ depends
on the parameter of the walls (the thickness, the shape, 
and the values of $Pr_S$) and on the fluid parameters (
$Re$ and $Pr_F$) . Therefore what has been described in  
figure \ref{fig15}a regarding the contributions to the Nusselt number
can not be directly extended to the heat fluxes, $Q_T$ and $Q_F$. The profiles
of the heat fluxes shown in figure \ref{fig15}c have a  complicated behavior, 
in fact, as for the Nusselt profiles, there is a growth up to
$Re=15$. At higher Reynolds number $Q$ decrease, therefore the 
growth of $Nu$ is due to the decrease of $Q_K$. From this
figure it can be further asserted that the change of slope
in figure \ref{fig14} occurs at the Rayleigh number
where the $Q_F$ start to decrease with the Reynolds number.

This behavior of the heat flux or Nusselt number contributions
is  indirectly linked to the values of $Pr_S$. 
The direct dependence is analysed in figure \ref{fig15}b, with 
the quantities calculated at three values of $Re$.
Regards to the contributions to the Nusselt number, figure \ref{fig15}b
shows that only at $Re\approx 50$
a detectable increase of $Nu$ is obtained
with materials of very high conductivity.
Highly conducting materials 
lead to get in figure \ref{fig14} 
the same $Nu(Ra)$ relationship, with the data aligned 
on the same $Ra^n$ fitting line.
The values of $Nu$ are  shifted on the 
right by  reducing  and on the left by increasing  $Pr_S$.
This can be appreciated in figure \ref{fig14} by the blue solid circle 
symbols obtained at $Re=250$, with $Pr_S=1.34$ on the left of the central
with $Pr_S=0.134$ and the one on the right with $Pr_S=0.0134$.
The comparison between figure \ref{fig15}b and figure \ref{fig15}d
shows that the  variation of the conductivity of the walls
at low Reynolds number does not largely affect the
components of the Nusselt number. On the other
hand, a decrease of $Pr_S$ leads to a large increase of
$Q_T$ and $Q_F$.  It is important to note that
this figure indicates a decrease of $Q$ when $Re$ increases.

\begin{figure}[ht]
\centering
\hskip -1.0cm
\psfrag{ylab}[][]{$ <u_1^{\prime 2>} $}
\psfrag{xlab}[][]{$ $}
\includegraphics[width=7.000cm]{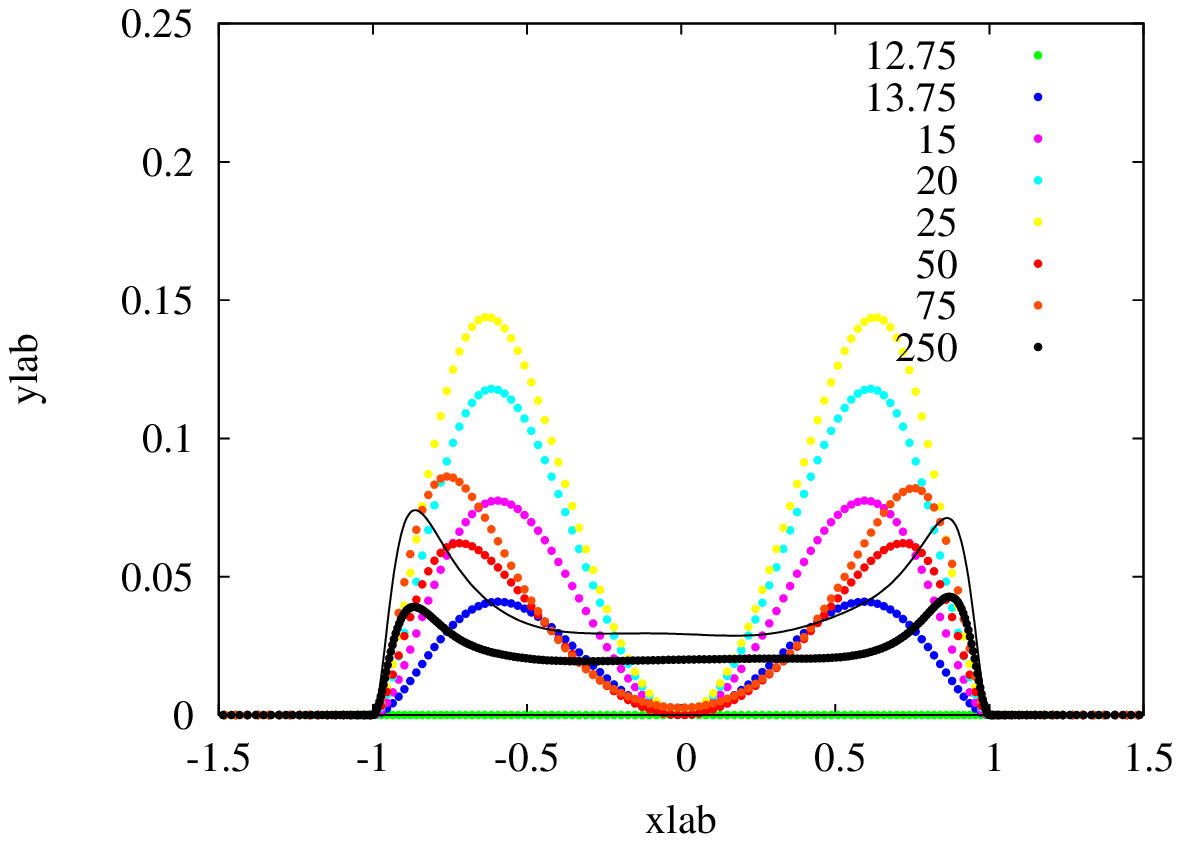}
\hskip -0.2cm
\psfrag{ylab}[][]{$ <u_2^{\prime 2>} $}
\psfrag{xlab}[][]{$ $}
\includegraphics[width=7.000cm]{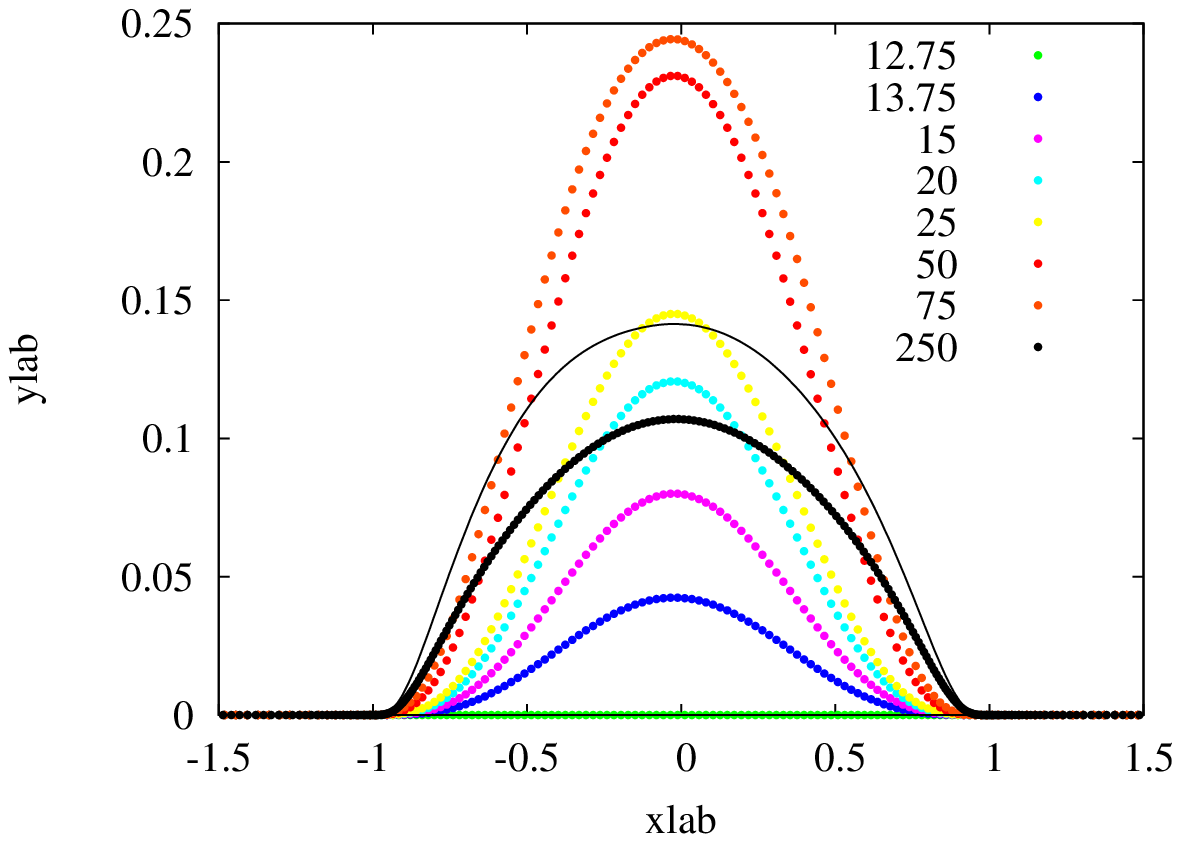}
\vskip -0.55cm
\hskip 5.25cm a) \hskip 6.5cm b)
\vskip 0.15cm
\hskip -1.0cm
\psfrag{ylab}[][]{$ <u_3^{\prime 2>} $}
\psfrag{xlab}[][]{$x_2 $}
\includegraphics[width=7.000cm]{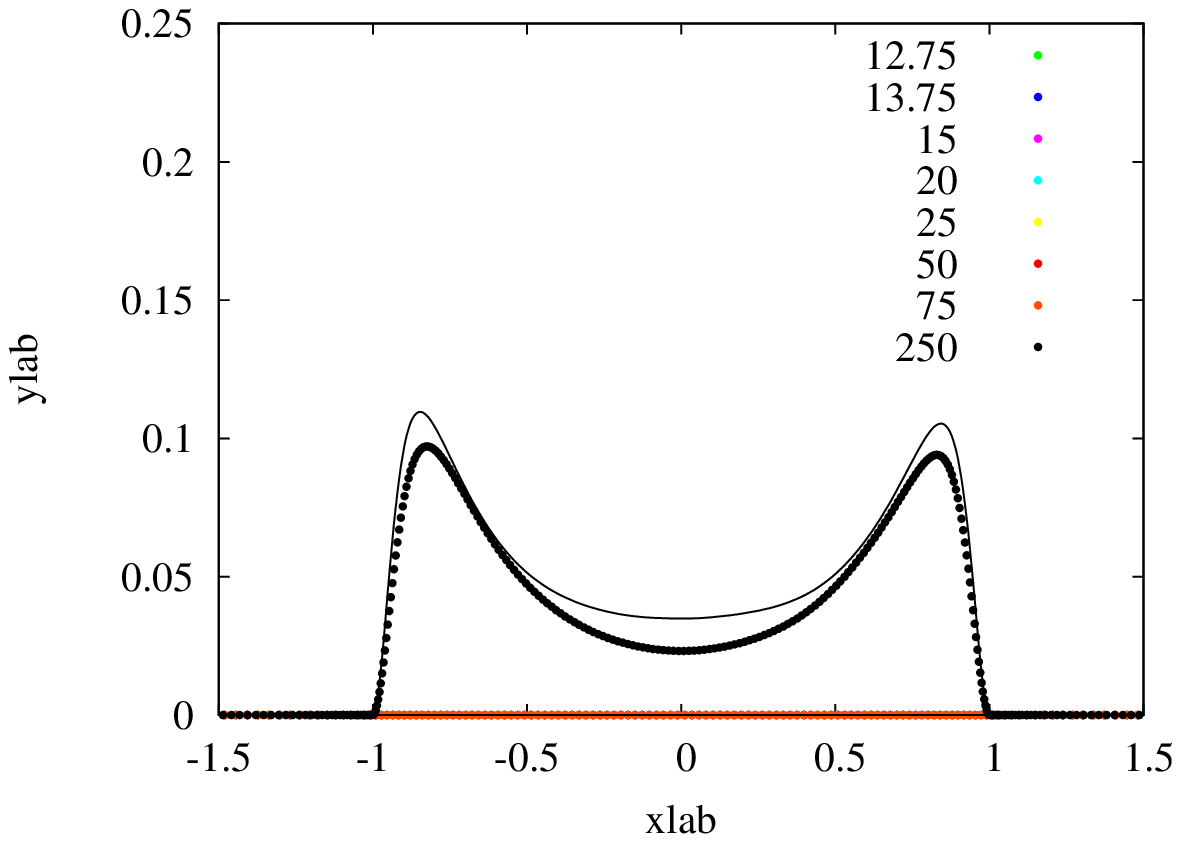}
\hskip -0.2cm
\psfrag{ylab}[][]{$ <\theta^{\prime 2}> $}
\psfrag{xlab}[][]{$x_2 $}
\includegraphics[width=7.000cm]{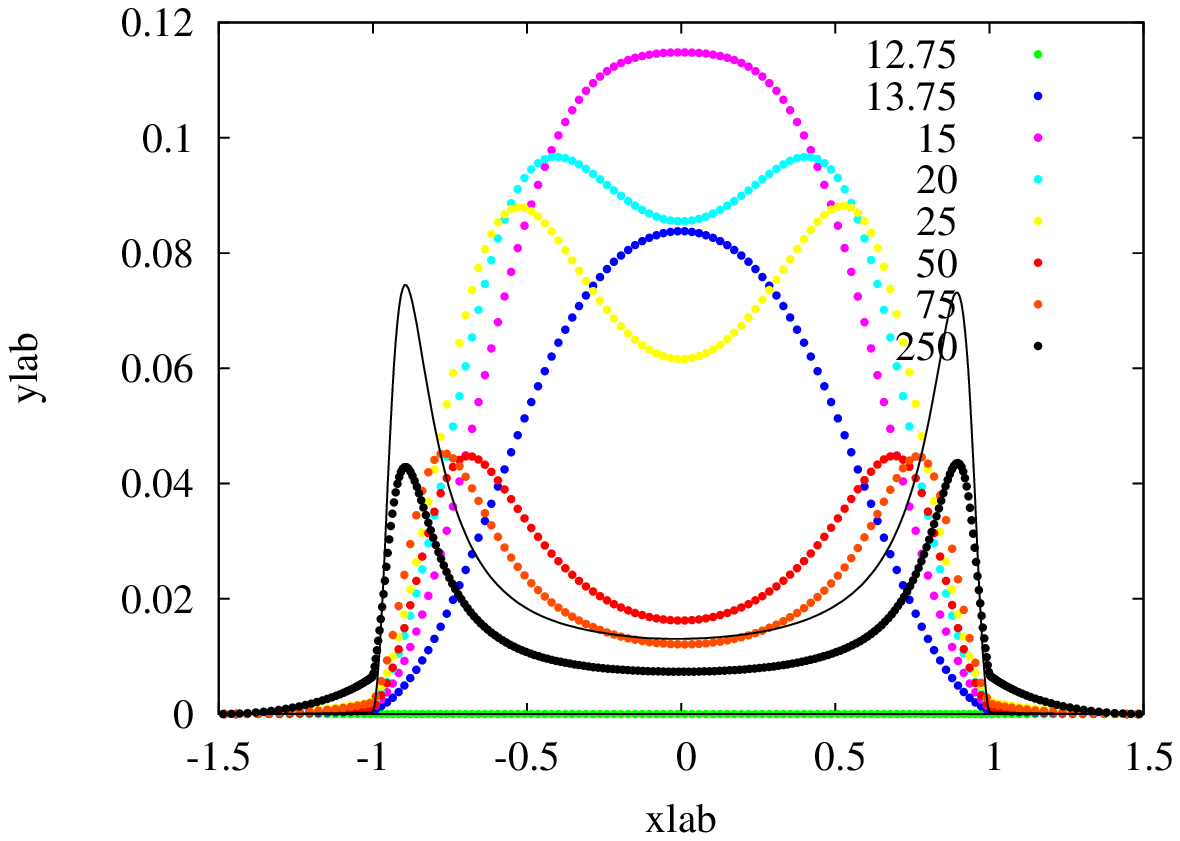}
\vskip -0.55cm
\hskip 5.25cm c) \hskip 6.5cm d)
\caption{Results for $SM$ surfaces: profiles of  velocity and temperature rms at 
different Reynolds number (as listed in the inset)
for $Pr_S=0.134$ (solid symbols), 
and for $Re=250$, $Pr_S=0.0134$ (solid line):
a) $ <u_1^{\prime 2>} $
b) $ <u_2^{\prime 2>} $
c) $ <u_3^{\prime 2>} $
d) $ <\theta^{\prime 2}> $.
}
\label{fig16}
\end{figure}

The two ranges with different growth of the $Nu(Ra)$ relationship
are characterised by different behavior of the profiles 
of $ <u_l^{\prime 2}> $ and $ <\theta^{\prime 2}> $.
The profiles of figure \ref{fig16}b show that for $Ra>Ra_c$,
$ <u_2^{\prime 2}> $ grows everywhere up to
$Re=75$ in a very consistent way. On the other hand,
the profiles of $ <\theta^{\prime 2}> $ in figure \ref{fig16}d
behave differently. In fact, up to $Re=15$ the maximum at the centre 
increases, and, in agreement with the arguments in figure \ref{fig15},
at $Re>15$ the thermal boundary layers are visible, 
with the thickness of the boundary layers 
reducing when $Re$ increases. For high conducting materials ($Pr_S=0.0134$) 
at $Re=250$ the amplitude of the fluctuations
inside the solid reduces and the intensity of the peak near the wall 
increases. Figure \ref{fig16}a shows the formation of
boundary layers for $<u_1^{\prime 2}>$ with thickness reducing by 
increasing the Reynolds number. The large decrease of the $<u_1^{\prime 2}>$
peak going from $Re=25$ to $Re=50$ is due to the formation
of two cells at $Re=50$. As above mentioned two cells are more
unsteady and these  begin to oscillate with an amplitude
linked to the Reynolds number. At $Re=75$ the oscillations 
(figure \ref{fig10}a and figure \ref{fig10}b) are rather
strong and squeeze the cells near the walls producing
an intense peak of $<u_1^{\prime 2}>$ closer to the walls
than that at $Re=50$.
Finally by increasing the Reynolds number
the formation of ordered spanwise disturbances (see figure \ref{fig16}c), 
depicted in figure \ref{fig12} and in figure \ref{fig13}, reach
the central region leading to a more uniform distribution of
$ <u_1^{\prime 2}> $. The analysis of the different behavior
for transitional natural convection in the presence of
smooth wall leads to conclude that the transition from
a pure conductive regime to a convective regime is due
to the formation of two-dimensional cells. In a short range of
Reynolds number the cells are not strong enough to largely
distort the temperature field and therefore there is a fast growth
of $Nu$ with $Ra$. At higher $Ra$ the increase of strength  of the cells,
their unsteadiness and the growth of spanwise disturbances 
produce boundary layers near the walls which  yield
the $Nu=0.21 Ra^{0.275}$ law.

\begin{figure}
\centering
\hskip -1.0cm
\psfrag{ylab}[][]{$ Nu_F, Nu_T  $}
\psfrag{xlab}[][]{$x_2 $}
\includegraphics[width=7.000cm]{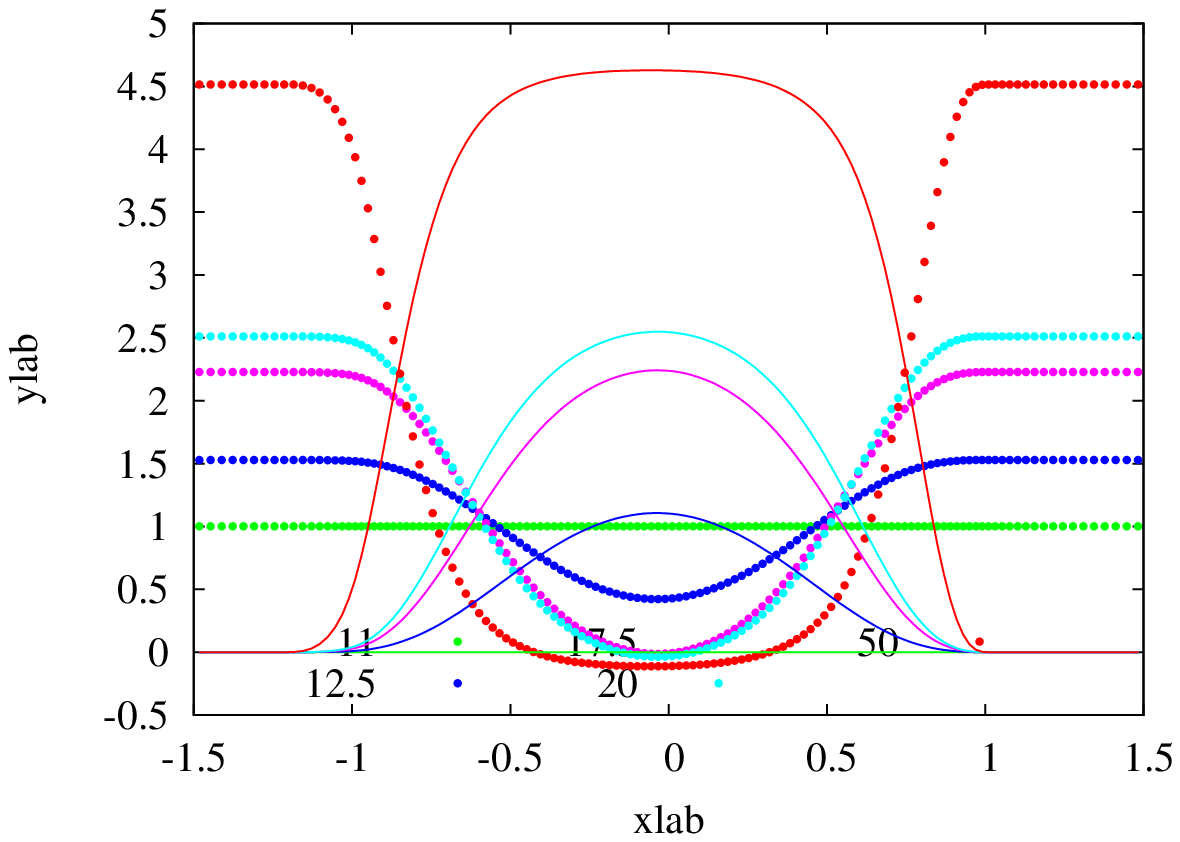}
\hskip -0.2cm
\psfrag{ylab}[][]{$   $}
\psfrag{xlab}[][]{$x_2 $}
\includegraphics[width=7.000cm]{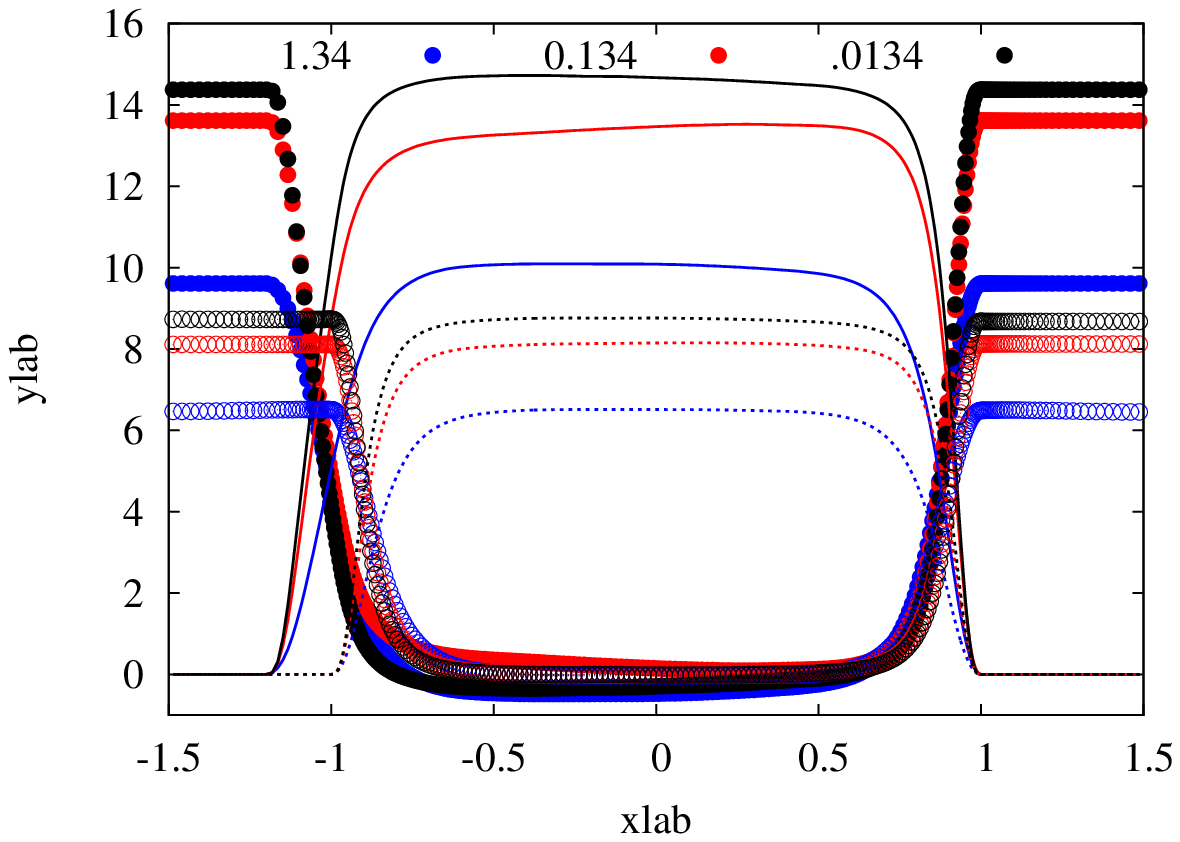}
\vskip -0.55cm
\hskip 5.25cm a) \hskip 6.5cm b)
\caption{a) Profiles of conductive ($Nu_F$, solid symbols) and turbulent ($Nu_T$, lines with same color) 
contribution to Nusselt number $WS$ surfaces at low 
Reynolds number (the values are given in the inset) for low-conducting walls with $Pr_S=0.134$,
b) 
profiles at the two
values of $Pr_S$, $Pr_S=0.134$ closed symbols $Nu_F$, lines with the 
same colour $Nu_T$, $Pr_S=0.0134$ open symbols $Nu_F$, lines with the
same colour $Nu_T$.
}
\label{fig17}
\end{figure}

To investigate why, in figure \ref{fig14}, for $WS$ 
a different relationship $Nu(Ra)$ 
with respect to that for $SM$ was obtained,
the profiles of the  contributions to $Nu$ at $Re$ numbers close
to that for $SM$ are plotted in figure \ref{fig17}a.
Since it has been observed that for $WS$ the same arguments
regard the different behaviors of $Q$ and $Nu$ hold, the figures
with the profiles of the contribution to $Q$ are not presented.
As shown in figure \ref{fig14}, at $Re=12.5$ ($Ra=2104)$ for $WS$ 
$Nu>1$, the  condition for which the convective cells appear.      
The heat fluxes are similar to those for $SM$ at $Re=13.75$ 
($Ra=2590$) in figure \ref{fig16}a.
Also for the $WS$ surface at this $Re$ there is no
change of sign in the mean temperature gradient at the center.
The change of sign occurs at about the same Rayleigh number
for both surfaces, confirming that changes of the slope
of the $Nu(Ra)$ relationship in figure \ref{fig14}, 
occur at the same Rayleigh numbers. From the profiles of the
heat fluxes we also observe that, as for $SM$,
$Q_F$ grows up to $Re=15$, and decreases at  higher $Re$.
At $Re=50$ ($Ra=25790$) figure \ref{fig17}a
further shows that the negative temperature gradient extends over
a wider region, suggesting  that the 
convective cells are stronger and greater than those
generated by smooth walls. The comparison between 
figure \ref{fig17}a and figure \ref{fig15}a highlights that, at $Re=50$, 
the increase from $Nu=3.6$ for $SM$ to $Nu=4.5$ for $WS$
is due to the  decrease of
$Q_T$ and of $Q_K$.
To investigate the effects of the $Pr_S$ of
the solid, and to show that for each value of $Pr_S$ there is
an increase of $Nu$ for $WS$ with respect to that for $SM$,
the components of $Nu$ for the two surfaces are compared, at $Re=250$,
in figure \ref{fig17}b.
The three values of $Pr_S$, previously mentioned, gave the 
blue solid circles for $SM$ and the blue squared symbols for $WS$
in figure \ref{fig14}.
Independently of the values of $Pr_S$, these points are 
aligned with $Nu\approx Ra^{0.275}$ for $SM$ and with
$Nu\approx Ra^{0.4}$ for $WS$ lines.  The profiles
in figure \ref{fig17}b   demonstrate that
at the centre of the box the temperature gradient does not
vary with the type of surface, therefore $Nu_T$ is the 
component driving the changes in $Nu$.

\begin{figure}[ht]
\centering
\hskip -1.0cm
\psfrag{ylab}[][]{$ <u_1^{\prime 2>} $}
\psfrag{xlab}[][]{$ $}
\includegraphics[width=7.000cm]{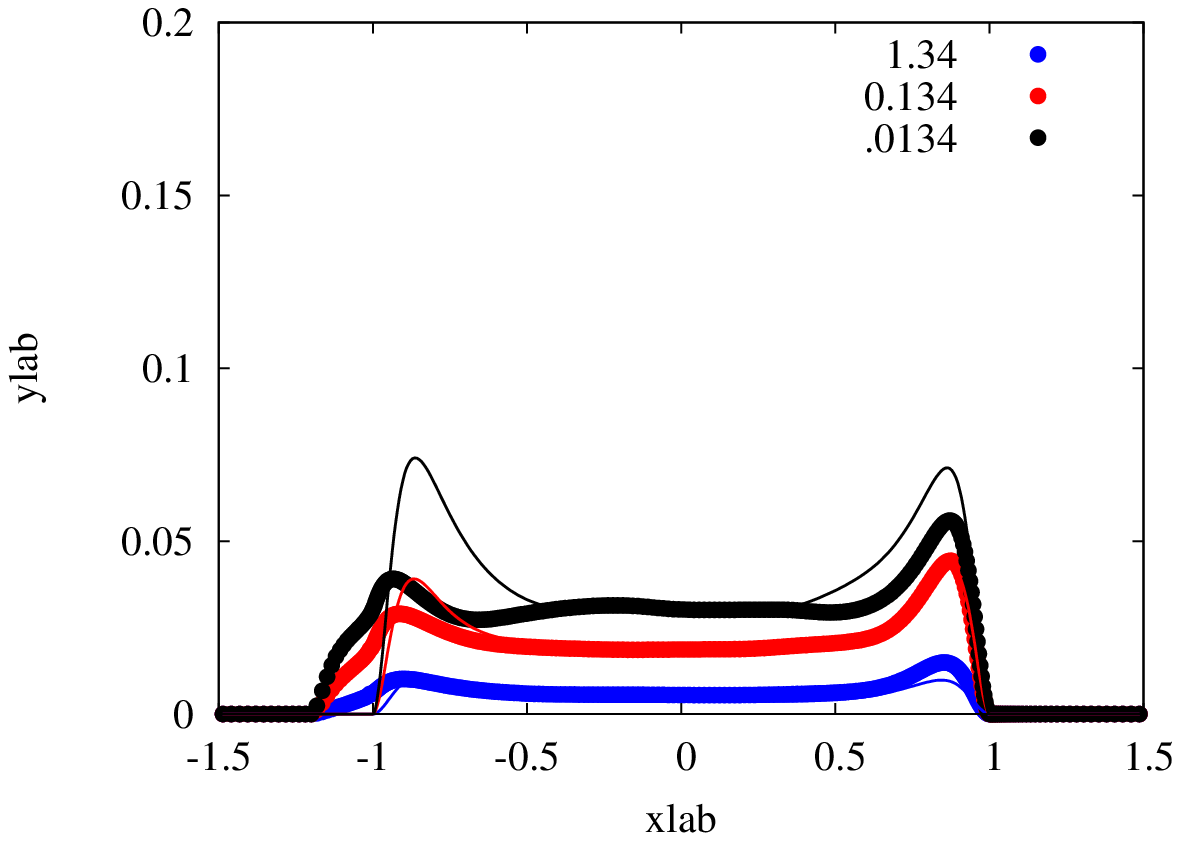}
\hskip -0.2cm
\psfrag{ylab}[][]{$ <u_2^{\prime 2>} $}
\psfrag{xlab}[][]{$ $}
\includegraphics[width=7.000cm]{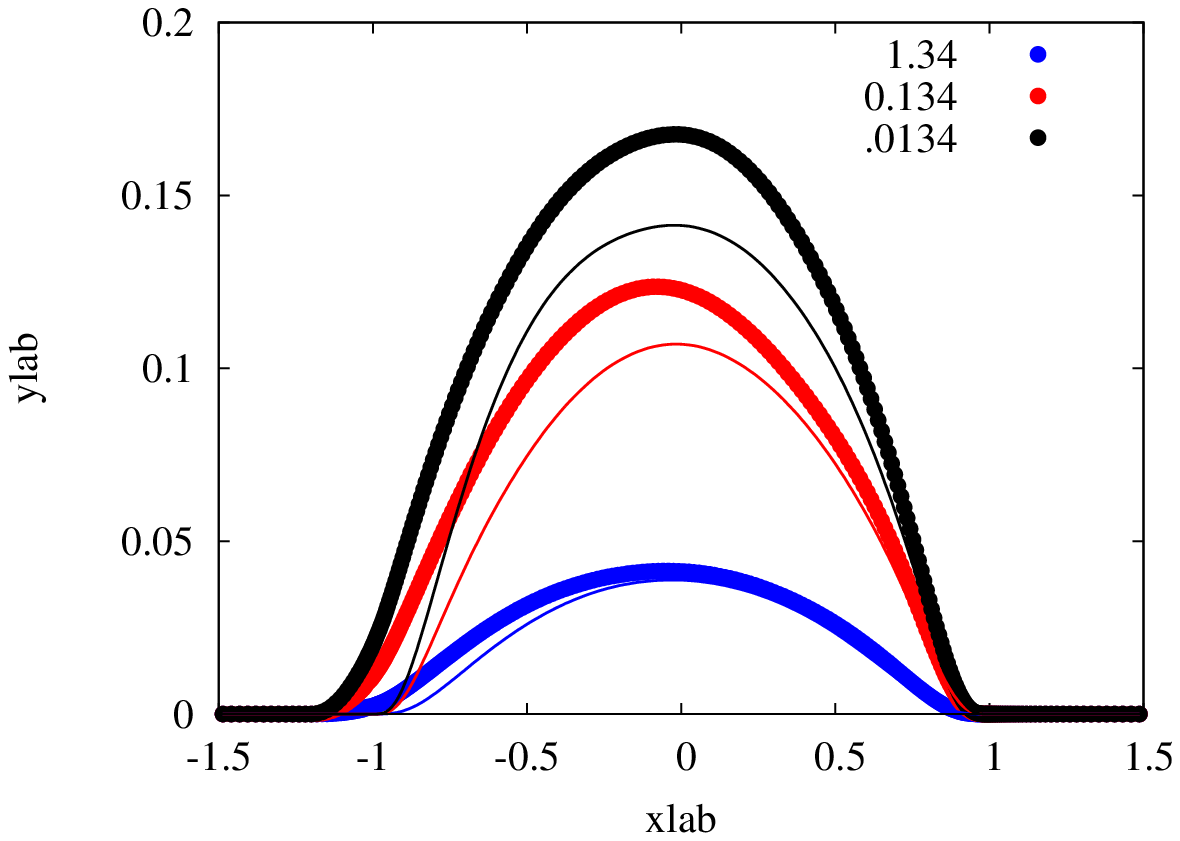}
\vskip -0.55cm
\hskip 5.25cm a) \hskip 6.5cm b)
\vskip 0.15cm
\hskip -1.0cm
\psfrag{ylab}[][]{$ <u_3^{\prime 2>} $}
\psfrag{xlab}[][]{$x_2 $}
\includegraphics[width=7.000cm]{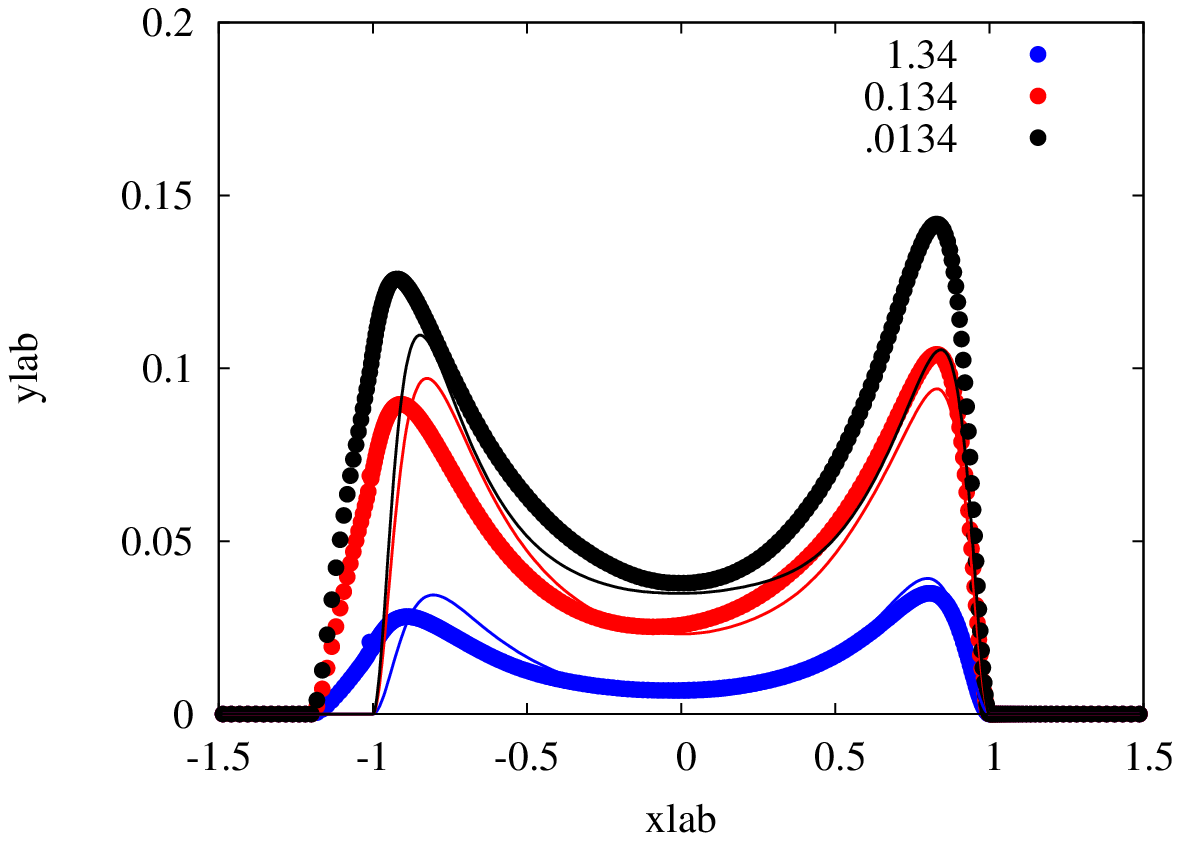}
\hskip -0.2cm
\psfrag{ylab}[][]{$ <\theta^{\prime 2}> $}
\psfrag{xlab}[][]{$x_2 $}
\includegraphics[width=7.000cm]{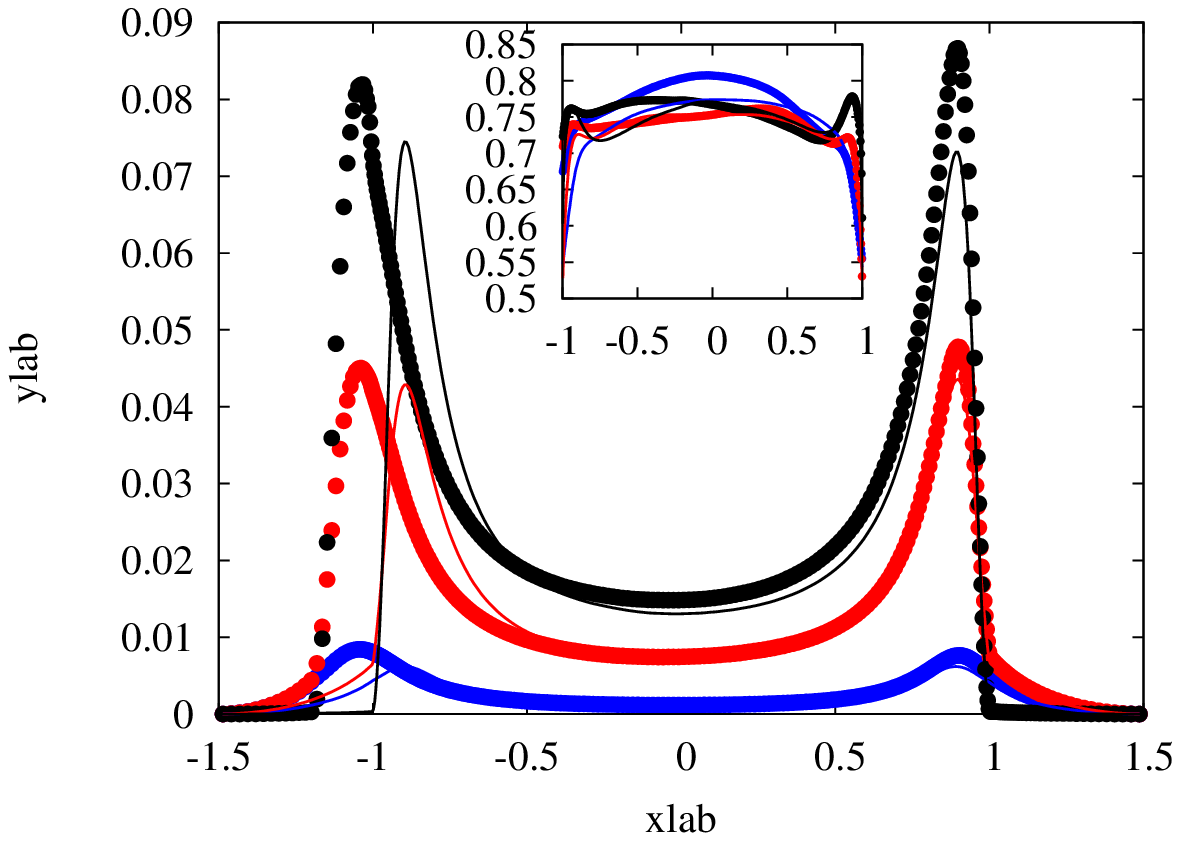}
\vskip -0.55cm
\hskip 5.25cm c) \hskip 6.5cm d)
\caption{
Profiles of velocity and temperature rms for $Re=250$,
at different values of $Pr_T$ (as listed in the inset), for 
$SM$ surfaces (lines) and $WS$ surfaces (solid symbols with same color):
a) $ <u_1^{\prime 2>} $
b) $ <u_2^{\prime 2>} $
c) $ <u_3^{\prime 2>} $
d) $ <\theta^{\prime 2}> $.
}
\label{fig18}
\end{figure}

The profiles of the rms of the velocity and temperature
components  for the $WS$ surfaces in the same range of 
Reynolds number  do not largely differ from those
given in figure \ref{fig16} for the $SM$ surface, 
therefore are not reported. Instead the profiles for both surfaces  at $Re=250$ 
and at different $Pr_S$ are compared in figure \ref{fig18}. The 
first observation is that for  walls made of low-conducting
materials ($Pr_S=1.34$, indicated by blue lines and symbols) 
the rms are rather small. In
particular at $Pr_S=1.34$  the profiles of $<\theta^{\prime 2}>$ 
(figure \ref{fig18}d) near the 
plane of the crest do  not largely  depend on the shape of the wall. 
For $Pr_S=1.34$ the more uniform profile of $<\theta^{\prime 2}>$
implies that the $\theta^{\prime}$
fluctuations are small, consequently small $u_2{\prime}$ fluctuations
are produced.
Despite this large decrease of the $\theta$ and $u_2$ fluctuations,
a large decrease of $Nu_T$ in figure \ref{fig17}b was not found.
This fact can be understood
by looking at the profiles of the correlation coefficients
$C_{v\theta}$ reported in the inset of figure \ref{fig18}d. 
These profiles  are almost constant in
the central region. It is also interesting to see
that the $WS$ surfaces produce greater correlation coefficients near the plane
of the crest, in fact the profiles with solid symbols do not
reduce, as those with lines, in a thin region near the boundary
at $x_2=-1$. The trend near the top wall suggests 
that for smooth wall a small thick layer is necessary to
have a high correlation  
between $\theta^\prime$ and $u_2$ fluctuations.
This was previously visualised in figure \ref{fig13}a and
in figure \ref{fig13}b.
The independence of $C_{v\theta}$ on the solid conductivity
indicates that small temperature fluctuations, for the effect
of the buoyancy term in the transport equation of $U_2$, 
generate $u_2$ fluctuations. This
process requires the formation of flow structures.
The flow around the roughness elements facilitates the formation of structures,
therefore at the plane of the crests 
$\theta^\prime$ and $u_2$ fluctuations are well correlated.
In presence of smooth walls the flow structures
form  in a thin thermal boundary layer.
The orientation of the large rollers near the surface causes the
formation of profiles of $<u_3^{\prime 2}> $ with amplitude
greater than that of $<u_1^{\prime 2}>$ in the presence of the
$WS$ surface, in fact as it was  observed in figure \ref{fig12}d
and in figure \ref{fig13}b through the fluctuating temperature
visualizations there is a preferential motion along the $x_3$ direction.

\section{Conclusions}

The interest in natural convention has been focused
on laboratory, numerical experiments and theory,
mainly to understand the variations of the Nusselt number
at high values of the Rayleigh number in the presence of smooth
walls. Also the difference between smooth and rough surfaces
was investigated at high values of the Rayleigh number. In the
experiments the control of the flow parameters, and the
measurements have limitations. On the other hand,
in the numerical simulations, once the numerics has been
validated, any quantity can be measured and all the
parameters can be exactly controlled. This is of
particular importance at low $Ra$, when
velocity and temperature fluctuations are small.
Usually in the numerical simulations the temperature
is assigned at the wall, therefore there is a
close relationship between the Nussuelt number and
the total heat flux. This is no longer true when
heat transfer within the solid walls is considered, and
therefore the difference between the hot and cold
walls varies with the Reynolds and Prandtl number of the
flow, and  with the thermal conductivity of the material
of the solid. In these condition is not granted that
to an increase of $Nu$ corresponds an increase of the
total heat flux. 

To understand in detail the effect of the shape of
surfaces is important to consider several geometries.
In this paper the results obtained by five different 
geometries have been compared with those 
of smooth walls at intermediate values of the Rayleigh
number ($Ra=O(10^7)$).
The thickness and the thermal conductivity
of the solid, here assumed to be that of the copper,
have been kept fixed. The conclusion has been that surfaces with
three-dimensional elements are more efficient to achieve
high $Nu$. In particular, it has been found
that surfaces constituted by staggered wedges are more efficient
than cubic staggered elements. Flow visualizations
allow to understand why different results were
achieved by changing the shape of the surfaces.
In addition velocity and temperature spectra in the homogeneous
direction demonstrated that the velocity spectra at the center of
the box did not differ from those of forced isotropic turbulence.
Differences, were found instead for the temperature
spectra. This is less important at low,
but it should be interesting at high  Rayleigh numbers.
It was also found that the shape of the
surfaces affects the small flow scales in a thin layer near the
plane of the crests. This effect is rapidly lost, and
at relatively short distance from the plane of the crests 
the small scale do not differ from those of isotropic turbulence.

The comprehension of the transition from a
pure conducting state, without flow and thermal structures, 
to an organized stage with, did not attract large interest in the past. 
This motivated
the present study to perform a large number of simulations for smooth
and for surfaces with three-dimensional wedges.
Flow visualizations and statistics 
of the conductive and turbulent contributions to the heat transfer 
have allowed to explain the observed
transition from an initial rapid to a reduced growth of $Nu(Ra)=Ra^n$.
The value of $Ra$ where $n$ sharply decreases does
not depend on the shape of the surfaces. Instead the
values of $n$ after this "secondary" transition 
slightly depends on the shape of the surfaces. In the range 
$4000<Ra<10000$ where this secondary transition occurs, spanwise
disturbances did not grow, therefore the small
differences between the two surfaces are due to the
three-dimensional disturbances present in the near-wall 
region not strong enough to modify the
shape of the large-scale roller filling the entire gap.
Further increasing the Rayleigh numbers a tertiary transition,
given by the formation of complex three-dimensional
structures, was observed. In this range of Rayleigh
numbers the values of $n$ for smooth walls is $n=0.275$,
that for the wedges is $n=0.4$. The main difference
between the two surfaces consists in the formation
of a thermal boundary layer in the case of smooth walls. In
this layer the turbulent heat flux $< u_2 \theta^\prime>$
grows and the conductive flux $\frac{1}{Re}
\frac{\partial <\theta>}{\partial x_2}$ is reduced.
In presence of rough surfaces this behavior
is no longer present, therefore the prevalence
of the turbulent heat flux promotes the higher value $n=0.4$.
It has been also found by simulations with different
values of the thermal conductivity of the solids that
the values of $n$ do not depend on this parameter.
On the other hand it has been observed that large effects on
the heat fluxes appear.

Simulations at much higher values of $Ra$ up to $Ra\approx 10^{11}$
have also been performed, but were not presented here because
our view is that large changes in the flow physics occur
in the transitional regime. The results at high $Ra$ will be reported
in a forthcoming paper.

\bibliography{references}

\end{document}